\documentstyle[onecolumn,epsf,afterpage]{mn}

\title[Peculiar Motions of Early-Type Galaxies. VI.]
{The Peculiar Motions of Early-Type Galaxies in Two Distant
Regions VI: The Maximum Likelihood Gaussian Algorithm}

\author[R.P. Saglia et al.]{R.P.~Saglia$^1$, Matthew Colless$^2$, 
David Burstein$^3$, Roger L. Davies$^4$, 
\newauthor 
Robert K. McMahan, Jr.$^5$ and Gary Wegner$^6$\\
$^1$ Institut f\"ur Astronomie und Astrophysik,
Scheinerstra\ss e 1, D-81679 Munich, Germany\\
$^2$ Research School of Astronomy and Astrophysics, The Australian National
University, Weston Creek, ACT 2611, Australia\\
$^3$ Dept of Physics and Astronomy, Arizona State University, 
Tempe, AZ 85287-1504\\
$^4$ Dept of Physics, University of Durham, South Road, Durham, DH1 3LE, UK\\
$^5$ Dept of Physics and Astronomy, University of North Carolina,
CB\#3255 Phillips Hall, Chapel Hill, NC 27599-3255\\
$^6$ Dept of Physics and Astronomy, Dartmouth College, Wilder Lab.,
Hanover, NH 03755}

\newlength{\plotwidth}
\newlength{\fullwidth}
\newcommand{\mgb}{\hbox{Mg$b$}}
\newcommand{\mgbp}{\hbox{Mg$b^\prime$}}
\newcommand{\mgtwo}{\hbox{Mg$_2$}}
\newcommand{\mgsig}{\hbox{Mg--$\sigma$}}
\newcommand{\mgbpsig}{\hbox{\mgbp--$\sigma$}}
\newcommand{\mgtwosig}{\hbox{\mgtwo--$\sigma$}}
\newcommand{\mgmg}{\hbox{\mgtwo--\mgbp}}
\newcommand{\Dnsig}{\hbox{$D_n$--$\sigma$}}
\newcommand{\logsig}{\mbox{$\log\sigma$}}
\newcommand{\logDn}{\mbox{$\log D_n$}}
\newcommand{\logRe}{\mbox{$\log R_e$}}
\newcommand{\logDw}{\mbox{$\log D_W$}}
\newcommand{\logDwO}{\mbox{$\log D_W^0$}}
\newcommand{\SBe}{\mbox{$\langle S\!B_e \rangle$}}
\newcommand{\meanRe}{\mbox{$\overline{\log R_e}$}}
\newcommand{\meansig}{\mbox{$\overline{\log\sigma}$}}
\newcommand{\meanSBe}{\mbox{$\overline{\langle S\!B_e \rangle}$}}
\newcommand{\sigone}{\hbox{$\sigma_1$}}
\newcommand{\sigtwo}{\hbox{$\sigma_2$}}
\newcommand{\sigthree}{\hbox{$\sigma_3$}}
\newcommand{\meanmgtwo} {\mbox{$\overline{\hbox{Mg}_2}$}}
\newcommand{\meanmgbp}{\mbox{$\overline{\hbox{Mg}b^\prime}$}}
\newcommand{\DWcut}{\mbox{$\log D_{W{\rm cut}}$}}
\newcommand{\FPcut}{\mbox{$F\!P_{\rm cut}$}}

\newcommand{\Deltadelta}{\mbox{$<\Delta\delta>$}}

\newcommand{\dj}{\mbox{$\delta_j$}}

\newcommand{\dex}{\mbox{${\rm dex}$}}

\newcommand{\plotone}[1]
           {\centering \leavevmode \epsfxsize=\plotwidth \epsfbox{#1}}

\begin{document}

\maketitle

\begin{abstract}
The EFAR project is designed to measure the properties and peculiar
motions of early-type galaxies in two distant regions.  Here we
describe the maximum likelihood algorithm we developed to investigate
the correlations between the parameters of the EFAR database. One-,
two-, and three-dimensional gaussian models are constructed to
determine the mean value and intrinsic spread of the parameters, and
the slopes and intrinsic parallel and orthogonal spread of the \mgmg,
\mgtwosig, \mgbpsig\ relations, and the Fundamental Plane.  In the
latter case, the cluster peculiar velocities are also determined. We
show that this method is superior to ``canonical'' approaches of
least-squares type, which give biased slopes and biased peculiar
velocities. We test the algorithm with Monte Carlo simulations of mock
EFAR catalogues and derive the systematic and random errors on the
estimated parameters. We find that random errors are always
dominant. We estimate the influence of systematic errors due to the
way clusters were selected and the hard limits and uncertainties in
the selection function parameters for the galaxies. We explore the
influence of uniform distributions in the Fundamental Plane parameters
and the errors. We conclude that the mean peculiar motions of the EFAR
clusters can be determined reliably. In particular, the placement of
the two EFAR sample regions relative to the Lauer \& Postman dipole
allows us to strongly constrain the amplitude of the bulk motion in
this direction. We justify a posteriori the use of a gaussian modeling
for the galaxy distribution in the Fundamental Plane space, by showing
that the mean likelihood of the EFAR sample is obtained in 10 to 30\%
of our simulations.  We derive the analytical solution for the maximum
likelihood gaussian problem in N dimensions in the presence of small
errors.
\end{abstract}


\section{Introduction}
\label{introduction}

The EFAR collaboration (Wegner et al. 1996; Paper I) has collected
photometric (Saglia et al. 1997a, 1997b; Paper III and IV) and
spectroscopic (Wegner et al. 1999; Paper II) data for galaxies in
clusters with the primary goal of using the tight correlations between
the global properties of early-types (the Fundamental Plane, hereafter
FP, and
the \Dnsig\ relation, Djorgovski \& Davis 1987, Dressler et al. 1987)
to measure the peculiar motions and the mass distribution on large
scales. Colless et al. (2000, Paper VII) discuss the results of these
investigations and the
interpretation of the peculiar motions in the context of cosmological
models. This paper presents the methodology adopted to derive these
results and the Monte Carlo simulations performed to test them and
estimate the uncertainties and residual systematic biases. Colless et
al. (1999, Paper V) discuss the implications of the EFAR \mgsig\
relation. 

An abundant literature is dedicated to the problem of how to best determine
regression lines from incomplete datasets subject to errors and
explicit selection criteria. 
Regression algorithms involving two variables are discussed
by Isobe et al. (1990) and Feigelson \& Babu (1992), who provide
least-squares fits and their uncertainties. Akritas \& Bershady (1996)
address the problem of linear regressions with errors and intrinsic
scatter, by considering the variance of the variables involved.  
La Barbera, Busarello and Capaccioli (2000) extend this work to the
analysis of the FP equation. 
The canonical methods used to find the coefficients of the
FP (or of correlations between two variables) are of the least-squares
type: one minimizes the orthogonal (absolute or squared) residuals
from the plane (J\o rgensen et al. 1996, J\o rgensen 1997, Pahre et
al. 1997, Scodeggio et al. 1997) or the (squared) residuals from one of
the variables (Hudson et al. 1997). Principal component analysis has
also been used (Carvalho \& Djorgovski 1992). Ample discussion of problems
related to selection is given by Willick (1994) and Teerikorpi (1997),
in the context of the
determination of peculiar velocity fields.  However, none of these
approaches is able to deal effectively with the multiple problems
that the EFAR and similar datasets pose: (i) the factor two in
redshift spanned,
(ii) the strong selection effects, (iii) the non-negligible and widely
varying measurement errors, (iv) the presence of defined selection
criteria, (v) the
intrinsic scatter of the relation. In this
paper we demonstrate that ``classical'' methods fail when confronted
with the EFAR dataset, and we quantify the biases they produce. We
construct an algorithm based on Maximum Likelihood (ML) gaussian modelling
and test it with Monte Carlo simulations. 
We demonstrate that it solves the problems posed by the dataset,
producing nearly unbiased estimates of the parameters involved.

The paper is organized as follows. \S\ref{mlalgo} discusses
critically the
results of linear regression analysis and presents the ML 
algorithm for the one-, two- and three-dimensional cases.  
\S\ref{montecarlo} describes the method to generate mock catalogues of
the EFAR database, the tests of the ``canonical'' least-square
methods and of the ML gaussian algorithm, and the estimates of the errors on
the derived parameters. Simulations backing the results presented in
Papers II, V and VII are discussed for the  \mgmg, \mgbpsig, \mgtwosig\
relations, the Fundamental Plane and the cluster peculiar velocities. 
Conclusions are drawn in 
\S\ref{conclusions}. The Appendices give the analytical solution of the 
N-dimensional gaussian maximum
likelihood problem in the presence of small errors and some results used in
the paper.

\section{The Maximum Likelihood Gaussian Algorithm}
\label{mlalgo}

After reviewing the results of the standard least-squares analysis 
in \S\ref{linear}, we describe the properties of the EFAR database in 
\S\ref{efarsample} to argue that a ML approach is needed
to study the sample. We set the equations of the problem for the
general case in \S\ref{general} and address the simplified cases
when one (\S\ref{onedim}), two (\S\ref{twodim}) or three (\S\ref{threedim}) 
variables are considered.

\subsection{Linear models}
\label{linear}

Before discussing the specific problems related to the EFAR database
and the solutions we have developed to address them, it is worth 
considering the more general question of how to model the correlations
existing between the structural parameters of galaxies. As usual
in astrophysics, logarithmic quantities are considered in order to
avoid scaling problems. In the simplest approach, one focuses on pairs
of datapoints $\{x_i,y_i\}$, looking for a linear relation $y=ax+b$
between them. Table 1 of Isobe et al. (1990) (see also
Eqs. \ref{eqyx}-\ref{eqao} below) summarizes the estimates
of the slopes $a$ and  zeropoints $b$, and their variances, derived using
five methods (Y-X and X-Y regressions, bisector, reduced major axis
and orthogonal
regression), when measurement errors are
negligible. Briefly, we recall that the Y-X regression derives the
slope $a_{YX}$ and zero point $b_{YX}$ of the line $y=a_{YX}x+b_{YX}$ 
by minimizing the quantity:
\begin{equation}
\label{eqchiyx}
\chi^2_{YX}=\Sigma_i (y_i-a_{YX}x_i-b_{YX})^2,
\end{equation}
and the X-Y regression by minimizing the analogous $\chi^2_{XY}$. The
bisector solution gives the slope and zeropoint of the line passing in
between the Y-X and X-Y lines. The reduced major axis slope is the
geometric mean of the Y-X and X-Y slopes. 
The orthogonal regression derives $a_O$
and $b_O$ by minimizing the orthogonal residuals from the line:
\begin{equation}
\label{eqchio}
\chi^2_{YX}=\Sigma_i \frac{(y_i-a_{O}x_i-b_{O})^2}{1+a_O^2}.
\end{equation}
If the differences in slopes and zeropoints between the
methods are smaller than the expected variance, a well-defined answer
is found. Unfortunately, this is rarely the case in
astronomy. Therefore, Isobe et al. (1990) recommend astronomers 
use the Y-X regression when it is clear that Y is the variable to be
``predicted'', and the bisector method when the functional relation
between the variables has to be investigated. They discourage the use
of the orthogonal regression, since its slope has greater
dispersion. The problem is complicated when errors (on one or both variables)
are present. In particular, Feigelson and Babu (1992) conclude that
``models incorporating both measurement errors and intrinsic scatter
are complex and not yet fully developed''. Akritas and Bershady (1996)
improve on this aspect, presenting an updated version of Table 1 of
Isobe et al. (1990), where corrections for measurement errors are
taken into account. Isobe et al. (1990), Feigelson and Babu (1992) and
Akritas and Bershady (1996) stress that ``there is no such thing as true
slope'', and that the regression method should be chosen according to
the problem to be solved.

These considerations become more transparent when the underlying
bivariate probability distributions $P(x,y)$ are examined. Assuming normality,
$P(x,y)$ can always be written as:
\begin{equation}
\label{eqbivariate}
P(x,y)=\frac{1}{2\pi\sigma_x\sigma_y\sqrt{1-\rho^2}}
\exp\left\{-\frac{1}{2(1-\rho^2)}\left[\frac{(x-\overline{x})^2}{\sigma_x^2}
-\frac{2\rho(x-\overline{x})(y-\overline{y})}{\sigma_x\sigma_y}+\frac{(y-\overline{y})^2}{\sigma_y^2}\right]
\right\} ,
\end{equation}
where $\overline{x}$ and $\overline{y}$ are the mean values of $x$ and
$y$, $\sigma_x$ and $\sigma_y$ the standard deviation (rms scatter), 
and $\rho$ the correlation
coefficient defined as $\sigma_{xy}=\rho\sigma_x\sigma_y$. 
Calling $\hat x=x-\overline{x}$ and $\hat
y=y-\overline{y}$, Eq. \ref{eqbivariate} can be cast as:
\begin{equation}
\label{eqbivlambda}
P(x,y)=\frac{|\Lambda|^{1/2}}{2\pi} 
\exp \left\{ -\frac{1}{2}
\left[ \Lambda_{xx}\hat x^2 +2\Lambda_{xy}\hat{x} \hat{y}
       +\Lambda_{yy}\hat y^2\right]
\right\} ,
\end{equation}
where $\Lambda=V^{-1}$,
$|\Lambda|=\Lambda_{xx}\Lambda_{yy}-\Lambda_{xy}^2=[(1-\rho^2)\sigma_x^2\sigma_y^2]^{-1}$ 
and $V$ is the covariance matrix:
\begin{equation}
\label{eqvariance}
V=\left(\begin{array}{cc}
\sigma_x^2 & \rho \sigma_x\sigma_y \\
\rho \sigma_x\sigma_y & \sigma_y^2 \\
\end{array}\right) , 
~~~\Lambda=\frac{1}{1-\rho^2}\left(\begin{array}{cc}
1/\sigma_x^2 & -\rho /(\sigma_x\sigma_y) \\
-\rho /(\sigma_x\sigma_y) & 1/\sigma_y^2 \\
\end{array}\right) .
\end{equation}
Eqs. \ref{eqbivariate} and \ref{eqbivlambda} are equivalent 
to the three following formulae:
\begin{equation}
\label{eqyx}
P(x,y)=\frac{1}{\sqrt{2\pi}\sigma_x}
\exp\left(-\frac{\hat x^2}{2\sigma_x^2}\right)
\frac{1}{\sqrt{2\pi}\sigma_y\sqrt{1-\rho^2}}
\exp\left[ -\frac{1}{2}
\frac{\left(\hat y-\frac{\rho\sigma_y}{\sigma_x}\hat x\right)^2}
{(1-\rho^2)\sigma_y^2} \right] ,
\end{equation}
\begin{equation}
\label{eqxy}
P(x,y)=\frac{1}{\sqrt{2\pi}\sigma_y}\exp\left(-\frac{
\hat y^2}{2\sigma_y^2}\right)
\frac{1}{\sqrt{2\pi}\sigma_x\sqrt{1-\rho^2}}\exp\left[ -\frac{1}{2}
\frac{\left(\hat x-\frac{\rho\sigma_x}{\sigma_y}\hat y\right)^2}
{(1-\rho^2)\sigma_x^2}\right] ,
\end{equation}
\begin{equation}
\label{eqortho}
P(x,y)=\frac{1}{\sqrt{2\pi}\sigma_1}\exp\left[-\frac{(\hat y-a_o\hat
x)^2}{2(1+a_o^2)\sigma_1^2}\right]
\frac{1}{\sqrt{2\pi}\sigma_2}\exp\left[-\frac{(a_o\hat y+\hat
x)^2}{2(1+a_o^2)\sigma_2^2}\right] ,
\end{equation}
where:
\begin{equation}
\label{eqao}
a_o=\frac{1}{2\rho\sigma_x\sigma_y}\left[\sigma_y^2-\sigma_x^2
+\sqrt{(\sigma_y^2-\sigma_x^2)^2+4\rho^2\sigma_x^2\sigma_y^2}\right] ,
\end{equation}
\begin{equation}
\label{eqsigma1}
\sigma_1^2=\frac{1}{2}(\sigma_x^2+\sigma_y^2+\sqrt{(\sigma_y^2-\sigma_x^2)^2+4\rho^2\sigma_x^2\sigma_y^2}) ,
\end{equation}
\begin{equation}
\label{eqsigma2}
\sigma_2^2=\frac{1}{2}(\sigma_x^2+\sigma_y^2-\sqrt{(\sigma_y^2-\sigma_x^2)^2+4\rho^2\sigma_x^2\sigma_y^2}).
\end{equation}

Eq. \ref{eqyx} shows that a gaussian bivariate probability
distribution can be generated extracting first the $x$ variable around
its mean with rms $\sigma_x$, then the residual 
$\xi=\hat y-\frac{\rho\sigma_y}{\sigma_x}\hat x$ around the $Y-X$ line with
slope equal to the regression of $Y$ on $X$ and rms 
$\sqrt{1-\rho^2}\sigma_y$. Similarly, Eq. \ref{eqxy} shows that the
same distribution can be generated first extracting the $y$ variable
and then the residual $\zeta=\hat x-\frac{\rho\sigma_x}{\sigma_y}\hat y$ 
around the $Y-X$ line with slope equal to the regression of $X$ on $Y$.
Finally, Eq. \ref{eqortho} generates the distribution extracting the
residuals around the eigenvectors of the covariance matrix. The
direction of the eigenvector with the largest eigenvalue is the slope
of the orthogonal regression. 
The zero points $b$ of the relevant linear
relations of slope $a$ are obtained from the mean values as:
\begin{equation}
\label{eqzeropoint}
b=\overline{y}-a\overline{x}.
\end{equation}

Do we therefore conclude, as Akritas and
Bershady (1996), that ``there is no such thing as true slope''? In
fact, the key of the problem is the covariance matrix, and therefore
the natural slope is the direction
of the principal axes (given by Eq. \ref{eqao}) with the largest
eigenvalue (given by Eq. \ref{eqsigma1}; Eq. \ref{eqsigma2} 
is the second eigenvalue of the covariance matrix).  
As a consequence, the orthogonal regression
determines the ``true slope'', not the bisector, which Isobe et
al. (1990) prefer for its lower variance.  However, what about
distance determination problems, where Eq. \ref{eqyx} should be preferred
according to Akritas and Bershady, if $y$ is the distance-dependent
quantity? Indeed, if one seeks for the distance shift $\delta$ of the
datapoint $(\hat y_i+\delta, \hat x_i)$, the most probable value
when Eq. \ref{eqbivariate} is given is:
\begin{equation}
\label{eqdshift}
\delta=-(\hat y_i-\frac{\rho \sigma_y}{\sigma_x}\hat x_i),
\end{equation}
which involves the $Y-X$ regression coefficient $V_{xy}/V_{xx}$.

Further complications arise when the datapoints $\{x_i,y_i\}$ are
affected by errors comparable to the rms spreads $\sigma_x$ and
$\sigma_y$, or when only a subset of the possible data volume is
available due to incompleteness. These affect the estimation of the
covariance matrix $V$ and the mean values of the variables.
For example, we estimate in Appendix \ref{twodimlim} that the
orthogonal regression solution ignoring the errors underestimates the
true slope if the error in the $X$ direction is larger than the one in $Y$.
These effects might be included in the regression
approach by careful treatement (see references given in Feigelson and
Babu 1992), but it is clear from the discussion
above that a maximum likelihood approach is superior, allowing  the natural and
simultaneous solution of all these difficulties.

\subsection{The EFAR database}
\label{efarsample}

The EFAR database comprises photometric and spectroscopic information
for a set of cluster galaxies spanning a factor of two in redshift, with
non-negligible measurement errors and selection effects.

The size of the errors can be quantified comparing the rms spread of
the variables with the mean of the errors. In this sense photometric
errors ($\approx 0.02$ mag, see Paper III and IV) are small in the
EFAR database. Unfortunately, this is not the case for the
spectroscopic data (see Paper II). For the case of the
velocity dispersions $\sigma$, we find that the mean value (25 km/s)
of the measurement errors on $\sigma$ is 36\% of the rms spread of the
values (71 km/s). For
the Mg$b$ index, the mean value (0.39 \AA) of the measurement errors
is 48\% of the rms spread of the values (0.83 \AA). 
In addition, we cannot measure
central velocity dispersions smaller than 100 km/s for lack of
spectral resolution. 

Selection effects are also severe.  The galaxy sample is more than
50\% complete only for galaxies with $D_W>20$ kpc (see Paper I), with
46\% of the galaxies having selection probabilities larger than 50\%.
The EFAR galaxy selection $S$ is described in Paper I and is a
function of the $D_{i,W}$ diameter measured in arcsec for galaxy $i$:
\begin{equation}
\label{eqsel}
S_{i,j}(\log D_{i,W})=0.5\left\{1+\hbox{\rm erf} \left[ \frac{\log D_{i,W}-\log
D^0_{W,j}}{\delta_{W,j}} \right]\right\} ,
\end{equation}
where $j$ is the cluster index, $\log D^0_{W,j}$ is the midpoint and 
$\delta_{W,j}$ the width of the cutoff in the selection function.
The diameter $D_{i,W}$ correlates with the $D_{i,n}$ diameter of
the sample as $\log D_{i,n} = 0.8*\log D_{i,W} +0.26$ with 0.09 dex
scatter in $D_{i,n}$ (see Paper III). Therefore we compute the
selection probabilities $S_{i,j}$ from the $D_{i,Wn}$ diameters
derived from the very accurate $D_{i,n}$ instead of using the values
of $D_{i,W}$ given in Paper I. In the following we shall indicate with
$S_i$ the selection probability for the galaxy $i$, dropping the
cluster index, and with $w_i=1/S_i$ the selection weight.

Due to the large spread in redshift,
the EFAR clusters have been sampled down to different limiting $D_W^0$(kpc).
However, 95\% of the early-types of the sample have $D_{Wn}>12.6$ kpc,
so that nearly unbiased estimates of the cluster peculiar velocities can
be obtained just by dropping from the sample objects with $\log D_{Wn}\le
d_{cut}=\DWcut$.

To our knowledge, no algorithm based on linear regressions has been
developed to model all these features at the same time.  In the
following we describe a parametric algorithm based on gaussian
modelling able with to deal with (i) a spread of
selection weights, (ii) sizeable measurement errors with a large
spread, (iii) possible explicit selection limits.

\subsection{The general case}
\label{general}

In the general case, we have N (logarithmic) data
$\vec{x_i}=(x_{i,1},\cdots,x_{i,N})$ for each galaxy $i$. We indicate
with $E_i$ the related $N\times N$ error matrix.
If the errors are uncorrelated,
$E_i$ is a diagonal matrix with
$E_{i,j,k}=\delta_{j,k}\sigma_{i,j}^2$, where $\delta_{j,k}=1$ if
$j=k$ and 0 otherwise. Each galaxy has been selected according to the
value of a selection diameter, which possibly could be expressed as 
a linear combination of a subsample of $x_{i,N}$, giving a selection
weight $w_i=1/S_i$. One variable $x_{i,1}$ may be
distance dependent. We want to estimate the vector of the mean values
of the data
$\overline{\vec{x}}=(\overline{x_1},\cdots,\overline{x_{N}})$, the
covariance matrix $V$ and, possibly, the vector
$\vec{\delta}=(\delta_1,\cdots,\delta_M)$ of the M cluster peculiar
velocity shifts. In this case we need an
additional constraint to fix $\overline{x_1}$, such as $\sum_j
\delta_j=0$. Other options are discussed in Paper VII. 
The direction of the eigenvector with the smallest
eigenvalue of the matrix V defines the minimum variance hyperplane 
describing the data
distribution, or the slope of the linear correlation in the two
dimensional case. We solve the problem assuming that both the
distributions of the galaxy parameters and of their errors are normal,
with covariance matrix $V$ and the error matrices $E_i$ respectively. 
The first
assumption can be verified {\it a posteriori}, estimating the
likelihood of the best-fitting gaussian model (see
Figs. \ref{figmg2sig}, \ref{figalllike} and \S\ref{alllike}). 
The second is based on
the error analysis performed in Papers II and III for the EFAR
database. Once the error convolution is performed, 
the probability density of the vector $\vec x_i$ is:
\begin{equation}
\label{eqprobndim}
P(\vec x_i)=
\frac{1}{(2\pi)^{N/2}|V+E_i|^{1/2}f_i}
\exp \left[-\frac{1}{2}\hat{\vec x_i}^T(V+E_i)^{-1}\hat{\vec x_i }
\right]
\hat{\theta}(A\hat{\vec x_i}-\vec{x}_{cut})
\end{equation}
where
$\hat{\vec{x}}_i=(x_{i,1}-\overline{x}_1+\delta_j,x_{i,2}-\overline{x}_2,
\cdots,x_{i,N}-\overline{x}_N)$. The function
$\hat{\theta}(\vec{y})=\prod_k \theta(y_k)$, where $\theta(y)=1$ if
$y\ge 0$ and 0 otherwise, takes into account that parts of the
parameter space might not be accessible because of selection effects
or explicit cuts. For simplicity we assume that these cuts are applied to
linear combinations of the variables, described by the appropriate
matrix $A$. The normalization factor $f_i$ is such that $\int
Pd^Nx=1$. Following Eadie et al. (1971), we write the likelihood of the
observed sample as:
\begin{equation}
\label{eqlikelihood}
{\cal L}=\prod_i P(\vec{x}_i)^{w_i},
\end{equation}
counting $w_i=1/S_i$ times galaxies with selection probability $S_i$
We determine $\overline{\vec{x}}$, $V$ and, possibly, $\vec{\delta}$
by minimizing $-\ln {\cal L}$. In the following sections we describe
how this general scheme is implemented in the different cases of interest.
In all cases the minimization is performed numerically, using the
simplex algorithm (Press et al. 1986). Appendix \ref{derivation}
discusses the analytical solution of Eq. \ref{eqlikelihood} 
for small errors, no explicit cuts and no peculiar velocities.

\subsection{The 1-dimensional Case}
\label{onedim}

The first simple application of the scheme described above is the
determination of the mean value $\mu$ and 
the (intrinsic) rms $\sigma$ of a set of data
$x_i$ with errors $\sigma_i$, selection weights $w_i$ and subject to
the censoring $x>x_{cut}$. 
Mean values and intrinsic spread of the
EFAR photometric (logRe, \SBe, \logDn, $(B-R)$) or
spectroscopic (\logsig, \mgb, \mgtwo) parameters, or of the
measured cluster peculiar velocities can be estimated in this way. 

The simplified version of Eq. \ref{eqlikelihood} reads now:
\begin{eqnarray}
\label{eqonedim}
{\cal L}=\prod_i \left( \frac{1}{f_{1i} \sqrt{2\pi (\sigma_i^2+\sigma^2})}
\exp \left[
-\frac{(x_i-\mu)^2}{2(\sigma_i^2+\sigma^2)}\right]\theta(x_i-x_{cut})\right)^{w_i},
\end{eqnarray}
Errors are taken
into account by convolving the intrinsic distribution with the
appropriate gaussian error distribution.  Selection effects are taken
into account in Eq. \ref{eqonedim} by counting $1/S_i$ times 
galaxies with selection probability $S_i$. 
The presence of explicit cuts
$x_{cut}$ is taken into account (for mild cases; see below) 
by the normalization correction:
\begin{equation}
\label{eqonenorm}
\begin{array}{ccl}
f_{1i} &=& \frac{1}{\sqrt{2\pi(\sigma_i^2+\sigma^2})}\int_{x_{cut}}^\infty
\exp \left( -\frac{(x-\mu)^2}{2(\sigma_i^2+\sigma^2)}\right)dx \\
    & =&  0.5\left(1-\hbox{\rm erf}\left(\frac{x_{cut}-\mu}{\sqrt{2(\sigma_i^2+\sigma^2)}}\right)\right).
\end{array}
\end{equation}
We expect our approach to work in the presence of {\it mild} cuts 
with $(x_{cut}-\mu)/\sigma<<0$, so that the cut is well away from the mean
value of the distribution (see \S \ref{onedimtest}).
Appendix \ref{onedimlim} gives the analytical solution of the problem 
for the limiting case of small errors.  

\subsection{The 2-dimensional Case}
\label{twodim}

We turn now to the two-dimensional case, which allows us to investigate
correlations between pairs of variables, such as \mgmg\ examined in
Paper II and the \mgbpsig\ and \mgtwosig\ discussed in Paper V.
Other relations that we shall not explicitly consider here are the 
$FP-\logsig$ (where $FP=\logRe-b\SBe$ with $b\approx 0.3$), 
the $(B-R)-\logsig$, $(B-R)-\mgb$
and $(B-R)-\mgtwo$ relations, the ``Kormendy'' relation $\logRe\ -
\SBe$, or the \logDn-\logsig\ relation. Note that the 
EFAR database measures the photometric parameters in the R band.

Eq. \ref{eqprobndim} reads in this case: 
\begin{equation}
\label{eqgausstwo}
P(\vec x_i=(x_{i,1},x_{i,2}))=\frac{|\Lambda_i|^{1/2}}{2\pi f_{2i}} \exp 
(-\frac{1}{2} \hat{\vec x}_i^T(V+E_i)^{-1} \hat{\vec x}_i)
\theta(x_{i,1}-x_{1cut})\theta(x_{i,2}-x_{2cut}),
\end{equation}
where the normalization factor $f_{2i}$ is
\begin{equation}
\label{eqtwonorm}
\begin{array}{ccl}
f_{2i} &=& \frac{|\Lambda_i|^{1/2}}{2\pi}
\int_{x_{1cut}}^\infty\int_{x_{2cut}}^\infty\exp (-\frac{1}{2}
\hat{\vec{x_i}}^T \Lambda_i \hat{\vec{x_i}})dx dy\\
&=&L(h,k,\rho).
\end{array}
\end{equation}
Here $\Lambda_i=(V+E_i)^{-1}$ and $L(h,k,\rho)$ is the bivariate
probability integral (Abramovitz
and Stegun 1971) with
$h=(x_{1cut}-\overline{x}_1)\sqrt{\Lambda_{i,22}(1-\rho^2)}$,
 $k=(x_{2cut}-\overline{x}_2)\sqrt{\Lambda_{i,11}(1-\rho^2)}$ and 
$\rho=-\Lambda_{i,12}/\sqrt{\Lambda_{i,11}\Lambda_{i,22}}$.
We set $x_{1cut}=-\infty$, $x_{2cut}=-\infty$ (and therefore
$f_{2i}=1$)
when considering the
the \mgmg\ relation, 
$x_{1cut}=-\infty$, $x_{2cut}=\log \sigma_{cut}$, when studying
the \mgbpsig\ and \mgtwosig\ relations, and 
$x_{1cut}=\FPcut$, $x_{2cut}=\log \sigma_{cut}$, when examining
the $FP-\sigma$ relation. In this case we derive \FPcut using its
relation with $D_{Wn}$: $FP=0.78\log D_{Wn}-6.14$ (see Paper III).
Finally, note that in every case we compute $\Lambda_i$ using the
simplifying assumption that the error matrix is diagonal with diagonal
terms given by the estimated total errors. For the spectroscopic data 
these involve two terms,
the uncorrelated statistical errors and the correlated errors coming
from run-to-run corrections (see Paper II).

Appendix \ref{twodimlim} gives the analytical solution
for the limiting case of small errors, no peculiar velocities and no cuts.  

\subsection{The 3-dimensional Case}
\label{threedim}

The three-dimensional case allows us to study the EFAR Fundamental
Plane.  As is well known, early-type galaxies do not fill the
three-dimensional space defined by the coordinates
$\vec{x}=(x_1=\logRe, x_2=\logsig, x_3=\SBe\ )$, but rather occupy a
narrow region around the Fundamental Plane defined by the equation:
\begin{equation}
\label{eqfp}
\logRe-a \logsig\ -b\SBe\ =c.
\end{equation}
Following \S\ref{general}, we determine the mean values
\meanRe, \meansig\ and \meanSBe, the covariance matrix $V$ and the
cluster peculiar velocity shifts $\delta_j$ (subject to the constraint
$\sum_j \delta_j=0$, see  \S\ref{general}) by maximizing the
likelihood ${\cal L}$:
\begin{eqnarray}
\label{eqmaxlike}
\ln {\cal L} & = & \sum_{\sigma>\sigma_{cut}, FP>\FPcut} \frac{1}{S_j(\log D_{Wn}^i)}\ln P(\vec{x}_i)\nonumber\\
& = & -\sum_{\sigma>\sigma_{cut}, FP>\FPcut} \frac{1}{S_j(\log D_{Wn}^i)}
\left[0.5\hat{\vec{x}}_i^T(V+E_i)^{-1}\hat{\vec{x}}_i
+\ln f_{3i}+1.5\ln(2\pi)+0.5\ln |V+E_i|\right],\nonumber
\end{eqnarray}
where
\begin{equation}
\label{eqnorm}
f_{3i}=\int P_i \theta (\log \sigma-\log \sigma_{cut})\theta (FP-\FPcut)d^3x
\end{equation}
is the fraction of galaxies with $\sigma>\sigma_{cut}$ and $FP>\FPcut$. 
The integral (Eq. \ref{eqnorm}) is performed in Appendix \ref{norm3dim}. 
The offset $\delta_j$ between true mean galaxy size \meanRe, and the
mean galaxy size observed for cluster $j$, $\meanRe-\delta_j$, is
related to the peculiar velocity of the cluster. In particular, 
the ratio of the true angular diameter
distance of a cluster, $D_j$, to the angular diameter distance
corresponding to its redshift, $D(z_j)$, is:
\begin{equation}
\label{eqn:dist}
\frac{D_j}{D(z_j)} = \frac{\dex(\meanRe)}{\dex(\meanRe-\dj)} = 10^{\dj} ~.
\end{equation}
The relation between angular diameter distance and redshift (Weinberg
1972) is given by
\begin{equation}
\label{eqn:daz}
D(z)=\frac{cz}{H_0(1+z)^2}\frac{1+z+\sqrt{1+2q_0z}}{1+q_0z+\sqrt{1+2q_0z}} ~.
\end{equation}
We assume $H_0$=50\,km\,s$^{-1}$\,Mpc, $q_0$=0.5, and compute all
redshifts and peculiar velocities in the CMB frame of reference. The
peculiar velocity of the cluster, $V_j$, is then obtained as
\begin{equation}
\label{eqn:pec}
V_j = \frac{cz_j-cz(D_j)}{1+z(D_j)},
\end{equation}
where $z(D_j)$ is the redshift corresponding to the true distance $D_j$
through the inverse of Eq. \ref{eqn:daz}. 

The error matrix $E_i$ is given by:
\begin{equation}
\label{eqerror}
E_i=\left( \begin{array}{ccc}
\delta r_i^2 & 0 &\frac{(1+\alpha^2)\delta r_i^2-\delta FP_i^2}{\alpha(1+\alpha^2)}\\
0 & \sigma^2_s  & 0 \\
\frac{(1+\alpha^2)\delta r_i^2-\delta FP_i^2}{\alpha(1+\alpha^2)} & 0 & 
\frac{(\alpha^2-1)\delta FP_i^2+(1+\alpha^2)\delta r_i^2}
{\alpha^2(1+\alpha^2)} +\delta ZP^2_i\\
          \end{array}
  \right),
\end{equation}
where $\delta r_i$ is the error on $\log R_{e,i}$, $\delta FP_i$ is the error
on the combined quantity $FP_i=\log R_{e,i}-\alpha \langle SB_{e,i} \rangle$,
with $\alpha\approx 0.3$, and $\delta ZP_i$ is the photometric zero-point
error (see Paper III and IV).  Note that in principle $b$ might be
different from $\alpha$. The quantity  
$\delta u_i=\sqrt{\frac{(\alpha^2-1)\delta FP_i^2+(1+\alpha^2)\delta r_i^2}
{\alpha^2(1+\alpha^2)}}$ is  the error on the effective surface
brightness coming from the fitting. See Appendix \ref{photerrmatrix} for a
derivation of these formulae. The error $\sigma_s$ on \logsig\
combines the uncorrelated and correlated errors (see Paper II). 

It is worth noting that the eigenvectors $\vec{v_1}$,
$\vec{v_2}$, $\vec{v_3}$
of the V matrix can be written to a close approximation 
as a function of the parameters $a$
and $b$ defining the Fundamental Plane of Eq. \ref{eqfp}: 
\begin{eqnarray}
\label{eqrot}
\vec{v}_1 & = & \hat{\vec{x}}_1-a\hat{\vec{x}}_2-b\hat{\vec{x}}_3,\nonumber \\
\vec{v}_2 & = &\hat{\vec{x}}_1+\hat{\vec{x}}_3/b,\\
\vec{v}_3 & = & -\hat{\vec{x}}_1/b-(1+b^2)\hat{\vec{x}}_2/(ab)+\hat{\vec{x}}_3,\nonumber
\end{eqnarray}
where $\vec{v}_1$ measures the displacement from the FP, and $\vec{v_2}$ and 
$\vec{v_3}$
are two orthogonal vectors in the FP. The absence of a
$\hat{\vec{x}}_2$ component in $\vec{v}_2$ is remarkable. In Paper VII
(Eq. 12) we fit for the coefficient of \logsig\ in the second principal
component and find that it is small. In \S\ref{secvariant}
(Eq. \ref{eqthird}) we show that it is compatible with zero within the 
statistical uncertainties.

In the following we indicate the rms
spread around the vectors $\vec{v}_1$, $\vec{v}_2$ and $\vec{v}_3$ with 
$\overline{\sigma_1}$, $\overline{\sigma_2}$, $\overline{\sigma_3}$,
respectively.

\section{Monte Carlo Simulations}
\label{montecarlo}

The algorithms described in the previous section have been extensively
tested on mock catalogues of the EFAR database. After describing how
these catalogues are generated (\ref{mock}), we proceed to test cases
of increasing complexity. We start with the one-dimensional $\sigma$
distribution (\ref{onedimtest}), as an example of the influence of
hard cuts on the mean and rms of the sample. Then we consider 
the \mgmg\ relation
(\ref{mg2mgbp}), where distance-independent quantities not subject to
explicit cuts are involved. We turn to the \mgsig\ relations
(\ref{mgsigma}), where no datapoints with central velocity dispersions
smaller than the resolution limit can be present.  Finally, we examine
the three-dimensional case of the Fundamental Plane (\ref{fp}), where
one distance-dependent quantity (\logRe) is correlated against two
distance-independent quantities (\logsig\ and \SBe), with
distance-dependent selection limits and the presence of cluster
peculiar velocities.

\subsection{Mock catalogues of the EFAR database}
\label{mock}

We generate mock catalogues of the EFAR database using the gaussian
distribution functions that are reconstructed from the data (see Paper
II, Paper V, Paper VII and Case 0 of Table \ref{tabsimumean}). We
specify the values of \meanRe, \meansig, \meanSBe, \meanmgtwo,
\meanmgbp, the coefficients of the FP and the slopes of the \mgtwosig\
and \mgbpsig\ relations, and the dispersions of the distributions in
the orthogonal directions. We generate one entry in the mock catalogue
for each early-type galaxy present in the database.  As a first step
we extract the triplet (\logRe, \logsig, \SBe), using Eq. \ref{eqrot}
to specify the directions of the principal components of the
distribution in the (\logRe, \logsig, \SBe) space.  We determine the
corresponding value of the $D_n$ diameter using the relation (a second
order approximation to the equation 4 of van Albada, Bertin and
Stiavelli 1993):
\begin{equation}
\label{eqdn}
      \log D_n = \log 2 R_e-0.289\Delta-0.019\Delta^2, 
\end{equation}
where $\Delta=\SBe-20$ matches the surface brightness used to
determine the EFAR $D_n$ diameters (see Paper III). Only galaxies with
a dominant exponential component in the luminosity profile deviate
strongly from Eq. (\ref{eqdn}).  We simulate the effects of cluster
peculiar velocities by adding cluster-dependent shifts to the
extracted values of \logRe\ and \logDn. We convert \logDn\ into $\log
D_W$ using the relation $\log D_W=(\logDn-0.27)/0.8$ (see Paper III)
and add the 0.09 dex rms random scatter of the relation.  We compute
the corresponding selection probability, according to the selection
parameters of the cluster being extracted and its (redshift) distance
and test if the extracted datapoints should be in the sample. If this
is the case, the values of \mgtwo\ and \mgbp\ are extracted following
the (two-dimensional) gaussian distribution computed at $\log \sigma$.
Measurement errors are assigned following the EFAR catalogue, so
extracted object N has the total errors of object N of the database,
taking into account the correlated terms of the total errors (see
Paper II) properly. Therefore, the error on $\sigma$, \mgtwo\ and
\mgbp\ for the extracted object N is computed considering the
spectroscopic runs where object N of the database was observed,
generating proper run correction errors and determining the final
error taking into account the run weights.

Finally, three additional sets of simulations (Cases 45, 46 and 47 of
Table \ref{tabsimumean}) are generated using uniform instead of
gaussian distributions, and zero input peculiar velocity fields. 
Following the procedure described above, we
consider the values of \meanRe, \meansig, \meanSBe,
$\sigma_1,\sigma_2,\sigma_3$ and of the coefficients of the FP as Case 0
of Table \ref{tabsimumean}, and Eq. \ref{eqrot} to specify the
directions of the principal components of the distribution in the
(\logRe, \logsig, \SBe) space. For Case 45 we add gaussian errors
generated as above, but we assume a uniform distribution of the FP
parameters, extending $\pm\sqrt{3} \sigma_1$, $\pm\sqrt{3} \sigma_2$,
$\pm\sqrt{3} \sigma_3$ in the $\vec{v}_1, \vec{v}_2, \vec{v}_3$
directions respectively from the mean values. This choice preserves
the covariance matrix of the gaussian case. For Case 46 we use
gaussian distributions of the FP parameters, but measurements errors drawn
using a uniform distribution extending $\pm \sqrt{3}$ the estimated
rms. For Case 47 we use uniform distributions for both the FP
parameters and the errors, with the $\sqrt{3}$ scaling as above.

\subsection{1-dimensional Distributions}
\label{onedimtest}

\begin{figure}
\plotone{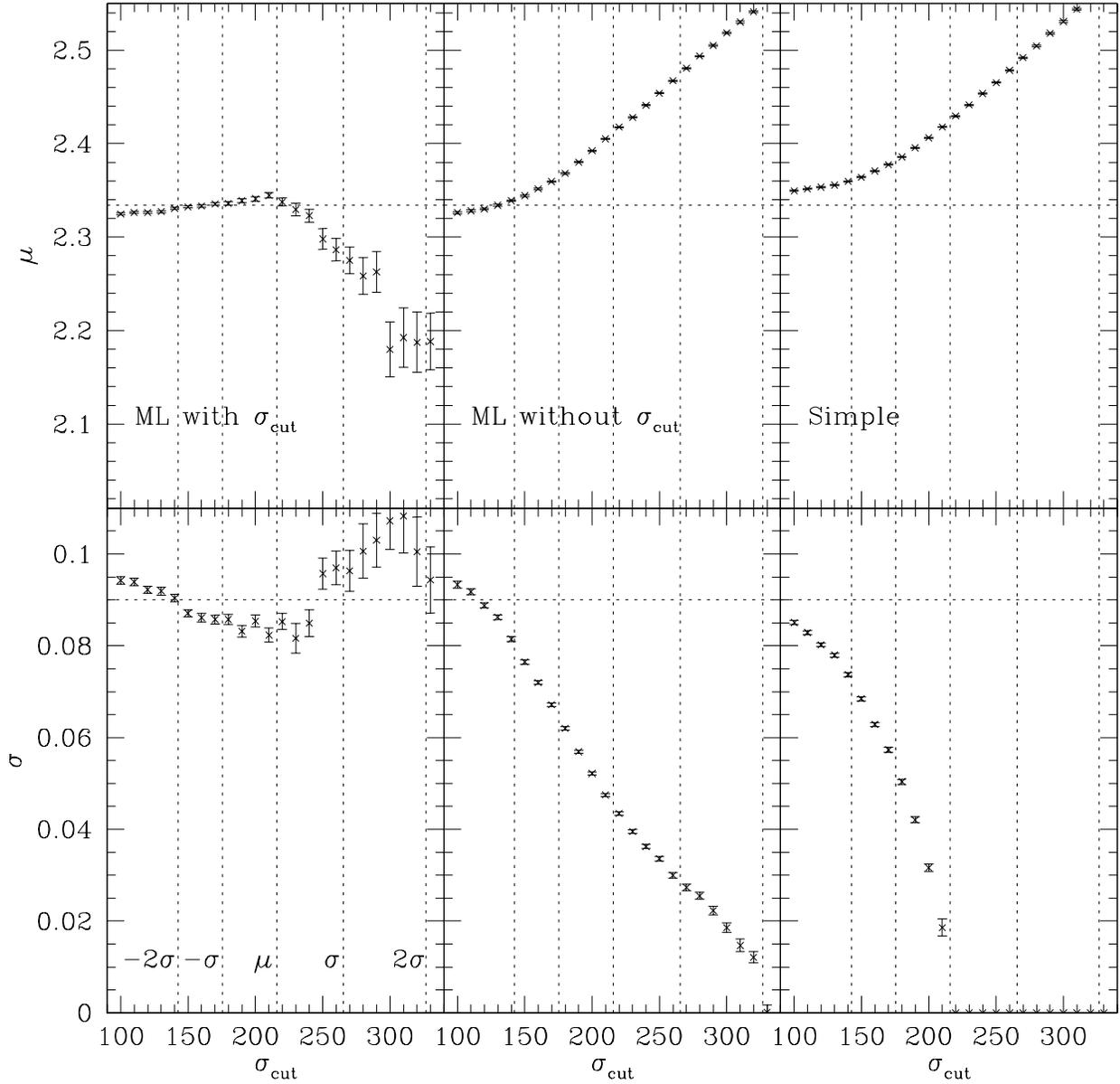}
\caption{The results of the simulations of the $\sigma$
distribution. The panels to the left show the mean (top panel) and rms
(bottom panel), averaged over 99 simulations, as derived
by the ML algorithm with normalization correction for
$\sigma_{cut}$. The errorbars are $1/\sqrt{99}$ smaller than the
random error expected for one simulation. 
The central panels show the mean (top) and rms
(bottom) as derived
by the ML algorithm without normalization correction for
$\sigma_{cut}$. The panels to the right show the simple mean (top) and rms
(bottom, corrected for measurement errors). The horizontal dotted
lines give the input values. The vertical dotted lines mark the
positions of the input mean value, and $\pm \sigma$, $\pm 2\sigma$
away from it.}
\label{figmlone}
\end{figure}

As a first example we discuss the use of the 1 Dimensional algorithm of \S 
\ref{onedim}. We generated 99 samples of the EFAR database (see Paper
V) and
considered the $\sigma$ distribution. Fig. \ref{figmlone} shows the
results as a function of the cut applied to the data. For cuts well
below the true mean of the distribution, the ML algorithm with or
without normalization correction estimates the mean and rms of the
distribution with small bias. The simple mean, which does not take into
account the selection weights of the data, slightly overestimates  the
true value, because objects with small $\sigma$ tend to have small
selection probabilities. The simple rms, corrected for measurement
errors as suggested by Arkitas and Bershadi (1996), underestimates
slightly the true values, not taking into account the selection
weights. As the cut is increased to values similar to
the true mean of the distribution, the ML algorithm with normalization
correction estimates the mean and rms with small bias, with the
statistical errors (for the single sample $\sqrt{99}$ larger than the
error bars shown) becoming increasingly large. With cuts one sigma or
more larger than the mean, the ML algorithm starts to fail, 
biasing the mean low and the sigma high. However, the bias is always
smaller than the random errors. Both the ML algorithm without
normalization correction and the simple estimates bias the mean to
larger values and the rms to smaller values. 

\begin{figure}
\plotone{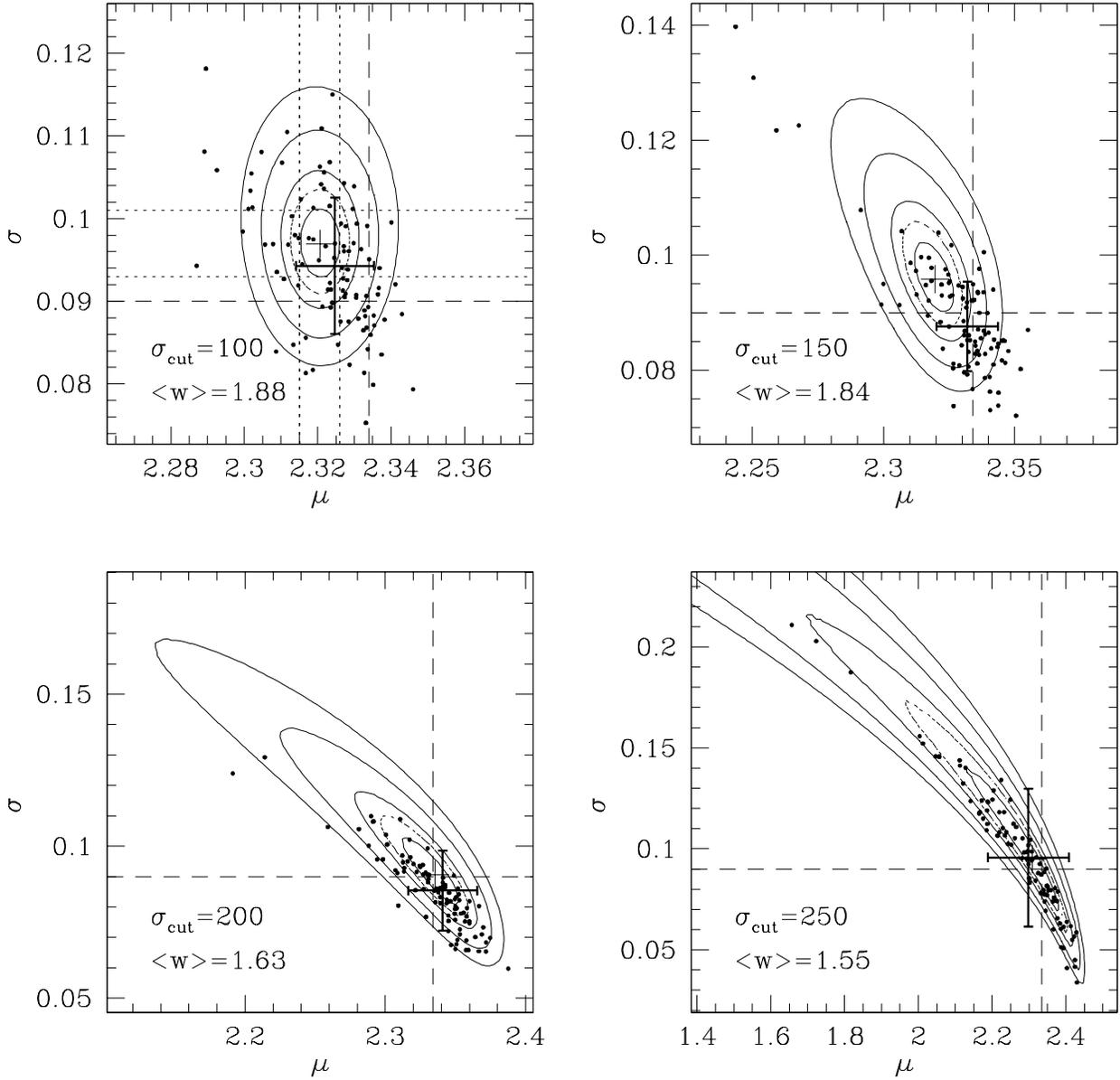}
\caption{The contours of constant likelihood in the $(\mu,\sigma)$
plane for one of the simulations of  Figure \ref{figmlone} analyzed
by the ML algorithm with normalization correction $\sigma_{cut}=100,
150, 200, 250$ km/s respectively. The projections (shown as dotted
lines for the 
$\sigma_{cut}=100$ case) of the four full-line
countours give the one-dimensional 1, 2, 3 and 4 sigma confidence
intervals. The dotted contours show the 1 sigma confidence regions in
two dimensions.  
The small crosses mark the maximum likelihood
solution. The applied mean selection weights $\langle w \rangle $ 
scaling factors are given in each panel (see
text). The dashed lines show the input values of the mean $\mu$ and
rms $\sigma$. The dots show the positions of maximum likelihood
solutions of the simulations of Figure \ref{figmlone}. 
The thick errorbars mark the position of $\mu$  and $\sigma$, 
averaged over the 99 simulations of Figure
\ref{figmlone}, and show their rms. }
\label{figlikeone}
\end{figure}

As it is well known (Press et al. 1986), 
the likelihood analysis offers in principle a
simple recipe to compute confidence intervals for the fitted
parameters. The $n\sigma$ one-dimensional confidence interval is fixed by the
contours of the likelihood function where: 
\begin{equation}
\label{eqlikeone}
\ln {\cal{L}}_{max}-\ln {\cal{L}}=0.5 n^2.
\end{equation}
Figure \ref{figlikeone} shows four contours of constant likelihood in
the $(\mu,\sigma)$ plane for one of the simulations of Figure
\ref{figmlone} analyzed by the ML algorithm with normalization
correction $\sigma_{cut}=100, 150, 200, 250$ km/s respectively.  The
thick errorbars mark the position of $\mu$ and $\sigma$, averaged over
the 99 simulations of Fig.  \ref{figmlone}, and show their rms. A
straight application of Eq. \ref{eqlikeone} would produce $n=1$
one-dimensional confidence intervals for $\mu$ and $\sigma$ a factor 
$\approx 2$
smaller than the rms measured by the simulations. Inspection of
Eq. \ref{eqlikelihood} reveals that since the mean weighting of each
galaxy is $\langle w\rangle =\Sigma_i w_i/N$, the $n\sigma$
one-dimensional confidence
interval is reached at a $0.5 \langle w \rangle n^2$ distance from the
maximum likelihood value. This factor is taken into account in Fig.
\ref{figmlone}, where the scaled $n=1$ one-dimensional confidence
intervals match
approximately the rms measured by the simulations. The two-dimensional
$1\sigma$ confidence region, set by 
$\Delta \ln {\cal{L}}=1.15 \langle w \rangle$, is slightly larger. 
However,
Fig. \ref{figalllike} shows that the mean weight fluctuates in the
range $1.5-2$, making the estimate of confidence intervals from the
likelihood function uncertain for a given dataset. In addition, this
analysis does not allow to study the small residual biases in the ML
estimates.  Therefore, in the following we will estimate confidence
intervals and residual biases using Monte Carlo simulations only.

\subsection{The \mgmg\ relation}
\label{mg2mgbp}

\begin{figure}
\plotone{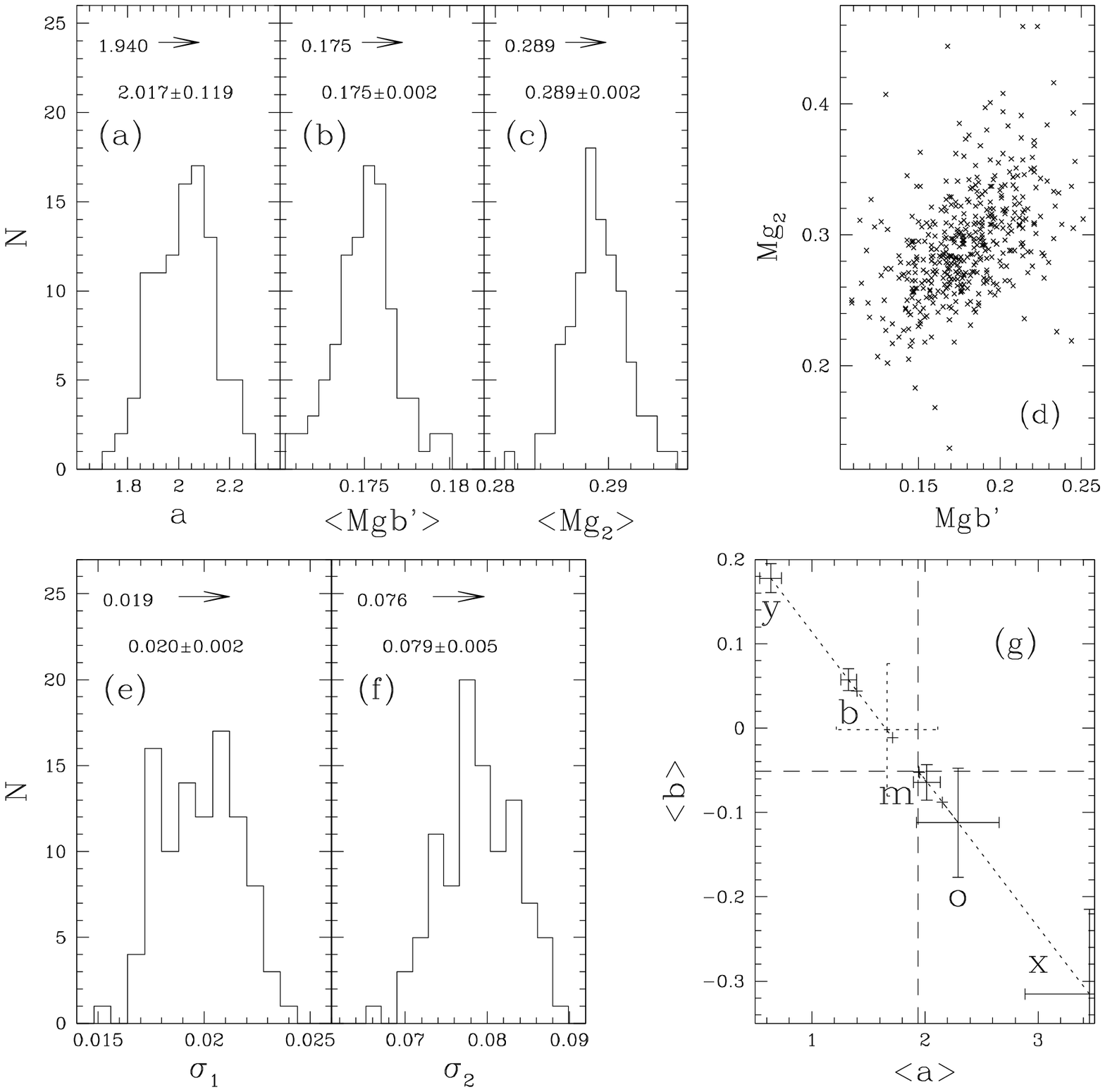}
\caption{The results of the \mgmg\ simulations. 
The panels show
the histograms of (a) the slopes,  the mean values of (b)
$<Mgb'>$ and (c) $<Mg_2>$,  the 
orthogonal and parallel spread (e) $\sigma_1$ and (f) $\sigma_2$ as derived
from the ML algorithm. Input values, means and rms of the 99
simulations are also given. Panel (d) shows one of the 99 \mgmg
realizations. Panel (g)  shows the values
of the slopes $a$ and zeropoints $b$, averaged over 99 simulations and
derived using the Y-X (y), the X-Y (x), bisector (b) and orthogonal
(o) regressions, plus the ML algorithm (m). The errorbars show the
random errors expected for one simulation. The dotted cross shows
the  values and the random errors
of the slopes $a$ and zeropoints $b$, averaged over 99 simulations,
derived using the Y-X regression, taking into account errors as in
Akritas and Berschady (1996).
The horizontal and vertical dashed lines show
the input values. The
other dotted lines of plot (g) terminating to the small crosses
show the values obtained when 
the errors are reduced by a factor 5. The crosses for the ML and the
orthogonal regression case overlap.}
\label{figmg2mgb}
\end{figure}

The simplest test of the 2-dimensional algorithm is performed considering the
\mgmg\ relation. We generated 99 samples without taking into
account the EFAR selection function. The input parameters are the ones
derived in Paper II. Two cases are examined, one with
a realistic error distribution and one with a factor 5 smaller errors. 
Figure \ref{figmg2mgb}(d) shows one of the realizations with
realistic errors. Figure \ref{figmg2mgb}(g)
shows the results averaged over the 99 simulations. 
In both the cases of small and realistic errors 
the ML algorithm retrieves the input parameters with small
biases and with small errors. The $Y-X$, $X-Y$ and bisector regressions
fail to derive the correct slopes and zeropoints, because of the
finite orthogonal scatter and the non-negligible errors. When the
method of Akritas and Bershady (1996) is used to correct for the
errors (dotted cross), the $Y-X$ result is less biased, but with
large random errors. Note that the method cannot be used when
selection weights are present. The
orthogonal residual fit gives a slightly biased answer as expected
from Eq. \ref{eqafirst}, with a nearly perfect result when the errors are
small. As expected from Eq. \ref{eqzeropoint}, the errors on $a$ and
$b$ are perfectly anticorrelated.

Figs. \ref{figmg2mgb}(a), (b), (c), (e) and (f) show 
the histograms of the results of
the ML parameters (for the case of realistic errors). Not only are the
slope and the mean values determined with high precision, but so also
is the intrinsic scatter (orthogonal and parallel to the relation),
despite the non-negligible measurement errors (the observed scatter
orthogonal to the relation is
more than a factor two larger than the intrinsic one). 

\subsection{The \mgtwosig\ and \mgbpsig\ relations}
\label{mgsigma}

\begin{figure}
\plotone{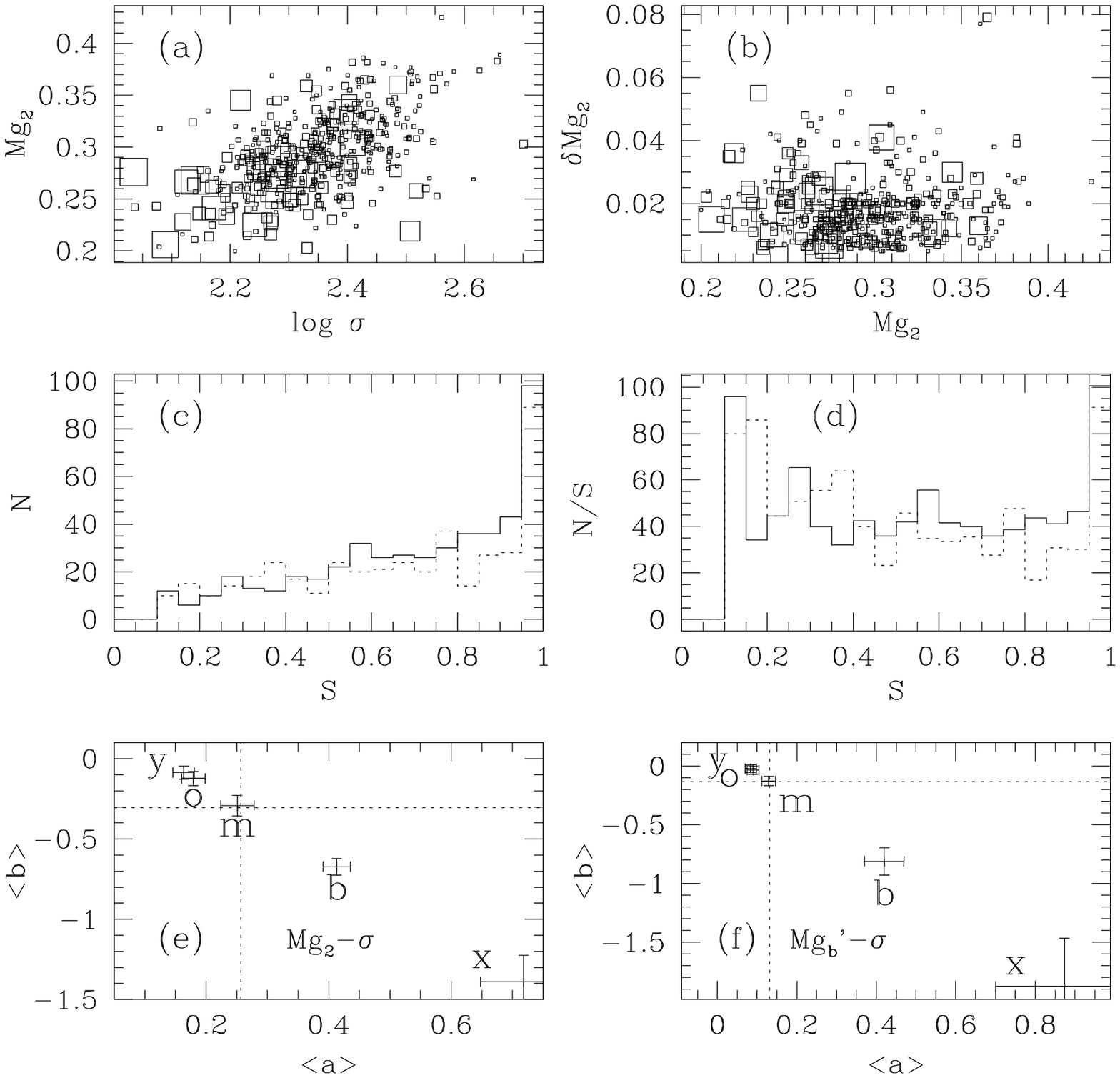}
\caption{The results of the \mgtwosig\ and \mgbpsig\ simulations. 
 Panel (a) shows one of the 99 \mgtwosig
realizations. The size of the points is proportional to their
selection weight $1/S$. Panel (b) shows the distribution of the Mg$_2$
errors. Panel (c) plots the
distribution of selection probabilities $S$ for the simulation shown
in (a) and (b) (full line) and for the EFAR database (dotted line)
analyzed in Paper V. 
Panel (d) plots the effective number of
data points $N/S$ as a function of $S$ for the simulation shown
in (a) and (b) (full line) and for the EFAR database (dotted line) 
analyzed in Paper V. Panel (e) and (f) show the values of the slopes
$a$ and zeropoints (b), averaged over 99 simulations and derived 
using the Y-X (y), the X-Y (x), bisector (b) and orthogonal
(o) regressions, plus the ML algorithm (m), for the \mgtwosig\ and
\mgbpsig\ relations, respectively. The errorbars show the random errors
expected for one simulation. The horizontal and vertical dotted lines show
the input values.}
\label{figmg2mgbsig}
\end{figure}

The tests on the \mgtwosig\ and \mgbpsig\ relations were
performed using the EFAR mock catalogues with realistic error
distributions. The input parameters are the ones
derived in Paper V. Figure
\ref{figmg2mgbsig}(a) shows one of these \mgtwosig\ realizations. 
Data at small $\sigma$ or Mg$_2$ have larger selection
weights. Fig. \ref{figmg2mgbsig}(b) shows that the spread in the
errors is large.
The distribution of the selection probabilities
(Fig. \ref{figmg2mgbsig}(c)) has a 
peak at the completeness value ($S=1$) with a prominent tail down to
the selection probability cut ($S=0.1$). 
The effective number of datapoints
per selection weight bin (Fig. \ref{figmg2mgbsig}(d)) is approximately 
constant. The
simulations (full lines) reproduce closely the observed dataset
(dotted lines) analyzed in Paper V. 

Fig. \ref{figmg2mgbsig}(e) shows the slopes and
zero-points derived from the various regressions and the ML algorithm,
considering points with selection probabilities larger than 0.1 for the
\mgtwosig\ relation. As already
seen in \ref{mg2mgbp}, the regressions give biased results due to the
errors (as discussed in \S \ref{linear}), while the ML algorithm is
nearly free
of biases and accurate. 

\begin{figure}
\plotone{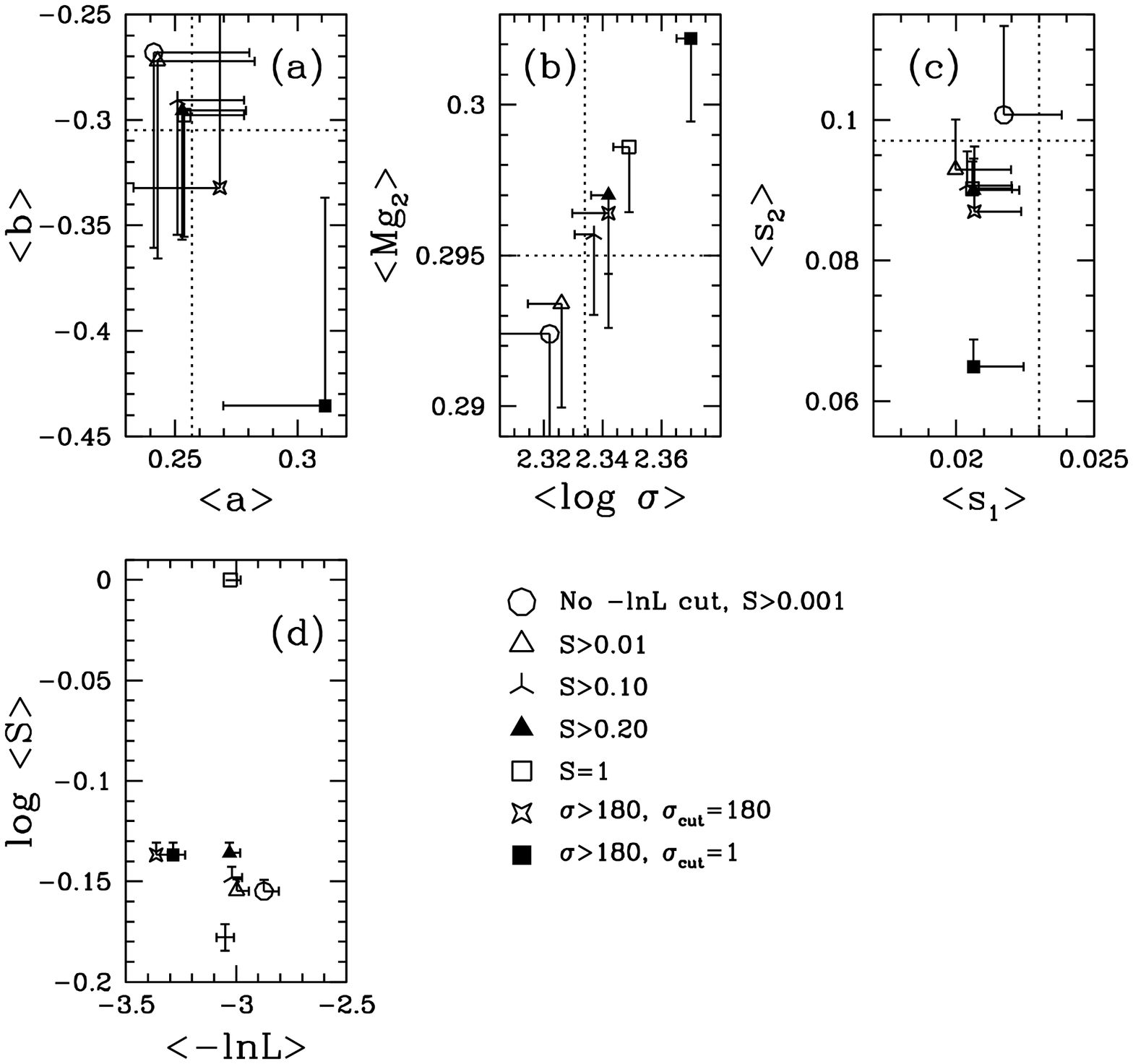}
\caption{The effects of the selection weighting and of the cut in
$\sigma$. The plots show the mean values (over 99 simulations) of (a) the
slope and  zeropoint, (b) the mean values, (c) the orthogonal and
parallel scatter as determined by the ML algorithm, (d) the mean
logarithmic likelihood and mean selection probability. 
The errorbars show the random errors expected for one simulation. 
The different types of points
refer to  the cases discussed in the text. The cross in panel (d)
shows the mean values derived for the EFAR sample examined in Paper V, 
with errorbars equal to their rms, divided by the square root of
number of the galaxies considered.}
\label{figmg2sig}
\end{figure}

Figure \ref{figmg2sig} illustrates the effects of the selection
weighting and of the cut in $\sigma$. When the analysis is performed
including all the available datapoints to the largest selection
weights and without any likelihood cut (open circle), the parallel and
orthogonal scatter are derived best. However, the rms on all the
parameters are rather large, because the
presence of some datapoints with very high weights can bias the
analysis of some realizations. It is therefore prudent to limit the
analysis to datapoints with weights not larger than 10 (e.g., $S_i>0.1$)
and clip points
with low likelihood ($-\ln {\cal L}>0$, triangular stars). This allows us 
to determine the slope and the zero-point of the relation with small biases and
with small variance.  The price is a slight underestimate of the 
orthogonal and parallel spread, $\sigma_1$ and
$\sigma_2$. Fitting points with weights not larger than 5 (filled
triangles) and the same likelihood clipping, one observes that the mean 
quantities \meanmgtwo\ and
\meansig\ are biased to systematically larger values, because
the small galaxies have on average larger selection weights (see
Fig. \ref{figmg2sig}).
The largest bias in the mean
quantities is obtained weighting all datapoints equally, independently
of the selection (open square). These tests justify the procedure
adopted in Paper V to study the \mgtwosig\ and \mgbpsig\
relations, where we consider galaxies with $S_i>0.1$ and apply
likelihood clipping.

To illustrate the effect of the cut in $\sigma$, we examine the 
extreme case where only galaxies with $\sigma>180$ km/s are considered
in the analysis.  When no correction is applied (filled square), the
slope and the zeropoint of the relation are slightly biased, the mean
quantities \meansig\ and \meanmgtwo\ are biased
to larger values, while the parallel spread is biased low (similar to
what is seen in Figure \ref{figmlone}).  The biases are fully corrected
for when we apply the normalization corrections appropriate for
$\sigma_{cut}=180$ km/s (open star), with a slight increase of the
variance. The cut applied to the EFAR database ($\sigma>100$ km/s) is
in any case low compared to the mean value of the sample. Finally,
plot (d) of Figure \ref{figmg2sig} shows the mean values and rms (over
the simulations) of the mean (over the datapoints) logarithmic
likelihood and selection probability. The cross shows the values
derived for the EFAR sample examined in Paper V, with errorbars equal
to their rms divided by the square root of
number of the galaxies considered.
The simulations match the mean likelihood of the EFAR
sample within the estimated errors, justifying {\it a posteriori} the
assumption of a gaussian distribution. The mean selection probability of 
the simulations 
(see Figure \ref{figmg2sig}) is only slightly larger than the 
ones of the real sample. 

Similar results are obtained when considering the \mgbpsig\
relation. Figure \ref{figmg2mgbsig}(f) shows again that the linear regressions
give biased results. The $Y-X$ and orthogonal fitting are very
similar, because of the shallow slope. The ML with probability cut
$S_i>0.1$ and likelihood clipping gives nearly unbiased results. 
Histograms similar to Fig. \ref{figmg2mgb}(a)-(f) for the
\mgtwosig\ and \mgbpsig\ relations have been already presented in
Paper V.

\subsection{The Fundamental Plane}
\label{fp}

In this section we describe the tests of the ML algorithm used to
derive the FP solution of Paper VII and the peculiar velocities of the
EFAR sample. In \S\ref{metcomp} we assess the superiority of the
ML algorithm with respect to  the linear regression methods, 
when the $a$ and $b$ parameters of the FP are determined. In
\S\ref{secvariant} we study the precision of the FP parameters
derived using the ML algorithm, justifying the strategy adopted in
Paper VII. 
\S\ref{alllike} justifies a posteriori the assumption of a
gaussian distribution of galaxies in the (\logRe, \logsig, \SBe) space. 
\S\ref{resbias} tests the bias correction scheme
for the peculiar velocities adopted in Paper
VII. Finally, \S\ref{meanmotions} investigates how well coherent and random
motions can be measured from the EFAR data sample. 

\subsubsection{Method comparison}
\label{metcomp}

As we have done above, we start comparing the performances of
the ``classical approaches'' (linear regressions and orthogonal
fitting) and the ML algorithm, focusing on mock catalogues generated
using parameters close to those derived for the EFAR data sample (see
Paper VII and Case 0 of Table \ref{tabsimumean}).  We consider 
the 29 best clusters of the EFAR sample (see Paper VII) and we do not add
peculiar velocities 
(i.e., the input peculiar velocity field is zero, $\delta_{input}=0$).
In each of the four methods considered we solve simultaneously
for the coefficients $a$ and $b$ (see Eq. \ref{eqfp}) 
and the vector $\delta_j$. The shifts
$\delta_j$ are determined by assuming that the mean of $\delta_j$ over
the 29 clusters is zero, which fixes the value of the parameter $c$. 
Objects with $\sigma<100$ km/s, or with
selection diameter $D_W$ less than 12.6 kpc (i.e., $\DWcut=1.1$),
or with selection probabilities less than
0.1, are excluded from the fits. Therefore, this procedure mimics
closely the ``fiducial'' solution of Paper VII (``Case 1'' of Table
4).

\begin{figure}
\plotone{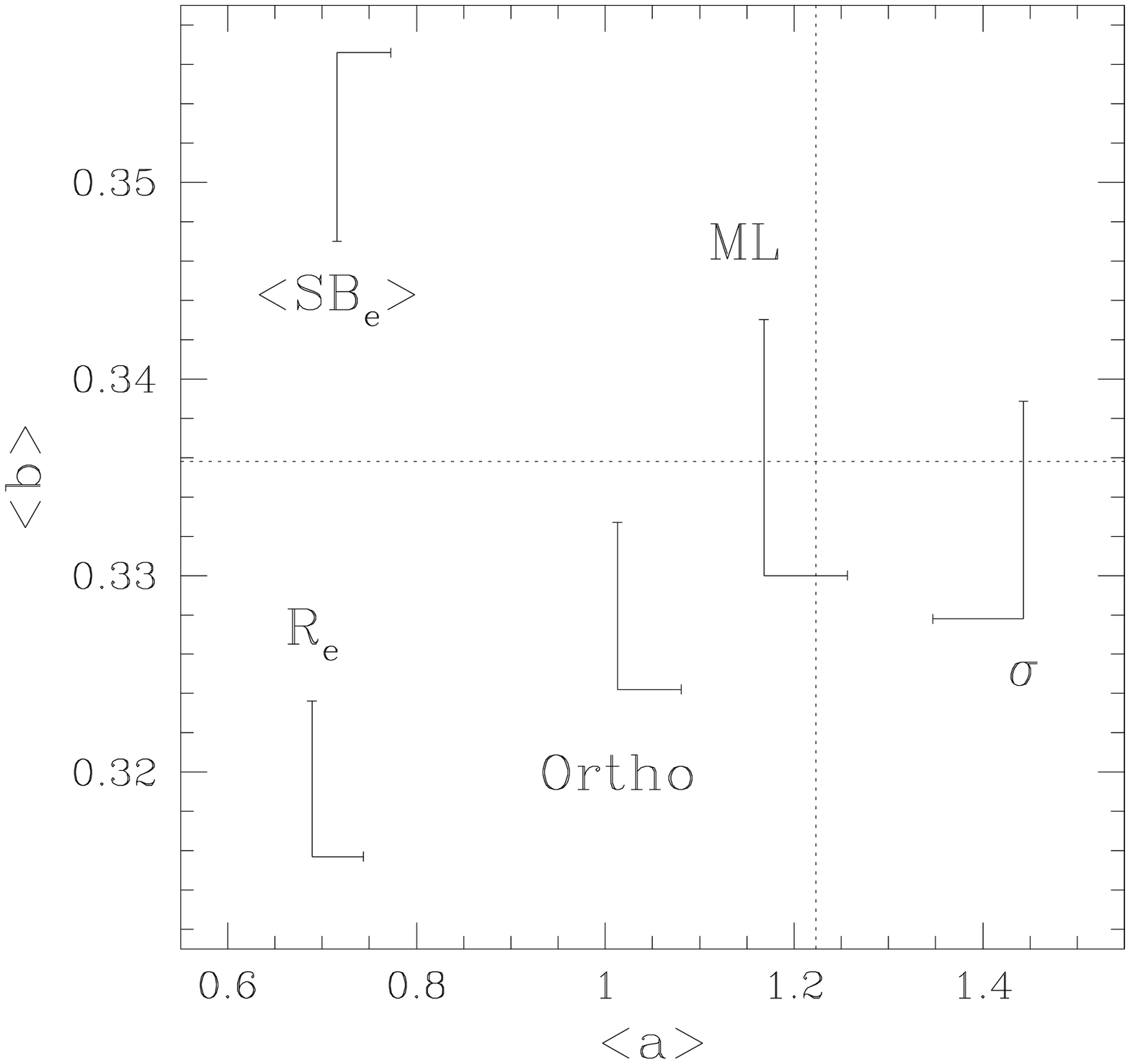}
\caption{The parameters of the FP relation. 
The plot shows the mean coefficients
$<a>$ and $<b>$ (over 99 simulations) derived from the linear
regressions on \logRe, \logsig, \SBe, the orthogonal minimization
and the ML algorithm. The errorbars show the rms. The dotted lines show the
input values.}
\label{figfp}
\end{figure}

Fig. \ref{figfp} shows the
mean values and rms (over 99 simulations) of the Fundamental Plane
coefficients $a$ and $b$ as determined using regressions on \logRe,
\logsig, \SBe, orthogonal fitting and ML.
As can be expected from the previous discussion, linear
regressions and orthogonal fitting are inadequate to determine
the coefficients of the Fundamental Plane.
All five methods determine $b$ quite accurately, with the \SBe\ and \logRe\
regressions giving the largest systematic deviations ($\approx
\pm 0.02$). Only the ML algorithm, however, is able to estimate $a$
with a bias smaller than the expected statistical uncertainty. 
As expected from Appendix \ref{twodimlim}, the \logRe\ and
\SBe\ regressions and (less severely) the orthogonal fitting
underestimate $a$, while the \logsig\ regression overestimates it. As
pointed out by Isobe et al. (1990), the \logRe\ regression gives the
smallest rms. The ML algorithm determines $a$ to 0.09 and $b$ to 0.013
statistical accuracy.

\begin{figure}
\plotone{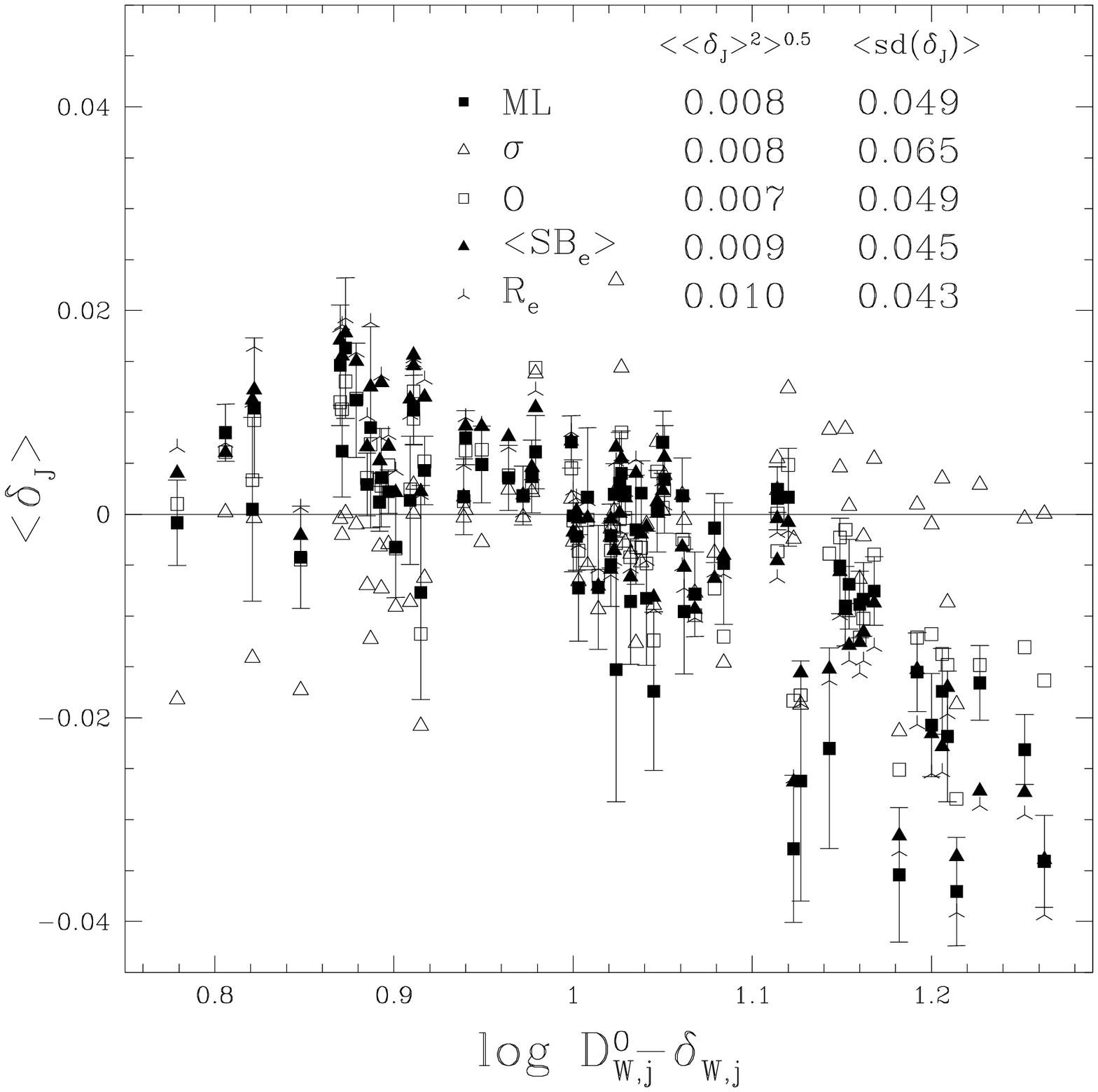}
\caption{The peculiar velocity shifts of the EFAR clusters.
The plot shows the mean 
(over 99 simulations) peculiar velocity shifts $<\delta_j>$
as a function of $\log D^0_{W,j}-\delta_{W,j}$ (see Eq. \ref{eqsel},
with $D_{W,j}^0$ in kpc). The methods used are the linear
regressions on \logRe\ (triangular crosses), \logsig\ (open
triangles), 
\SBe\ (filled triangles), the orthogonal minimization (open squares)
and the ML algorithm (filled squares). 
The errorbars show the errors on the mean, the
rms are $\approx 10$ times larger. The numbers give the rms
over the EFAR clusters of the
mean differences $<\delta_j>$ over the 99
simulations and the mean
over the 99 simulations of the standard deviation (over the EFAR
clusters) of the $\delta_j$ values with $\log
D^0_{W,j}-\delta_{W,j}<1.18$. The input peculiar velocity shifts are
all zero.}
\label{figpec}
\end{figure}

Figure \ref{figpec} shows how well the five methods measure the
input peculiar velocity shifts $\delta_j$ (all equal to zero) of the
EFAR clusters. Similarly to the procedure described in \S\ref{resbias}
for the ML algorithm, the parameters $a$, $b$ and $c$ of the FP 
determined from the regressions on the subset of 29 best clusters 
for each separate simulation  
are used to compute the peculiar velocities of the remaining
clusters. 

Let us first focus on the clusters with $\log
D^0_{W,j}-\delta_{W,j}<1.18$. These are the clusters that statistically
have at least one galaxy with $\logDw \approx 1.1$ and selection
probability larger than 0.1. In this case, the
biasing influence of errors and selection effects is small for all 
the methods. The absolute values of the mean $\delta_j$ are typically 
smaller than 0.02 dex. The rms over the EFAR clusters of the
mean $\delta_j$ over the 99 simulations is $\approx 0.008$. The \logRe\
regression gives slightly poorer results. The statistical errors on the
single cluster measurements are in any case larger, in the range
0.02-0.04 dex for the best-populated clusters, and up to 0.15 dex for
the poorer populated clusters. The mean
over the 99 simulations of the standard deviation (over the EFAR
clusters) of the $\delta_j$ values differs for the various methods.
The \logRe\ regression gives the
smallest value (0.043 dex, or a typical 10.4\% accuracy on the
determination of the distance of a single clusters), similar to
the \SBe\ regression (0.045 dex, 10.9\% distance error).
The ML algorithm and the orthogonal regression perform only slightly
worse (0.049 dex, or 11.9\% distance error).
The \logsig\ regression gives the largest (0.065 dex, or 16\% distance
error) random error. 

\begin{figure}
\plotone{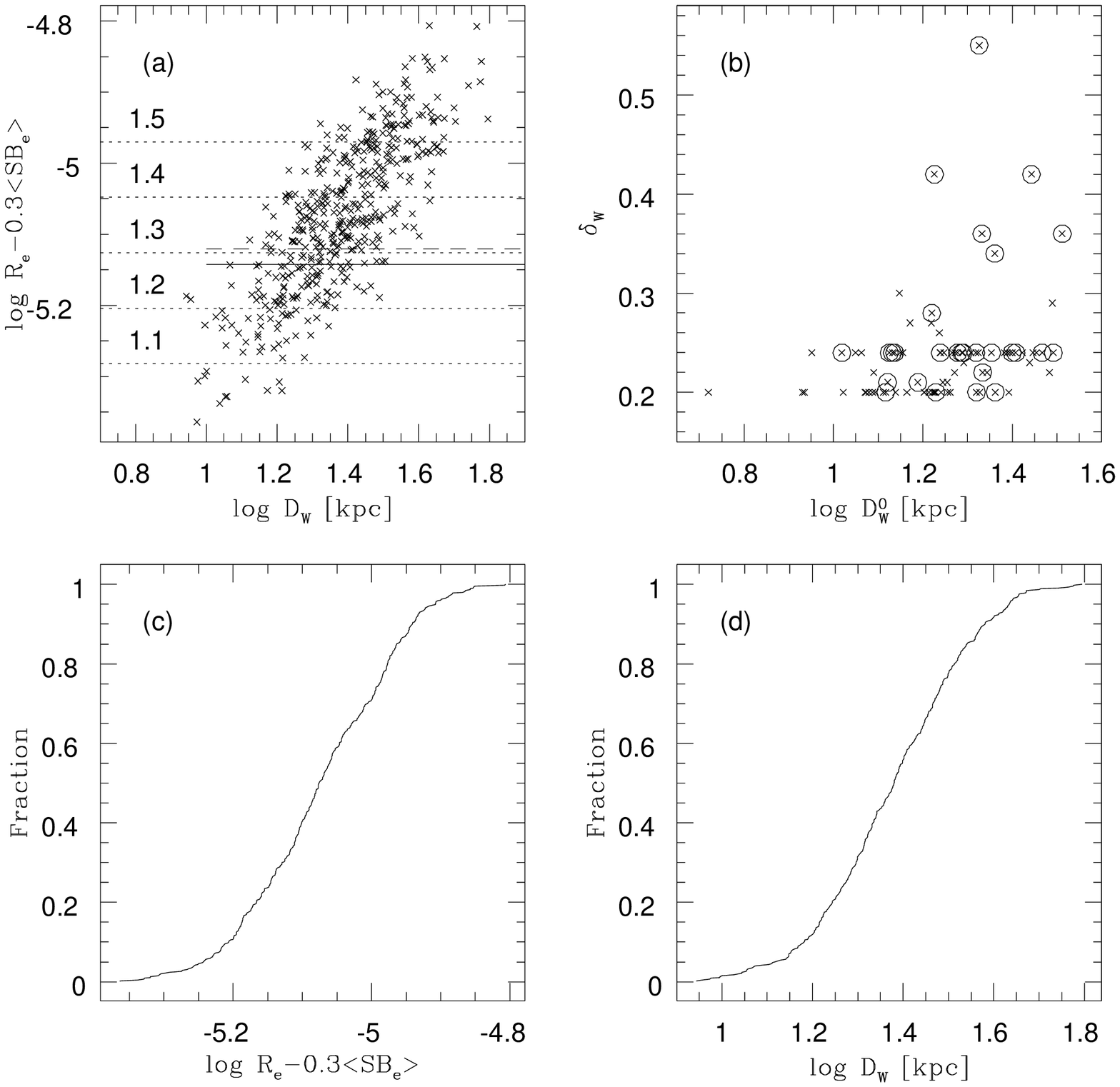}
\caption{One of the simulations of \S\ref{metcomp}. Panel (a)
shows the correlation between \logRe-0.3\SBe\ and \logDw. The full line
shows the input mean value \meanRe-0.3\meanSBe, the dashed dashed line
the value derived by the ML algorithm. The dotted
lines correspond to the given values of \DWcut. 
Panel (b) shows the values of the selection
parameters $\log D^0_{W,j}$ and $\delta_{W,j}$. The 29 best clusters
are shown circled.
Panel (c) shows the cumulative distribution of
\logRe-0.3\SBe\ values. Panel (d) shows the cumulative distribution of
\logDw. }
\label{figfpsim}
\end{figure}

The ML algorithm (and in order of increasingly importance the orthogonal, \SBe\
and \logRe\ regressions) derives $\delta_j$ biased to low values when
$\log D^0_{W,j}-\delta_{W,j}\ge 1.18$. The clusters where this
happens are so distant that their smallest galaxies in the EFAR sample
have $\logDw>1.1$. Fig. \ref{figfpsim} shows that these are 
clusters where we sample less than half of the galaxy
distribution. From what discussed in \S\ref{onedimtest} we expect the ML
algorithm to progressively fail in this regime. 
In \S\ref{resbias} we 
correct for this residual bias due to the difference in the way the FP
galaxy distribution is sampled in different clusters through
simulations. In any case, the standard deviations (over the EFAR
clusters) of the $\delta_j$ values do not change, remaining $\approx
4$ times larger that the typical systematic errors. 

Therefore we conclude that overall the ML method provides the best solution,
keeping low the systematic and random errors on both the FP parameters
and the $\delta_j$ values. In addition, it is more robust against
outliers (since we can identify and remove the extreme low-likelihood objects,
see Paper VII).

\subsubsection{Variant cases and systematic errors}
\label{secvariant}

\begin{figure}
\plotone{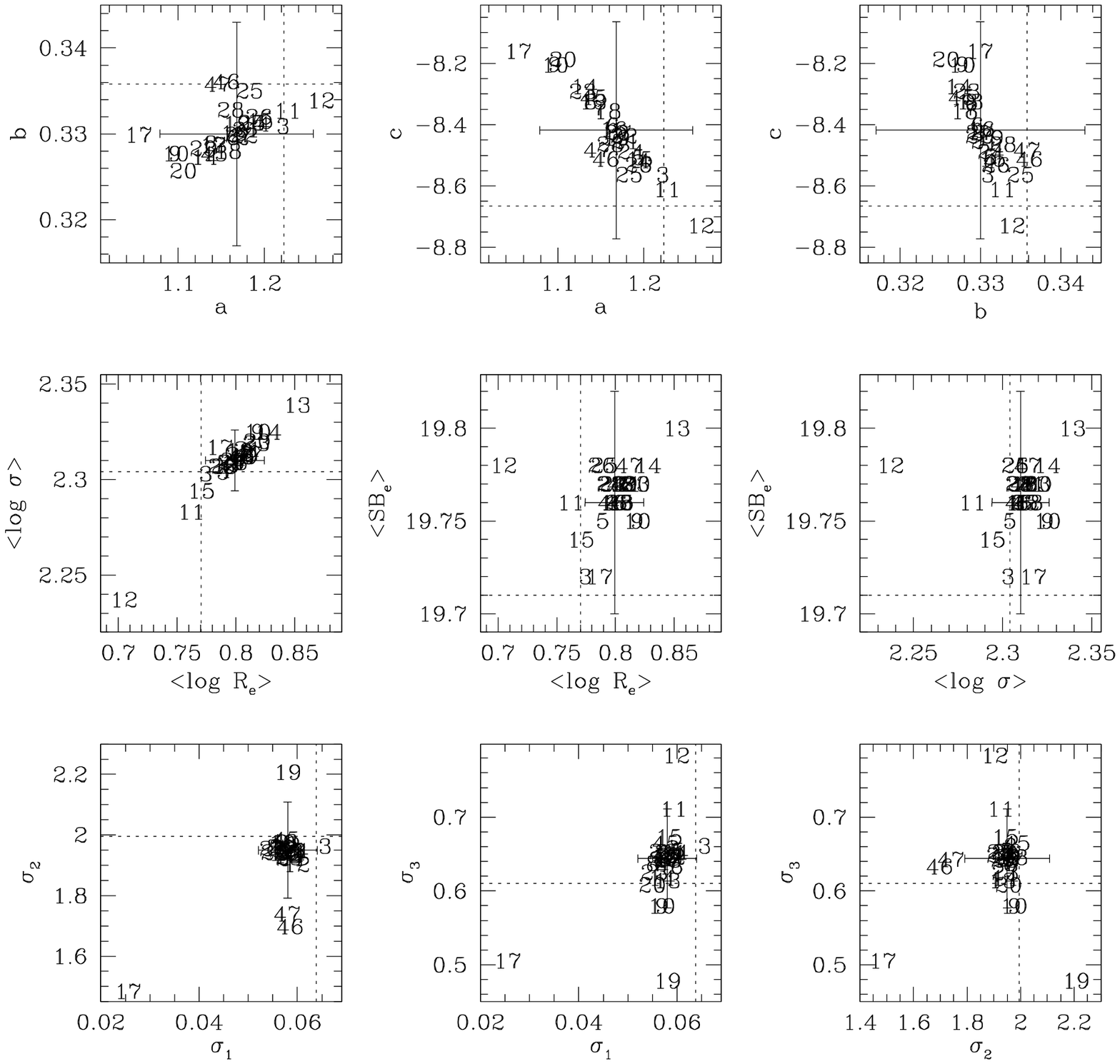}
\caption{The fitted FP parameters for each case in Table
\ref{tabsimumean}, showing the distributions and correlations for various
pairs of parameters. Each case is numbered as in the Table. The dotted
lines show the input parameters. The errorbars show the rms over 1000
simulations derived for Case 1.}
\label{figparvar}
\end{figure}

In this section we study the precision of the FP parameters
derived using the ML algorithm, to justify the strategy adopted in
Paper VII. Table \ref{tabsimumean} summarizes the tests performed,
giving the mean values (over 99
simulations) of the number of clusters and galaxies (rounded to
integer numbers) and the parameters of the Fundamental Plane derived for 28
cases. Table \ref{tabsimurms} lists the corresponding rms. 
Figure \ref{figparvar} shows the results graphically. 

Case 0 gives the parameters of the Fundamental Plane as derived
in Paper VII for the standard fit. These are the input parameters used
to generate the simulations considered here.
Case 1 is the standard fit discussed
above. Fig. 2 of Paper VII presents the histograms of 1000 simulations 
for this case. The mean values derived from the 99 simulations used
here do not differ by more than 1.5\% from those derived from
the 1000 simulations. As discussed in Paper VII, there are small
residual biases in the fitted parameters: a is biased low by 6\%
(i.e., since the mean value of $a$ recovered from the simulations is smaller
than the input value,  we infer that the true value of $a$ for the
EFAR sample must be larger by $\approx 6$\% than what found in Case 0), b is
biased low by 2\%; c is biased high by 4\%; \meanRe, \meansig\ and
\meanSBe\ are all biased high, by 0.036 dex, 0.07 dex and 0.05 mag
respectively; the scatter about the FP \sigone\ is under-estimated by 0.006
dex, or 1.4\%; the widths \sigtwo\ and \sigthree\ of the galaxy distribution
in the FP are biased by 0.049 (low) and 0.009 (high) dex respectively.
These biases are all less than or comparable to the rms width of the
distribution, so that although they are statistically significant
(i.e. much greater than the standard error in the mean), they do not
dominate the random error in the fitted parameters. Therefore in Paper
VII we do not correct for these biases, since they are small and have
negligible impact on the derived distances and peculiar velocities
(see \S\ref{resbias}).

Case 2 is similar to Case 1: we do not add
peculiar velocities and we do not fit
objects with $\sigma<100$ km/s, or with
selection diameter $D_W$ less than 12.6 kpc (i.e., $\DWcut=1.1$),
or with selection probabilities less than
0.1. However, in each simulation the list of fitted clusters 
is not fixed to the 29 clusters considered in Case 1, rather 
it is restricted to the clusters that have 6 or more galaxies.
The number of fitted clusters fluctuates from 20 to 29 from simulation
to simulation. The mean FP parameters do not show differences from
Case 1. 

Cases 3 to 8 examine fits performed on the subset of the clusters 
having 10 or more
galaxies (Case 3) to clusters having 3 or more galaxies (Case 8).
The mean FP parameters derived for Case 3 are the ones with the
smallest biases. However, only 5-6 clusters for $\approx 70-80$  galaxies are
fit. As a consequence, the statistical error on the parameters is
nearly a factor two larger than for Case 1. On the contrary, Case 8 
fits a larger number of galaxies and clusters than Case 1, and
therefore gives slightly smaller ($\approx 20$\%) 
statistical errors. However, the mean values of the FP parameters are
biased more than for Case 1. In particular, the value of \sigone\ is
spuriously small, as offsetting the FP with spurious peculiar velocity 
suppresses the apparent scatter. We conclude that the list of clusters
adopted in Paper VII is a reasonable compromise between the need of
obtaining bias-free parameters and minimizing the statistical errors.

Cases 9 to 12 explore the effects of different \DWcut. Low \DWcut\
give slightly smaller $a$ coefficients; in addition, \meanRe\ is too
large, since essentially no normalization correction is
applied. However, the statistical errors are smaller, because the
sample size is maximized. In contrast, large \DWcut\ give larger $a$
coefficients and too small \meanRe. In addition, the statistical
errors are larger, because the sample size is reduced. We conclude
that again Case 1 is the reasonable compromise both in terms of bias
and statistical errors. This choice, however, forces us to apply
additional corrections when deriving the peculiar velocities of the whole
cluster sample (see \S\ref{resbias}).

Case 13 ignores selection probabilities altogether and applies a
uniform weight to all galaxies, resulting in an effective
over-weighting of the larger galaxies. As a consequence, the mean
values of \meanRe, \meansig, and \meanSBe\ are biased to higher values.
A similar effect, but of reduced amplitude, is observed for Case 14,
where galaxies with selection probabilities lower than 0.2 are
excluded. On the contrary, Case 15 includes essentially all 
galaxies in the fit,
irrespective of their selection probabilities. The biases in the
recovered parameters are reduced, but the statistical errors increase,
because of the presence of a few (deviant) points with large weights.
To conclude, weighting the data points with their inverse selection
probability, and excluding galaxies with selection probabilities lower
than 0.1 is a reasonable compromise that minimizes the biases and
the statistical errors of the recovered parameters.

Case 16 shows that excluding the (few) galaxies with large errors on
$\sigma$ does not affect the results of the fits, both the mean and
rms values of the parameters.

Case 17 excludes not only the galaxies rejected from the standard fit,
but also galaxies with low likelihoods ($\ln {\cal L}<0$); this
results in a highly biased fit, with low $a$ coefficient, 
an artificially lowered FP scatter and a substantially narrower 
distribution in the FP. 
Case 18 replaces the individual error estimates for all measured
quantities with uniform (average) errors. This has little effect on
the derived parameters, but the statistical errors are larger.

\begin{figure}
\plotone{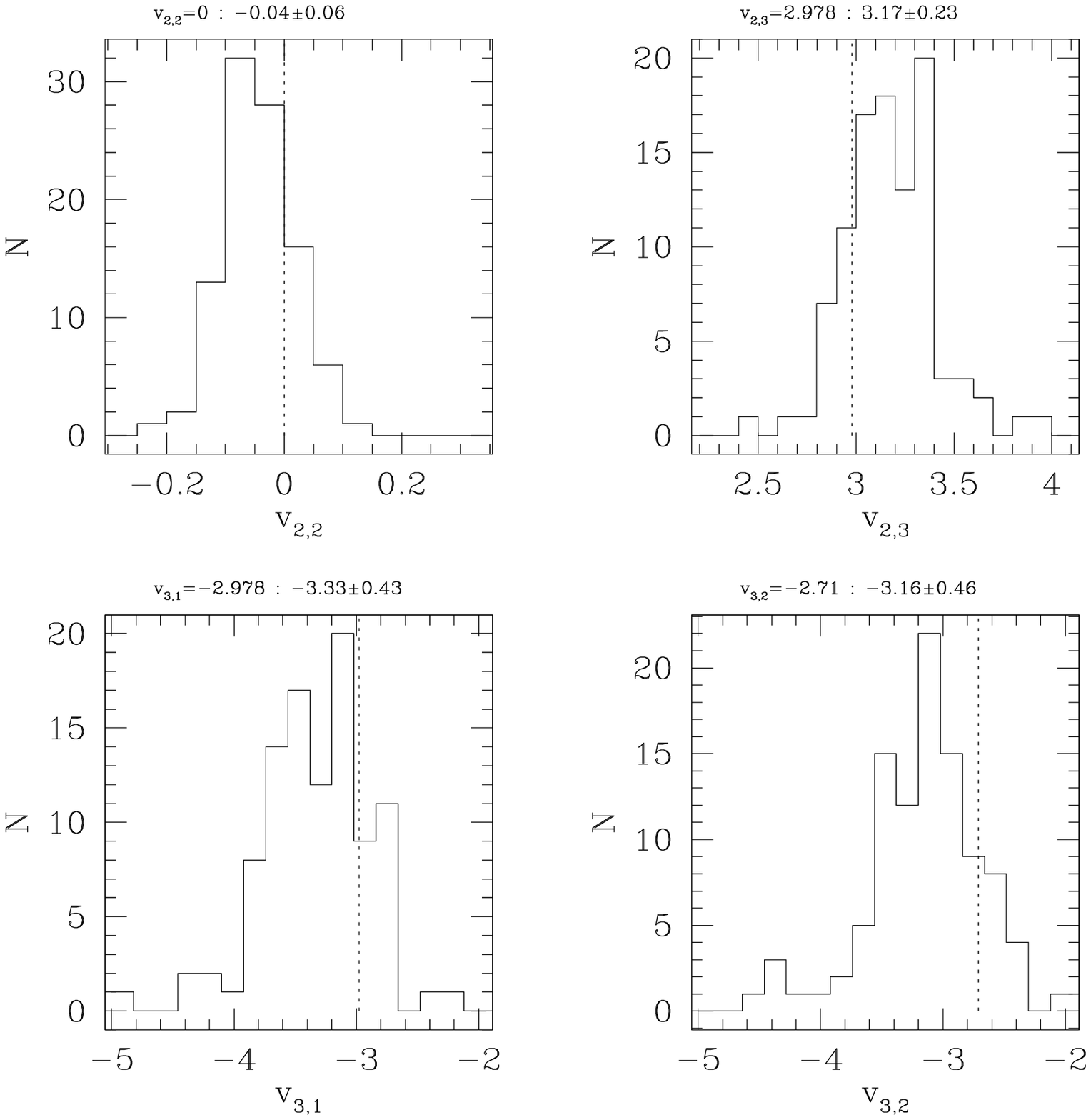}
\caption{The distributions of the second and
third component of vector $\vec{v}_2$, and first and second component
of vector $\vec{v}_3$ (see Eq. \ref{eqthird}) resulting from fitting 99
simulations allowing the orientation of the major axis of the galaxy 
distribution within the FP to be a free parameter (Case 19). 
The input parameters of the simulations are given at the head of each
panel (and indicated by the vertical dotted line), followed by the
mean and rms of the fits to the simulations.}
\label{figthird}
\end{figure}

Case 19 allows an extra degree of freedom by permitting the orientation
of the major axis of the galaxy distribution within the FP to be
fitted, rather than specified a priori. The vectors $\vec{v}_2$ and
$\vec{v}_3$ of Eq. \ref{eqrot} read now:
\begin{eqnarray}
\label{eqthird}
\vec{v}_2 & = &\hat{\vec{x}}_1+v_{2,2}\hat{\vec{x}}_2+v_{2,3}\hat{\vec{x}}_3,\\
\vec{v}_3 & = & v_{3,1}\hat{\vec{x}}_1+v_{3,2}\hat{\vec{x}}_2+\hat{\vec{x}}_3.\nonumber
\end{eqnarray}
The only parameters affected
are \sigtwo\ and \sigthree\, which are biased high and low respectively.
Fig. \ref{figthird} shows the histogram of the recovered second and
third component of vector $\vec{v}_2$, and first and second component
of vector $\vec{v}_3$. The input parameters (derived from the input $a$ and
$b$ values and Eq. \ref{eqrot}) are recovered with residual biases
that are smaller than the statistical error. Therefore in Paper VII we
use the simplifying approximation of Eq. \ref{eqrot}.

Cases 20 to 23 explore the effects of errors on the parameters of the
selection functions determined in Paper I. Case 20 considers mock
catalogues where the selection probabilities used in the fitting
procedure have been computed adding +0.1 dex to the true \logDw. As a
consequence, the selection weights are systematically higher than the
true ones (see Fig. \ref{figalllike}) and the mean number of selected
galaxies (i.e. with selection probabilities larger than 0.1) is
systematically smaller (see Table \ref{tabsimumean}). 
In Case 21 we added -0.1 dex to
the true \logDw, getting selection weights systematically smaller than
the true ones (see Fig. \ref{figalllike}) and therefore a larger mean
number of selected galaxies. Cases 22 and 23 use
selection weights distorted by adding +0.1 dex (Case 22) and -0.1 dex 
(Case 23) to the selection probability widths $\delta_W$. The
largest systematic effects are observed for Case 20, where
systematically smaller values of $a$, $b$, \meansig, and
systematically larger values of $c$ and \meanRe\ (because more small
galaxies are excluded from the fit) are
obtained. See \S\ref{alllike} and Fig. \ref{figalllike} 
for a discussion of the effects on the distribution of 
likelihood probabilities.

\begin{figure}
\plotone{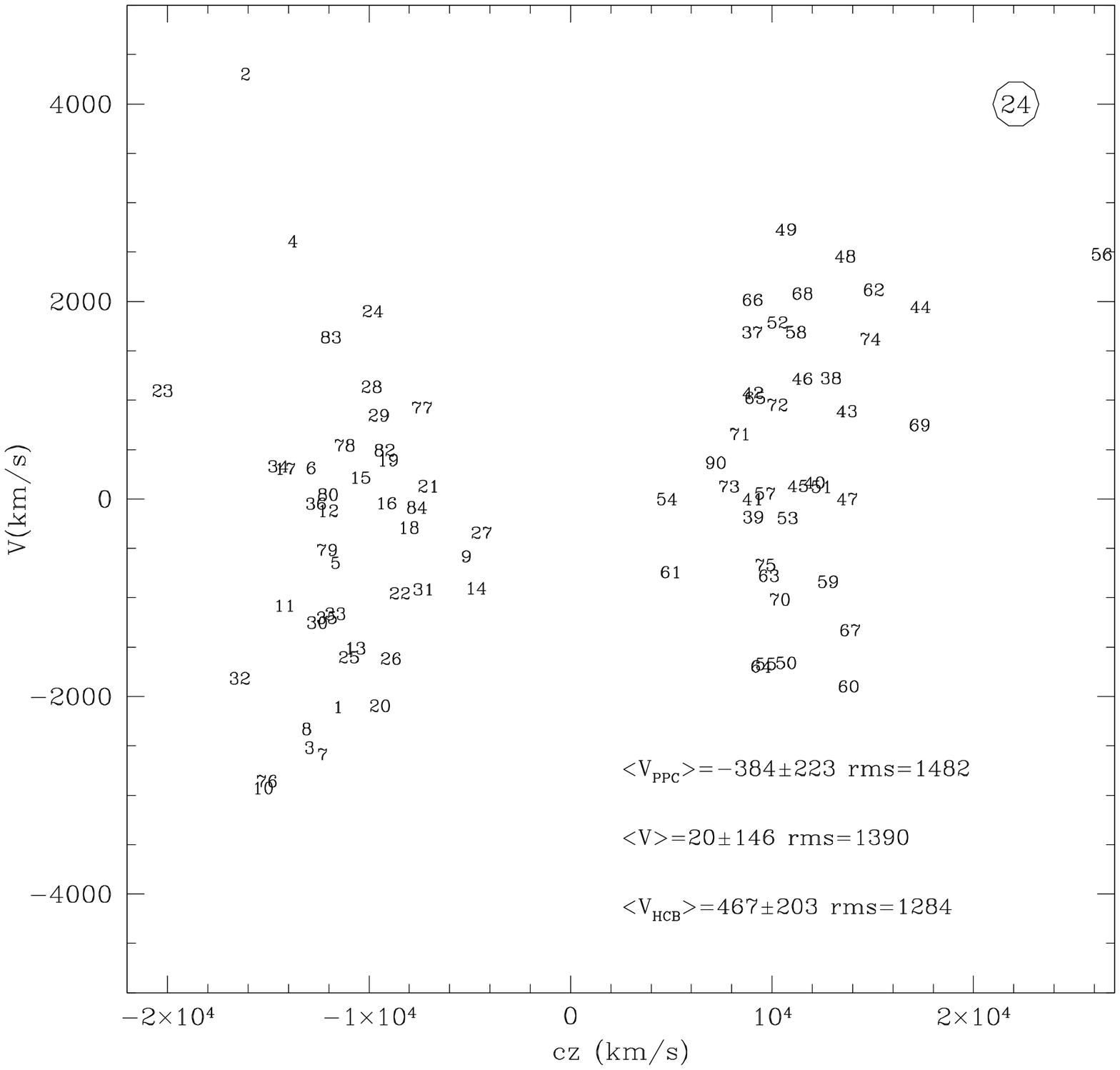}
\caption{The input peculiar velocity field of the simulations of Case 24
 of Table \ref{tabsimumean}. Each cluster is identified by its Cluster
Assignment Number (hereafter CAN, see Paper II). 
Clusters in the southern Galactic hemisphere are plotted at
negative redshifts. The resulting mean velocities (with statistical
errors) for the whole
sample, the clusters at positive (the Hercules
Corona Borealis sample, HCB) and negative Galactic longitudes (the Pisces
Perseus Cetus sample, PPC) are given with their rms.}
\label{figpecinput}
\end{figure}

\begin{figure}
\plotone{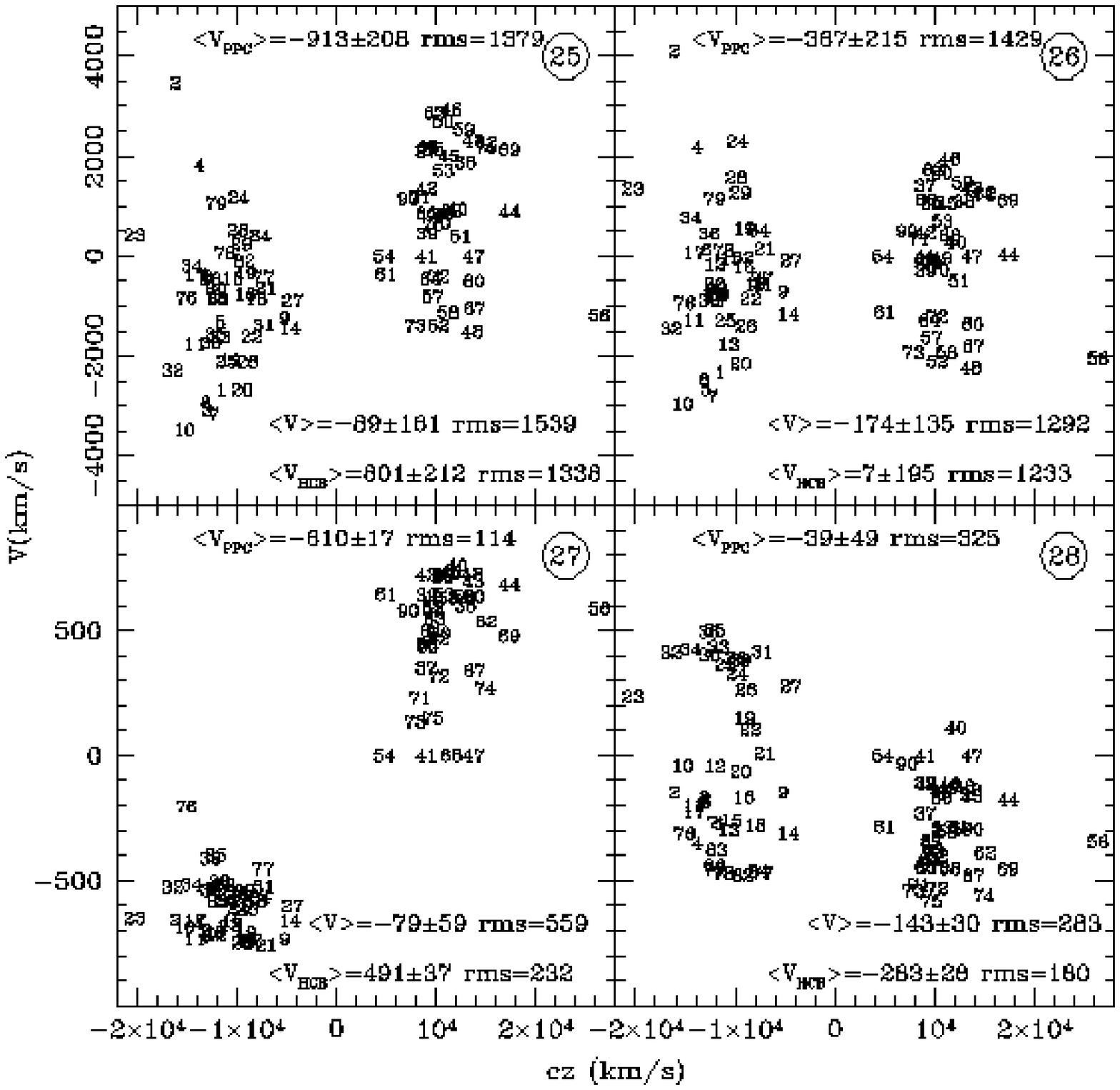}
\caption{The input peculiar velocity fields of the simulations of
Cases 25 to 28 of Table \ref{tabsimumean}. Labels as in Fig. 
\ref{figpecinput}.}
\label{figlpsmacinput}
\end{figure}

Cases 24 to 28 refer to mock catalogues of the EFAR sample with
peculiar velocity fields. Case 24 has a random peculiar velocity field
(always the same for the performed 99 simulations) 
of 0.05 dex rms amplitude. We considered this particular field 
(see Fig. \ref{figpecinput}, where we plot the peculiar velocities of
all clusters, the 29 FP plus the remaining ones) 
because its mean bulk motions 
of the clusters at positive (the Hercules
Corona Borealis sample, HCB) and negative Galactic longitudes (the Pisces
Perseus Cetus sample, PPC) are slightly non-zero and of opposite sign,
reminiscent of what measured in Paper VII. In
addition, its mean velocities projected along the LP and SMAC dipole 
directions (see Fig. \ref{figlpsmac} for the peculiar velocity field
of all clusters) are also similar to what seen in 
Paper VII.
Case 25 has a random field (always the same for the performed 99
simulations) of the same rms amplitude as Case 24 (0.05 dex)
plus the bulk flow motions determined by Lauer and Postman (1994,
hereafter LP, see Fig. \ref{figlpsmacinput} for the peculiar velocity
field of all clusters).  
LP measured the bulk motion of the Abell Clusters with
$cz<15000$ km/s with respect to the Cosmic Microwave Background to be
764 km/s in the direction of $l=341, b=44$ (reanalysis by Colless
1995), using the brightest cluster galaxies $L-\alpha$ relation. 
The resulting peculiar velocity field has a prominent asymmetry of
the mean bulk motions of the HCB and PPC clusters (see Fig. 
\ref{figlpsmacinput}), which gives an apparently larger dipole motion
in the LP direction  (see Fig. \ref{figlpsmac}).
Case 26 adds on top of a random
field of 0.05 rms amplitude (always the same for the performed 99 
simulations) the bulk motion determined by Hudson et
al. (1999, hereafter SMAC).  SMAC determined the motions of the Abell
Clusters with $cz<12000$ km/s using the FP, finding $V_{bulk}=630$
km/s towards $l=260, b=-1$. 
The resulting peculiar velocity field has 
small mean bulk motions of the HCB and PPC clusters (see Fig. 
\ref{figlpsmacinput} for the peculiar velocity field of all clusters), 
and an approximately zero
dipole motion in the SMAC direction (see Fig. \ref{figlpsmac}).
Case 27 has a pure LP bulk motion without
a random component.  Case 28 has a pure SMAC bulk motion without a
random component. In all five cases the FP parameters are recovered
without additional biases with respect to Case 1. The statistical
errors are slightly larger, especially for the \meanRe\ parameter.
The mean number of galaxies varies slightly, as galaxies near the
selection cut might get included or not in the sample accordingly to
the positive or negative peculiar velocity of the clusters they belong
to (see Sect. \ref{mock}). 
We conclude that the ML algorithm recovers the FP parameters correctly
also when peculiar velocity fields are present.

The last three sets of simulations (Cases 45, 46 and 47) refer to
input distributions uniform in the FP parameters, in the errors, or
both (see Sect. \ref{mock}), with zero  peculiar velocity fields. 
Inspection of Tables \ref{tabsimumean}
and \ref{tabsimurms} and Fig. \ref{figparvar} shows that also in these
cases the input parameters are recovered with small biases. The
largest effects are seen for $\sigma_2$, where the ML gaussian
algorithm overestimates the effects of the uniform error
distribution, biasing $\sigma_2$ low. The rms values are similar to
the ones derived for the previous cases: $da, db, dc$ are marginally
larger, $d\sigma_2$ marginally smaller.

Summarizing, with only a few exceptions (noted above), the
systematic differences in the fits derived for different cases are
comparable to the random errors in the determination of the parameters
for the standard case. Therefore, the uncertainties in our best-fit FP
parameters are dominated by the random errors and not by systematic
effects from the fitting method. In addition, the systematic effects
explored in the simulations (Tables \ref{tabsimumean} and
\ref{tabsimurms}) 
reproduce, at least qualitatively, the trends observed in the fits to
the actual EFAR data (Table 4 of Paper VII).

\begin{table*}
\caption[simu]{The mean values (over 99 simulations) of the parameters of the Fundamental Plane derived for
various cases.}
\begin{tabular}{rrrrrrrrrrrrl}
\noalign{\smallskip}
Case &	$N_{cl}$ & $N_{gal}$ &    $a$  &	$b$    &     $c$    &  \meanRe    & \meansig  & \meanSBe    &	 \sigone	   &   \sigtwo	&   \sigthree   & Notes \\
0    &	  29  &       255  &     1.223 &     0.336 &     -8.666 &     0.770  &     2.304 &      19.71  &   0.064   &    1.995   &   0.610  & standard fit Paper VII\\ %
1    &	  29  &       240  &     1.150 &     0.329 &     -8.360 &     0.808  &     2.311 &      19.77  &   0.058   &    1.946   &   0.643  & standard fit 1000 simulations \\ 
1    &	  29  &       240  &     1.168 &     0.330 &     -8.418 &     0.799  &     2.310 &      19.76  &   0.058   &    1.950   &   0.644  & standard fit 99 simulations \\
2    &	  26  &       224  &     1.175 &     0.330 &     -8.427 &     0.797  &     2.308 &      19.76  &   0.058   &    1.954   &   0.648  & uses clusters with $N_{gal}\ge 6$\\
3    &	   5  &        80  &     1.222 &     0.331 &     -8.563 &     0.774  &     2.303 &      19.72  &   0.066   &    1.963   &   0.661  & uses clusters with $N_{gal}\ge 10$\\
4    &	  13  &       144  &     1.200 &     0.331 &     -8.517 &     0.791  &     2.305 &      19.76  &   0.061   &    1.948   &   0.653  & uses clusters with $N_{gal}\ge 8$\\ 
5    &	  18  &       176  &     1.194 &     0.331 &     -8.506 &     0.789  &     2.304 &      19.75  &   0.060   &    1.948   &   0.655  & uses clusters with $N_{gal}\ge 7$\\ 
6    &	  35  &       269  &     1.164 &     0.330 &     -8.401 &     0.803  &     2.311 &      19.76  &   0.057   &    1.955   &   0.643  & uses clusters with $N_{gal}\ge 5$\\ 
7    &	  47  &       317  &     1.148 &     0.329 &     -8.345 &     0.809  &     2.314 &      19.77  &   0.056   &    1.959   &   0.637  & uses clusters with $N_{gal}\ge 4$\\ 
8    &    62  &       364  &     1.138 &     0.329 &     -8.325 &     0.810  &     2.314 &      19.77  &   0.055   &    1.960   &   0.638  & uses clusters with $N_{gal}\ge 3$\\ 
9    &	  29  &       247  &     1.096 &     0.328 &     -8.203 &     0.818  &     2.325 &      19.75  &   0.057   &    1.976   &   0.580  & no $D_W$ cut is applied \\
10   &	  29  &       247  &     1.097 &     0.328 &     -8.205 &     0.818  &     2.325 &      19.75  &   0.057   &    1.975   &   0.580  & $D_{Wcut}=6.3$ kpc      \\ 
11   &	  29  &       233  &     1.227 &     0.333 &     -8.611 &     0.762  &     2.283 &      19.76  &   0.059   &    1.926   &   0.711  & $D_{Wcut}=14.1$ kpc     \\ 
12   &	  29  &       219  &     1.266 &     0.334 &     -8.729 &     0.705  &     2.237 &      19.78  &   0.060   &    1.903   &   0.783  & $D_{Wcut}=15.9$ kpc      \\ 
13   &	  29  &       240  &     1.141 &     0.329 &     -8.325 &     0.852  &     2.339 &      19.80  &   0.058   &    1.932   &   0.614  & uses no selection weighting \\     
14   &	  29  &       231  &     1.131 &     0.327 &     -8.275 &     0.828  &     2.325 &      19.78  &   0.057   &    1.941   &   0.618  & uses galaxies with $S_i>0.2$\\   
15   &    29  &       243  &     1.195 &     0.332 &     -8.512 &     0.771  &     2.294 &      19.74  &   0.058   &    1.943   &   0.673  & uses galaxies with $S_i>0.001$ \\ 
16   &	  29  &       223  &     1.165 &     0.330 &     -8.414 &     0.799  &     2.309 &      19.76  &   0.058   &    1.944   &   0.649  & uses galaxies with $\delta\sigma<0.1$\\  
17   &	  29  &       171  &     1.054 &     0.330 &     -8.160 &     0.786  &     2.317 &      19.72  &   0.026   &    1.484   &   0.506  & excludes galaxies with $\ln {\cal L}<0$\\
18   &    29  &       240  &     1.158 &     0.328 &     -8.356 &     0.803  &     2.315 &      19.76  &   0.057   &    1.978   &   0.646  & uses uniform errors for all galaxies\\
19   &    29  &       240  &     1.168 &     0.331 &     -8.443 &     0.807  &     2.312 &      19.77  &   0.058   &    2.207   &   0.478  & also fit third axis of FP\\ 
20   &	  29  &       228  &     1.106 &     0.326 &     -8.187 &     0.818  &     2.320 &      19.77  &   0.055   &    1.955   &   0.608  & all $\log D_W^0$ shifted by +0.1\\  
21   &    29  &       243  &     1.179 &     0.330 &     -8.454 &     0.810  &     2.316 &      19.77  &   0.059   &    1.943   &   0.652  & all $\log D_W^0$ shifted by -0.1\\  
22   &	  29  &       243  &     1.177 &     0.330 &     -8.437 &     0.803  &     2.313 &      19.76  &   0.059   &    1.947   &   0.646  & all $\delta_W$ shifted by +0.1\\ 
23   &	  29  &       229  &     1.129 &     0.328 &     -8.291 &     0.817  &     2.319 &      19.77  &   0.055   &    1.943   &   0.625  & all $\delta_W$ shifted by -0.1\\ 
24   &	  29  &       239  &     1.185 &     0.331 &     -8.488 &     0.797  &     2.309 &      19.77  &   0.059   &    1.923   &   0.648  & random peculiar velocity field\\ 
25   &	  29  &       237  &     1.183 &     0.335 &     -8.563 &     0.791  &     2.307 &      19.78  &   0.058   &    1.926   &   0.653  & LP dipole plus random component\\ 
26   &	  29  &       238  &     1.194 &     0.332 &     -8.531 &     0.788  &     2.307 &      19.78  &   0.057   &    1.950   &   0.654  & SMAC dipole plus random component\\ 
27   &	  29  &       240  &     1.170 &     0.330 &     -8.425 &     0.798  &     2.310 &      19.77  &   0.057   &    1.961   &   0.643  & pure LP dipole\\ 
28   &	  29  &       240  &     1.161 &     0.333 &     -8.464 &     0.796  &     2.309 &      19.77  &   0.058   &    1.940   &   0.642  & pure SMAC dipole  \\ 
45   &    29  &       240  &     1.142 &     0.328 &     -8.311 &     0.804  &     2.312 &      19.76  &   0.058   &    1.986   &   0.665  & FP uniform distribution\\
46   &    29  &       240  &     1.156 &     0.336 &     -8.512 &     0.805  &     2.312 &      19.77  &   0.059   &    1.698   &   0.633  & Error uniform distribution\\
47   &    29  &       239  &     1.146 &     0.336 &     -8.482 &     0.812  &     2.314 &      19.78  &   0.058   &    1.740   &   0.643  & FP and Error uniform distribution\\

\end{tabular}
\label{tabsimumean}
\end{table*}

\begin{table*}
\caption[simu]{The rms values (over 99 simulations) of the parameters of the Fundamental Plane derived for
various cases.}
\begin{tabular}{rrrrrrrrrrrrl}
\noalign{\smallskip}
Case &	$dN_{cl}$ & $dN_{gal}$ & d$a$ &	d$b$    &    d$c$ & d\meanRe&d\meansig& d\meanSBe  & d\sigone	&   d\sigtwo & d\sigthree  & Notes\\
1    & 0    &  4  &  0.089 &  0.013  &   0.354 &   0.025 &   0.016 &   0.060 &  0.006  &   0.158  &  0.067   & standard fit 1000 simulations \\ 
1    & 0    &  4  &  0.089 &  0.013  &   0.347 &   0.027 &   0.018 &   0.048 &  0.006  &   0.168  &  0.066   & standard fit 99 simulations \\
2    & 1    & 10  &  0.093 &  0.014  &   0.360 &   0.027 &   0.018 &   0.048 &  0.006  &   0.172  &  0.068   & uses clusters with $N_{gal}\ge 6$\\
3    & 1    &  8  &  0.141 &  0.021  &   0.533 &   0.047 &   0.028 &   0.087 &  0.011  &   0.246  &  0.108   & uses clusters with $N_{gal}\ge 10$\\
4    & 1    & 10  &  0.117 &  0.016  &   0.445 &   0.038 &   0.025 &   0.063 &  0.008  &   0.198  &  0.089   & uses clusters with $N_{gal}\ge 8$\\ 
5    & 1    &  9  &  0.106 &  0.015  &   0.411 &   0.033 &   0.022 &   0.057 &  0.007  &   0.179  &  0.079   & uses clusters with $N_{gal}\ge 7$\\ 
6    & 2    &  9  &  0.085 &  0.013  &   0.338 &   0.026 &   0.017 &   0.044 &  0.006  &   0.156  &  0.062   & uses clusters with $N_{gal}\ge 5$\\ 
7    & 1    &  7  &  0.079 &  0.012  &   0.328 &   0.022 &   0.014 &   0.043 &  0.005  &   0.143  &  0.054   & uses clusters with $N_{gal}\ge 4$\\ 
8    & 2    &  6  &  0.074 &  0.012  &   0.309 &   0.022 &   0.014 &   0.041 &  0.005  &   0.136  &  0.054   & uses clusters with $N_{gal}\ge 3$\\ 
9    & 0    &  3  &  0.067 &  0.011  &   0.274 &   0.019 &   0.009 &   0.048 &  0.005  &   0.162  &  0.039   & no $D_W$ cut is applied \\
10   & 0    &  3  &  0.067 &  0.011  &   0.275 &   0.018 &   0.009 &   0.046 &  0.005  &   0.162  &  0.039   & $D_{Wcut}=6.3$ kpc      \\ 
11   & 0    &  5  &  0.097 &  0.014  &   0.380 &   0.041 &   0.028 &   0.048 &  0.007  &   0.167  &  0.075   & $D_{Wcut}=14.1$ kpc     \\ 
12   & 0    &  6  &  0.097 &  0.014  &   0.366 &   0.068 &   0.055 &   0.059 &  0.007  &   0.157  &  0.109   & $D_{Wcut}=15.9$ kpc      \\ 
13   & 0    &  4  &  0.068 &  0.011  &   0.281 &   0.016 &   0.010 &   0.037 &  0.005  &   0.122  &  0.048   & uses no selection weighting \\     
14   & 0    &  5  &  0.077 &  0.013  &   0.323 &   0.021 &   0.012 &   0.046 &  0.006  &   0.149  &  0.057   & uses galaxies with $S_i>0.2$\\   
15   & 0    &  3  &  0.101 &  0.014  &   0.385 &   0.047 &   0.032 &   0.064 &  0.006  &   0.179  &  0.104   & uses galaxies with $S_i>0.001$ \\ 
16   & 0    &  3  &  0.093 &  0.013  &   0.362 &   0.029 &   0.019 &   0.048 &  0.006  &   0.171  &  0.070   & uses galaxies with $\delta\sigma<0.1$\\  
17   & 0    & 10  &  0.150 &  0.026  &   0.667 &   0.030 &   0.017 &   0.064 &  0.010  &   0.268  &  0.080   & excludes galaxies with $\ln {\cal L}<0$\\
18   & 0    &  4  &  0.106 &  0.014  &   0.385 &   0.031 &   0.024 &   0.047 &  0.009  &   0.175  &  0.114   & uses uniform errors for all galaxies\\
19   & 0    &  4  &  0.087 &  0.013  &   0.333 &   0.025 &   0.017 &   0.049 &  0.006  &   0.161  &  0.040   & also fit third axis of FP\\ 
20   & 0    &  5  &  0.082 &  0.014  &   0.359 &   0.022 &   0.014 &   0.054 &  0.006  &   0.172  &  0.059   & all $\log D_W^0$ shifted by +0.1\\  
21   & 0    &  3  &  0.086 &  0.012  &   0.325 &   0.027 &   0.018 &   0.042 &  0.006  &   0.145  &  0.069   & all $\log D_W^0$ shifted by -0.1\\  
22   & 0    &  3  &  0.087 &  0.012  &   0.329 &   0.027 &   0.018 &   0.041 &  0.006  &   0.146  &  0.067   & all $\delta_W$ shifted by +0.1\\ 
23   & 0    &  5  &  0.086 &  0.015  &   0.379 &   0.026 &   0.016 &   0.056 &  0.006  &   0.174  &  0.064   & all $\delta_W$ shifted by -0.1\\ 
24   & 0    &  4  &  0.087 &  0.014  &   0.357 &   0.032 &   0.020 &   0.059 &  0.006  &   0.127  &  0.068   & random peculiar velocity field\\ 
25   & 0    &  4  &  0.096 &  0.014  &   0.410 &   0.030 &   0.019 &   0.058 &  0.006  &   0.156  &  0.072   & LP dipole plus random component\\ 
26   & 0    &  4  &  0.078 &  0.013  &   0.354 &   0.029 &   0.016 &   0.063 &  0.005  &   0.139  &  0.065   & SMAC dipole plus random component\\ 
27   & 0    &  3  &  0.087 &  0.016  &   0.398 &   0.028 &   0.019 &   0.057 &  0.006  &   0.170  &  0.070   & pure LP dipole\\ 
28   & 0    &  3  &  0.096 &  0.012  &   0.378 &   0.029 &   0.019 &   0.056 &  0.005  &   0.147  &  0.079   & pure SMAC dipole  \\ 
45   & 0    &  4  &  0.096 &  0.011  &   0.356 &   0.027 &   0.020 &   0.045 &  0.005  &   0.126  &  0.072   & FP uniform distribution\\
46   & 0    &  3  &  0.093 &  0.016  &   0.402 &   0.022 &   0.015 &   0.045 &  0.006  &   0.138  &  0.058   & Error uniform distribution\\
47   & 0    &  4  &  0.100 &  0.017  &   0.438 &   0.031 &   0.021 &   0.053 &  0.005  &   0.116  &  0.073   & FP and Error uniform distribution\\
\end{tabular}
\label{tabsimurms}
\end{table*}

\subsubsection{The likelihood distribution}
\label{alllike}

\begin{figure}
\plotone{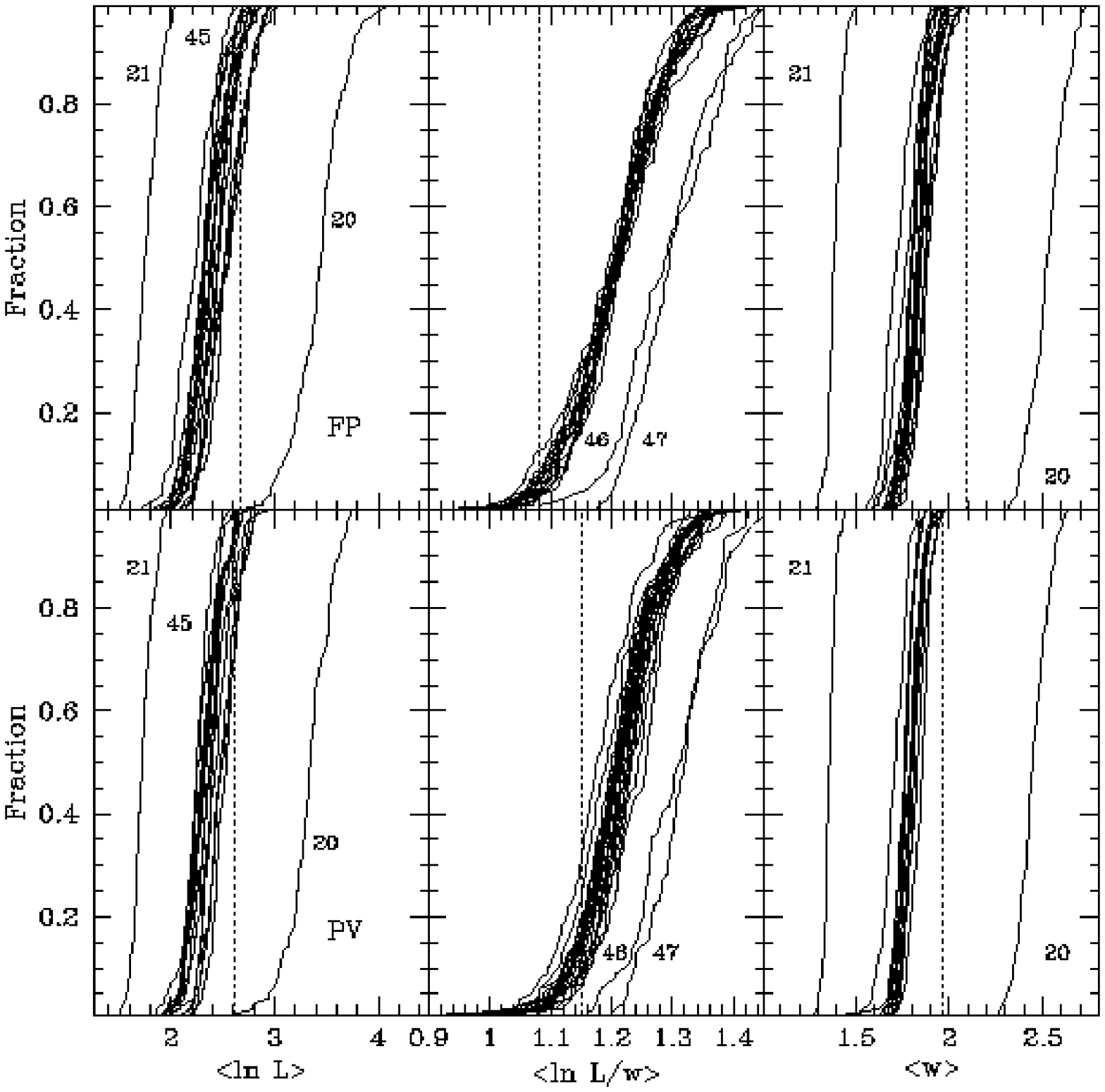}
\caption{The cumulative distributions of the mean likelihoods (left),
mean unweighted likelihoods (center), mean selection weights (right).
The top row refers to the galaxies of the clusters used in the
determination of the FP
parameters, the bottom row to all galaxies of the clusters for which peculiar
velocities are computed. The vertical dashed lines show the values of
EFAR sample, as determined in Paper VII. The numbers identify the
Cases discussed in the text.}
\label{figalllike}
\end{figure}

We now investigate 
whether the assumption of a gaussian distribution of galaxies in the
FP space is compatible with
the EFAR data sample. Fig. \ref{figalllike} (left panel) shows the 
cumulative distributions of the mean (i.e. normalized to the number of
galaxies fitted) likelihoods of the simulations of Cases 1
to 8, 18 to 28, and 45 to 47 (top row, only galaxies of clusters used to
determine the FP parameters) and  Cases 1 and 20 to 47 (bottom row,
all galaxies of the
clusters for which peculiar velocities have been computed). 
Note that Cases 29 to 44, not listed in Tables \ref{tabsimumean} and 
\ref{tabsimurms}, are discussed in detail in \S\ref{resbias}. They are
perturbed versions of Case 1. 
 
Except for Cases 20 and 21, all simulations (including the uniform
distribution cases 45, 46 and 47) give similar cumulative
distributions of the mean likelihoods. Depending on the exact case, a
mean likelihood equal to or larger than that of the EFAR sample is
observed in up to 30\% of the simulations. Case 45 (a uniform
distribution of the FP parameters and gaussian errors) gives only slightly
smaller mean likelihoods. Case 20 gives
systematically higher mean likelihoods, and Case 21 systematically
lower, than that of the EFAR sample. Inspection of the central and
left panel plots of Fig.  \ref{figalllike} clarifies that the
cumulative distribution of the mean unweighted likelihoods are similar
for all the simulations with gaussian error distributions, predicting
that values larger than that of the EFAR sample are observed in 70 to
95\% of the cases. The Cases 46 and 47, where the error distributions
are uniform, produce systematically 20-30\% larger values, because the
ML gaussian algorithm is overstressing the role of errors. As
noticed before (see Fig. \ref{figmg2sig}), the mean selection weights
of all simulations, except Case 20, are smaller than that of the EFAR
sample. 

Therefore, we conclude that the gaussian modeling is a
reasonable description of the distribution of the EFAR galaxies in the 
(\logRe, \logsig, \SBe) space and of the error distribution. 
The EFAR data set does not allow to discriminate between a gaussian or
a uniform distribution of the FP parameters. Error distributions with
tails slightly stronger than gaussians are hinted. 
Values of \logDwO\ slightly larger (by
$\approx 0.04$ dex) than that measured in Paper I are needed
to match the mean selection weight of the EFAR sample.

\subsubsection{Peculiar velocities and bias corrections}
\label{resbias}

Once the parameters of the best-fitting gaussian distribution are
determined using a subset of clusters (hereafter, the FP clusters), 
the ML algorithm is used
with these parameters fixed, to compute the peculiar velocities of the
remaining clusters.  Tables \ref{tabdelta}, \ref{tabdelta2938} and
\ref{tabdelta3947} give the average residuals (over 99 simulations) 
from the input peculiar velocities $\Deltadelta =
<\delta-\delta_{input}>$ for a number of cases.  Table \ref{tabdelta}
considers Cases 1 and 20 to 28 of Table \ref{tabsimumean}, listing
also the values of \Deltadelta\ given in Table 7 of Paper VII, based on
1000 simulations of Case 1, and the statistical precision reached for
Case 1 and 99 simulations. Tables \ref{tabdelta2938} and
\ref{tabdelta3947} give the results for 19 additional Cases, from 29
to 47. The 16 Cases from 29 to 44 are obtained by perturbing one of
the FP parameters by
plus or minus one sigma, keeping the others fixed to the best solution
value. Cases 29 and 30 perturb the $a$ coefficient, 31 and 32 $b$, 33
and 34 \meanRe, 35 and 36 \meansig, 37 and 38 \meanSBe, 39 and 40
\sigone, 41 and 42 \sigtwo, 43 and 44 \sigthree, respectively. Cases
45, 46 and 47 refer to the simulations with uniform distributions in
the FP parameters, the errors or both. 

\begin{figure}
\plotone{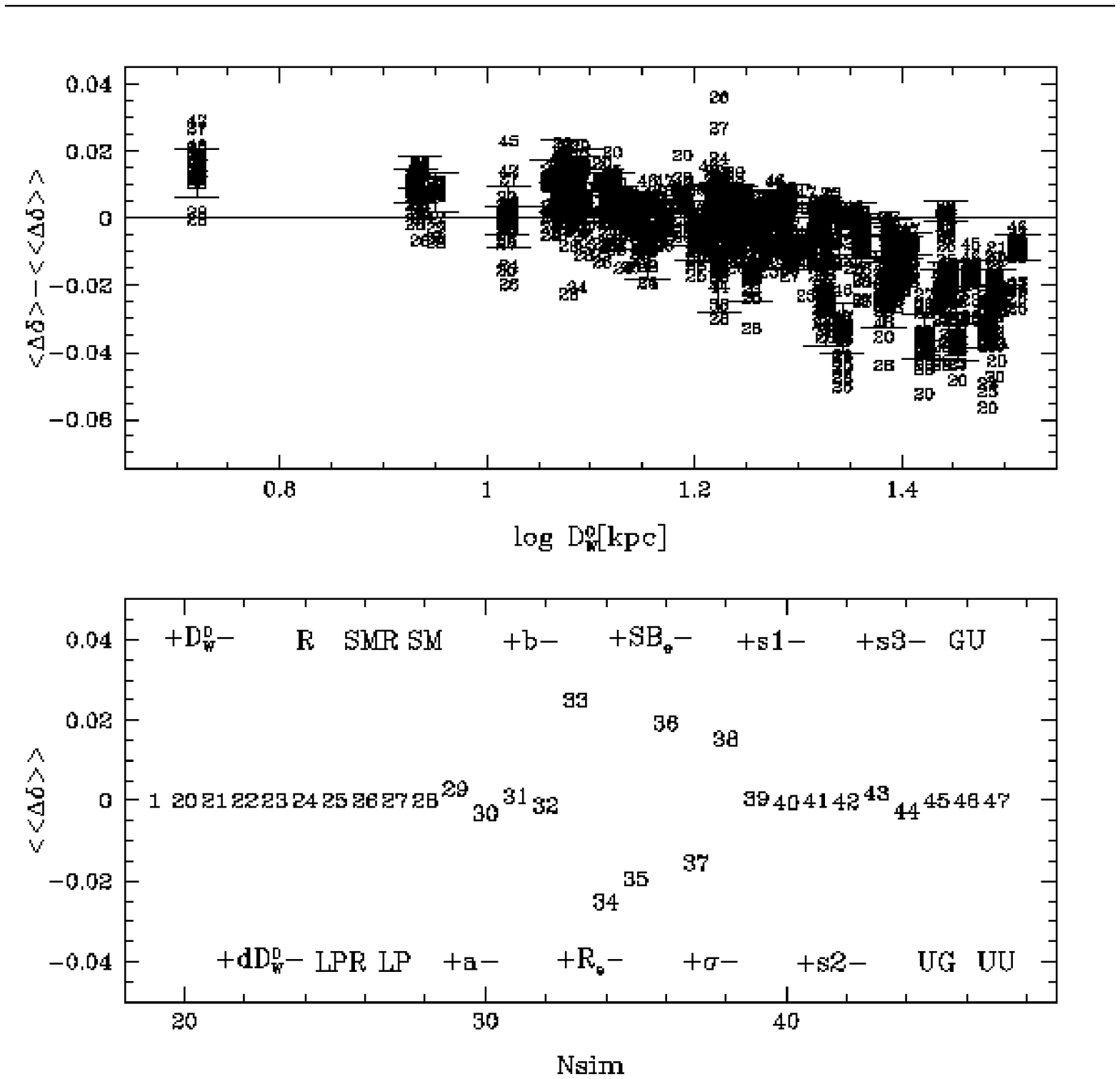}
\caption{The residuals selection bias determined from 99 simulations
and Cases 1 and 20 to 28 of Tables \ref{tabsimumean} and
\ref{tabdelta}, and Cases 29 to
44 of Tables \ref{tabdelta2938} and \ref{tabdelta3947}. The top panel
shows the residuals 
$<\Delta\delta>=<\delta-\delta_{input}>$ from the input peculiar
velocity field $\delta_{input}$ ($\delta_{input}=0$ for Cases 1, 20 to
23, and 29
to 44; see text, Table \ref{tabsimumean} and Figs. 
\ref{figpecinput}-\ref{figlpsmacinput} 
for Cases 23 to 28), with
the average residual over the FP clusters 
subtracted off, as a function of \logDwO . Each Case is identified by
its number (see Tables \ref{tabsimumean}, \ref{tabdelta2938}
and \ref{tabdelta3947}). The error bars show the
statistical precision reached for Case 1 with 99 simulations.
The bottom panel shows the average residual measured for the FP clusters
 as a function of the Case number. The labels provide
an additional mnemonic identifier.}
\label{figpecvar}
\end{figure}

Fig. \ref{figpecvar} shows \Deltadelta-$<\Deltadelta>$ as a
function of \logDwO.  Here $<\Deltadelta>$ is the average residual
over the FP clusters. It is zero by construction for Cases 1,  20
to 28 and 45 to 47. The general trend of increasingly negative
residuals at large values of \logDwO\ is easily explained. The
clusters at higher redshifts (large \logDwO) lack small galaxies and
therefore have mean values of \logRe\ larger than \meanRe, or 
$\delta=\meanRe-\logRe$ progressively more negative.

Once the zeropoint offset (given by the residual \Deltadelta averaged
over the FP clusters) is subtracted off, the simulations
29 to 44 give residuals compatible with Case 1 within the statistical
errors. Therefore, the residual bias corrections are robust against
small errors in the FP solution. The largest variations of the 
zeropoint are observed when the values of \meanRe, \meansig\ and
\meanSBe\ are perturbed. As expected, one finds
$<\Deltadelta>\approx \pm0.025=\pm d\meanRe$, when \meanRe\ is perturbed, 
$<\Deltadelta>\approx \pm0.02 =\mp a~d\meansig$, when
\meansig\ is perturbed, and $<\Deltadelta>\approx \pm0.02 =\mp b~d\meanSBe$
when \meanSBe\ is perturbed. Errors in \logDwO\ affect the selection
weighting scheme and therefore the determination of
\Deltadelta\ for the clusters (at higher redshifts) that have the
largest \logDwO. In some cases the errors in \Deltadelta\ are larger
than the statistical precision of Case 1. Positive errors ($+D_W^0$,
Case 20) produce too negative \Deltadelta, since more small galaxies
than with the correct values of \logDwO\ are excluded from the fits. 
The contrary happens for
negative errors ($-D_W^0$, Case 21). Errors in $\delta_W$ affect
also \Deltadelta, especially at higher redshifts. In this case
negative errors (-d$D^0_W$, Case 23) produce too negative \Deltadelta,
because a sharper (i.e., a smaller $\delta_W$) selection function
eliminates more small galaxies.
Positive errors (d$D^0_W$, Case 22) produce the reverse. 

The presence of a peculiar velocity field (Cases 24 to 28) can also affect
\Deltadelta, because redshift distances are used to convert the
measured selection cuts from arcsec to kpc. In particular, clusters
sampled by a small number of galaxies are prone to small number
statistics fluctuations, as galaxies near the selection cutoff might
not be always included in the simulation catalogues.

Uniform distributions of the FP parameters, of the errors or both do
not affect \Deltadelta\ systematically, within the statistical errors. 

\begin{figure}
\plotone{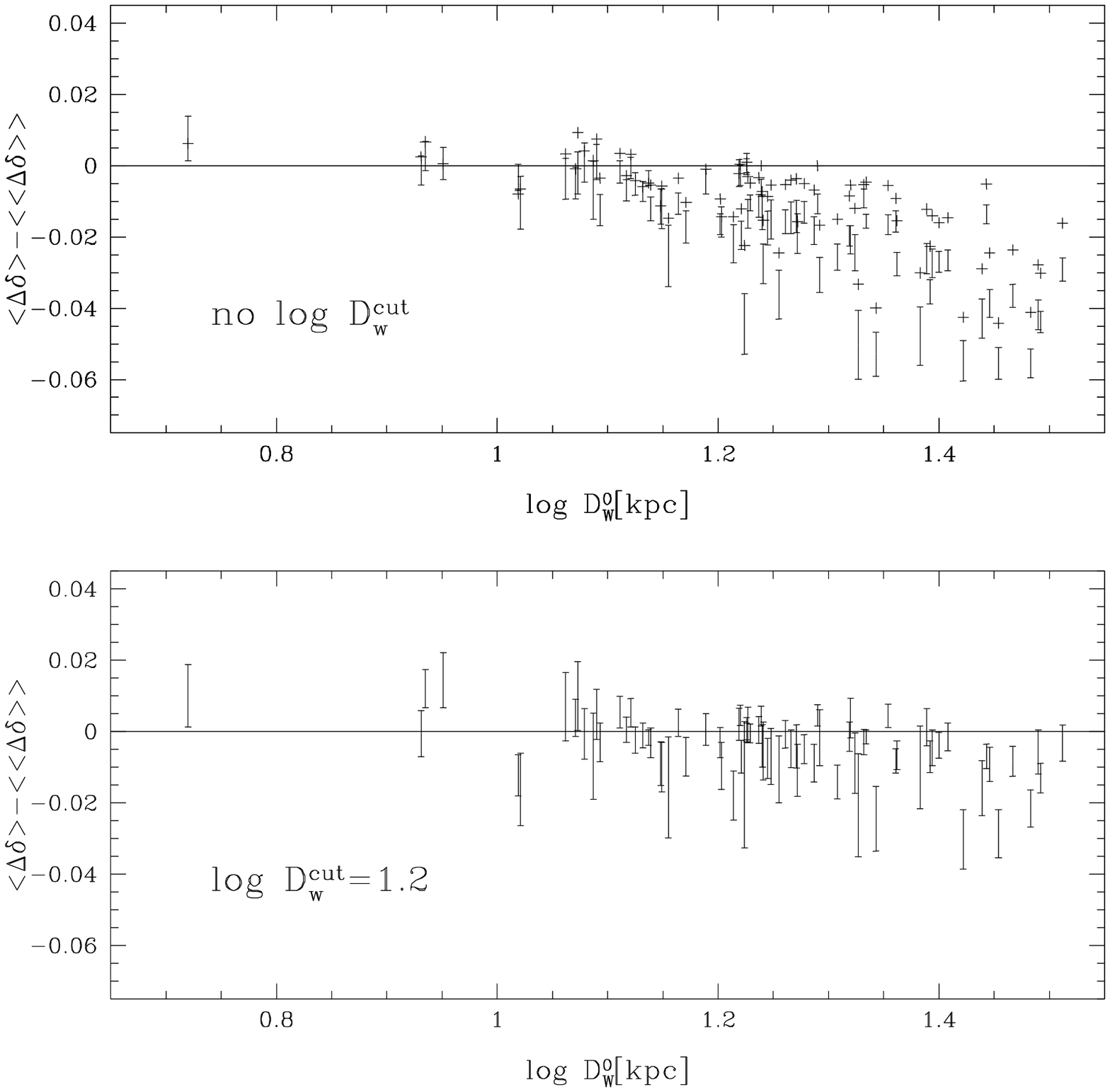}
\caption{The residual selection bias
$<\Delta\delta>-<<\Delta\delta>>$ determined for Case 9 (no \DWcut\
applied, top) and Case 12 (\DWcut=1.2, bottom), as a function of
\logDwO . The error bars show the
statistical precision reached with 99 simulations. The crosses show
the mean values for Case 1.}
\label{figpecdw}
\end{figure}

Finally, Fig. \ref{figpecdw} illustrates the effect of the \DWcut\ value on the
required peculiar velocities. When no \DWcut\ is applied (Case 9), 
the residual biases at high \logDwO\ are larger than when \DWcut\ is
equal to 1.1 (Case 1). The biases  are reduced if a larger 
\DWcut\ (1.2, Case 12) is chosen, but the statistical errors
increase. In addition, as discussed in
\S\ref{secvariant}, this choice gives larger
systematic and statistical errors for the FP parameters.

We conclude that the peculiar velocity biases are much smaller than
the random errors in the peculiar velocities. The uncertainties in the
peculiar velocity bias corrections as computed from the 1000
simulations of Case 1 used in Paper VII do not increase the random
errors in the peculiar velocities. However, residual systematic
effects at the level of 0.01 dex might still be present, especially
for the clusters at higher redshifts, due to the uncertainties in the
determination of the selection function.

\afterpage{

\begin{table*}
\caption[delta]{The mean residual selection bias $<\Delta\delta>$ for Cases
1 (for 1000 and 99 simulations, with the rms error of the 99
simulations), and 20 to 28 of Table \ref{tabsimumean}.}
\begin{tabular}{rrrrrrrrrrrrr}
\noalign{\smallskip}
CAN& 1 (1000)  & 1 (99)    &d$\delta$&    20    &  21     &   22    &  23     &    24    &    25   &   26    &   27    &   28 \\
\noalign{\smallskip}
 1 & $-$0.0182 &   0.0017 &  0.0048  & -0.0119 & -0.0028 & -0.0014 & -0.0081  &  0.0007 & -0.0051 & -0.0173 & -0.0121 & -0.0103\\
 2 & $-$0.0310 &  -0.0371 &  0.0053  & -0.0484 & -0.0299 & -0.0320 & -0.0435  & -0.0348 & -0.0350 & -0.0376 & -0.0436 & -0.0352\\
 3 & $-$0.0204 &  -0.0166 &  0.0037  & -0.0301 & -0.0137 & -0.0141 & -0.0244  & -0.0191 & -0.0303 & -0.0294 & -0.0156 & -0.0196\\
 4 & $-$0.0103 &  -0.0096 &  0.0061  & -0.0092 & -0.0110 & -0.0110 & -0.0088  & -0.0091 & -0.0136 & -0.0155 & -0.0174 & -0.0153\\
 5 & $-$0.0142 &  -0.0153 &  0.0130  & -0.0146 & -0.0117 & -0.0134 & -0.0151  &  0.0179 &  0.0100 &  0.0362 &  0.0269 & -0.0302\\
 6 & $-$0.0021 &   0.0036 &  0.0032  &  0.0081 &  0.0042 &  0.0042 &  0.0061  &  0.0021 & -0.0074 & -0.0051 & -0.0037 & -0.0051\\
 7 & $-$0.0218 &  -0.0174 &  0.0042  & -0.0306 & -0.0147 & -0.0155 & -0.0254  & -0.0275 & -0.0374 & -0.0346 & -0.0221 & -0.0215\\
 8 & $-$0.0107 &   0.0034 &  0.0053  &  0.0035 &  0.0012 &  0.0012 &  0.0057  &  0.0012 & -0.0002 & -0.0051 & -0.0109 & -0.0031\\
 9 & $+$0.0069 &   0.0085 &  0.0099  &  0.0110 &  0.0112 &  0.0094 &  0.0120  & -0.0207 & -0.0026 & -0.0018 & -0.0028 &  0.0154\\
10 & $-$0.0052 &   0.0018 &  0.0029  &  0.0053 &  0.0017 &  0.0007 &  0.0074  & -0.0074 & -0.0098 & -0.0125 & -0.0059 & -0.0066\\
11 & $-$0.0073 &  -0.0072 &  0.0052  & -0.0021 & -0.0074 & -0.0068 & -0.0062  & -0.0092 & -0.0050 & -0.0070 & -0.0065 & -0.0028\\
12 & $-$0.0024 &   0.0037 &  0.0036  &  0.0081 &  0.0028 &  0.0028 &  0.0077  & -0.0014 & -0.0044 & -0.0122 &  0.0020 & -0.0041\\
13 & $-$0.0011 &   0.0061 &  0.0036  &  0.0187 &  0.0034 &  0.0057 &  0.0118  &  0.0069 & -0.0044 & -0.0054 &  0.0041 &  0.0001\\
14 & $+$0.0033 &  -0.0015 &  0.0053  &  0.0047 & -0.0021 & -0.0015 &  0.0010  &  0.0041 &  0.0008 &  0.0056 &  0.0021 &  0.0037\\
15 & $+$0.0007 &   0.0112 &  0.0056  &  0.0214 &  0.0117 &  0.0137 &  0.0126  & -0.0015 &  0.0052 & -0.0227 &  0.0020 & -0.0083\\
16 & $+$0.0019 &   0.0029 &  0.0031  &  0.0105 &  0.0027 &  0.0037 &  0.0053  & -0.0001 & -0.0094 & -0.0085 & -0.0017 &  0.0018\\
17 & $-$0.0074 &  -0.0022 &  0.0033  &  0.0086 & -0.0042 & -0.0021 &  0.0023  & -0.0096 & -0.0172 & -0.0140 & -0.0126 & -0.0069\\
18 & $-$0.0001 &  -0.0032 &  0.0050  &  0.0027 & -0.0029 & -0.0028 & -0.0005  &  0.0037 & -0.0040 & -0.0081 &  0.0013 & -0.0001\\
19 & $+$0.0009 &   0.0005 &  0.0090  &  0.0033 &  0.0030 &  0.0011 &  0.0040  & -0.0142 &  0.0022 & -0.0196 &  0.0118 & -0.0033\\
20 & $-$0.0005 &   0.0071 &  0.0026  &  0.0135 &  0.0053 &  0.0062 &  0.0120  &  0.0013 &  0.0001 & -0.0084 &  0.0025 & -0.0020\\
21 & $+$0.0014 &   0.0043 &  0.0034  &  0.0109 &  0.0043 &  0.0050 &  0.0063  & -0.0018 & -0.0080 & -0.0080 & -0.0042 & -0.0003\\
22 & $+$0.0008 &   0.0036 &  0.0048  &  0.0089 &  0.0046 &  0.0047 &  0.0054  & -0.0025 & -0.0092 & -0.0053 &  0.0016 & -0.0112\\
23 & $-$0.0077 &   0.0019 &  0.0030  & -0.0056 &  0.0017 &  0.0000 &  0.0008  & -0.0011 & -0.0125 & -0.0081 & -0.0048 & -0.0050\\
24 & $-$0.0044 &   0.0015 &  0.0031  &  0.0004 &  0.0018 &  0.0011 &  0.0019  & -0.0008 & -0.0126 & -0.0151 & -0.0075 & -0.0012\\
25 & $-$0.0127 &  -0.0089 &  0.0033  & -0.0195 & -0.0067 & -0.0077 & -0.0169  & -0.0136 & -0.0208 & -0.0139 & -0.0152 & -0.0218\\
26 & $-$0.0073 &  -0.0072 &  0.0061  & -0.0032 & -0.0074 & -0.0070 & -0.0051  & -0.0025 & -0.0065 & -0.0062 & -0.0029 & -0.0049\\
27 & $+$0.0070 &   0.0133 &  0.0073  &  0.0179 &  0.0162 &  0.0142 &  0.0177  &  0.0179 &  0.0212 & -0.0004 &  0.0269 &  0.0016\\
29 & $-$0.0070 &  -0.0083 &  0.0066  & -0.0081 & -0.0073 & -0.0069 & -0.0092  & -0.0129 & -0.0015 & -0.0149 & -0.0042 & -0.0017\\
30 & $-$0.0135 &  -0.0174 &  0.0078  & -0.0241 & -0.0140 & -0.0158 & -0.0214  & -0.0080 & -0.0152 & -0.0329 & -0.0173 & -0.0158\\
31 & $+$0.0052 &   0.0096 &  0.0050  &  0.0116 &  0.0112 &  0.0104 &  0.0110  &  0.0130 &  0.0017 & -0.0018 &  0.0040 &  0.0094\\
32 & $-$0.0071 &   0.0018 &  0.0037  & -0.0009 & -0.0012 &  0.0002 & -0.0008  & -0.0140 & -0.0045 & -0.0127 & -0.0087 & -0.0065\\
33 & $-$0.0114 &  -0.0049 &  0.0060  & -0.0106 & -0.0058 & -0.0063 & -0.0069  & -0.0315 & -0.0197 & -0.0274 & -0.0207 & -0.0227\\
34 & $-$0.0219 &  -0.0090 &  0.0039  & -0.0272 & -0.0040 & -0.0071 & -0.0215  & -0.0206 & -0.0269 & -0.0238 & -0.0196 & -0.0214\\
35 & $-$0.0059 &   0.0024 &  0.0022  & -0.0029 &  0.0003 &  0.0005 & -0.0006  & -0.0008 & -0.0137 & -0.0136 & -0.0030 & -0.0073\\
36 & $-$0.0146 &  -0.0084 &  0.0036  & -0.0100 & -0.0072 & -0.0072 & -0.0069  & -0.0183 & -0.0250 & -0.0153 & -0.0243 & -0.0190\\
37 & $-$0.0018 &   0.0049 &  0.0038  &  0.0111 &  0.0048 &  0.0050 &  0.0088  & -0.0049 & -0.0027 & -0.0074 &  0.0061 & -0.0023\\
38 & $-$0.0054 &   0.0031 &  0.0050  &  0.0005 & -0.0000 &  0.0008 &  0.0025  &  0.0014 & -0.0122 & -0.0039 &  0.0006 & -0.0086\\
39 & $+$0.0004 &   0.0102 &  0.0034  &  0.0196 &  0.0098 &  0.0113 &  0.0126  &  0.0077 & -0.0034 & -0.0083 &  0.0063 &  0.0016\\
40 & $-$0.0054 &  -0.0002 &  0.0055  &  0.0011 & -0.0009 & -0.0011 &  0.0053  & -0.0087 & -0.0065 & -0.0100 & -0.0074 & -0.0018\\
42 & $-$0.0039 &  -0.0050 &  0.0041  &  0.0011 & -0.0052 & -0.0047 & -0.0017  &  0.0038 & -0.0045 & -0.0082 & -0.0030 &  0.0032\\
43 & $-$0.0272 &  -0.0207 &  0.0051  & -0.0473 & -0.0115 & -0.0147 & -0.0377  & -0.0283 & -0.0292 & -0.0345 & -0.0252 & -0.0379\\
44 & $-$0.0438 &  -0.0341 &  0.0045  & -0.0566 & -0.0230 & -0.0254 & -0.0494  & -0.0247 & -0.0517 & -0.0342 & -0.0271 & -0.0349\\
45 & $+$0.0004 &   0.0106 &  0.0037  &  0.0163 &  0.0113 &  0.0119 &  0.0119  & -0.0011 & -0.0110 & -0.0132 &  0.0017 & -0.0077\\
46 & $-$0.0089 &  -0.0014 &  0.0034  & -0.0042 & -0.0045 & -0.0042 & -0.0029  & -0.0100 & -0.0134 & -0.0128 & -0.0084 & -0.0001\\
48 & $-$0.0070 &   0.0003 &  0.0040  &  0.0001 & -0.0025 & -0.0023 &  0.0012  & -0.0049 & -0.0095 & -0.0095 & -0.0034 & -0.0054\\
49 & $-$0.0160 &  -0.0051 &  0.0047  & -0.0110 & -0.0012 & -0.0025 & -0.0075  & -0.0061 & -0.0120 & -0.0073 &  0.0003 & -0.0172\\
50 & $-$0.0033 &   0.0071 &  0.0031  &  0.0077 &  0.0046 &  0.0048 &  0.0091  &  0.0017 & -0.0079 & -0.0069 &  0.0038 & -0.0037\\
51 & $-$0.0031 &   0.0076 &  0.0059  &  0.0102 &  0.0101 &  0.0086 &  0.0108  &  0.0081 & -0.0028 & -0.0062 & -0.0030 & -0.0072\\
52 & $-$0.0067 &   0.0163 &  0.0069  &  0.0224 &  0.0180 &  0.0179 &  0.0187  & -0.0040 &  0.0050 &  0.0011 &  0.0154 &  0.0008\\
53 & $-$0.0357 &  -0.0231 &  0.0034  & -0.0427 & -0.0088 & -0.0127 & -0.0374  & -0.0245 & -0.0273 & -0.0234 & -0.0213 & -0.0275\\
55 & $-$0.0024 &   0.0014 &  0.0063  &  0.0055 &  0.0019 &  0.0018 &  0.0039  & -0.0035 & -0.0127 & -0.0125 & -0.0093 & -0.0083\\
56 & $-$0.0316 &  -0.0329 &  0.0072  & -0.0449 & -0.0327 & -0.0338 & -0.0410  & -0.0312 & -0.0502 & -0.0414 & -0.0346 & -0.0475\\
57 & $-$0.0032 &   0.0017 &  0.0068  &  0.0065 & -0.0003 &  0.0006 &  0.0041  &  0.0011 & -0.0039 & -0.0050 &  0.0004 &  0.0011\\
58 & $-$0.0165 &  -0.0075 &  0.0034  & -0.0173 & -0.0053 & -0.0063 & -0.0134  & -0.0129 & -0.0190 & -0.0168 & -0.0109 & -0.0071\\
59 & $-$0.0047 &  -0.0021 &  0.0031  & -0.0011 & -0.0025 & -0.0039 &  0.0018  &  0.0013 & -0.0107 & -0.0089 & -0.0072 & -0.0012\\
60 & $-$0.0251 &  -0.0155 &  0.0039  & -0.0233 & -0.0126 & -0.0139 & -0.0197  & -0.0179 & -0.0243 & -0.0210 & -0.0165 & -0.0268\\
61 & $+$0.0083 &   0.0062 &  0.0045  &  0.0099 &  0.0084 &  0.0074 &  0.0088  &  0.0013 & -0.0016 &  0.0025 & -0.0041 & -0.0013\\
62 & $-$0.0066 &   0.0040 &  0.0041  &  0.0057 &  0.0010 &  0.0027 &  0.0044  & -0.0015 & -0.0099 & -0.0121 &  0.0019 & -0.0053\\
63 & $+$0.0010 &  -0.0077 &  0.0105  & -0.0055 & -0.0042 & -0.0065 & -0.0042  & -0.0192 & -0.0096 & -0.0192 & -0.0044 & -0.0146\\
64 & $-$0.0309 &  -0.0218 &  0.0064  & -0.0266 & -0.0150 & -0.0182 & -0.0221  & -0.0304 & -0.0425 & -0.0439 & -0.0306 & -0.0383\\
65 & $-$0.0015 &   0.0080 &  0.0028  &  0.0130 &  0.0085 &  0.0078 &  0.0126  &  0.0061 & -0.0031 & -0.0053 & -0.0004 &  0.0006\\
\end{tabular}
\label{tabdelta}
\end{table*}
\addtocounter{table}{-1}
\begin{table*}
\caption{\it Continued.}
\begin{tabular}{rrrrrrrrrrrrr}
\noalign{\smallskip}
CAN& 1 (1000)  & 1 (99)    &d$\delta$&    20    &  21     &   22    &  23     &    24    &    25   &   26    &   27    &   28 \\
\noalign{\smallskip}
66 & $+$0.0001 &   0.0075 &  0.0027  &  0.0140 &  0.0065 &  0.0070 &  0.0118  &  0.0006 & -0.0077 & -0.0098 &  0.0040 &  0.0010\\
67 & $-$0.0198 &  -0.0069 &  0.0044  & -0.0185 & -0.0067 & -0.0075 & -0.0150  & -0.0238 & -0.0233 & -0.0207 & -0.0212 & -0.0230\\
68 & $-$0.0060 &  -0.0009 &  0.0042  &  0.0063 &  0.0014 &  0.0014 &  0.0023  & -0.0066 & -0.0155 & -0.0075 & -0.0009 & -0.0043\\
69 & $-$0.0217 &  -0.0230 &  0.0099  & -0.0358 & -0.0114 & -0.0142 & -0.0262  & -0.0181 & -0.0215 & -0.0439 & -0.0231 & -0.0250\\
70 & $-$0.0026 &   0.0021 &  0.0032  &  0.0048 & -0.0013 & -0.0004 &  0.0050  &  0.0003 & -0.0074 & -0.0121 &  0.0056 &  0.0010\\
71 & $+$0.0025 &   0.0146 &  0.0059  &  0.0216 &  0.0152 &  0.0158 &  0.0165  &  0.0017 & -0.0037 &  0.0060 & -0.0001 &  0.0044\\
72 & $-$0.0096 &  -0.0086 &  0.0062  & -0.0078 & -0.0095 & -0.0089 & -0.0101  &  0.0043 & -0.0160 & -0.0084 & -0.0071 &  0.0026\\
73 & $+$0.0013 &   0.0105 &  0.0069  &  0.0144 &  0.0125 &  0.0121 &  0.0122  & -0.0028 & -0.0051 &  0.0122 &  0.0035 &  0.0001\\
74 & $-$0.0329 &  -0.0354 &  0.0066  & -0.0524 & -0.0264 & -0.0282 & -0.0450  & -0.0360 & -0.0348 & -0.0423 & -0.0229 & -0.0440\\
75 & $+$0.0001 &  -0.0042 &  0.0050  &  0.0035 & -0.0038 & -0.0036 &  0.0002  &  0.0068 &  0.0026 & -0.0105 & -0.0040 & -0.0009\\
76 & $-$0.0306 &  -0.0262 &  0.0118  & -0.0356 & -0.0196 & -0.0217 & -0.0301  & -0.0228 & -0.0135 & -0.0336 & -0.0235 & -0.0170\\
77 & $+$0.0026 &   0.0138 &  0.0046  &  0.0160 &  0.0156 &  0.0150 &  0.0158  &  0.0038 &  0.0042 & -0.0067 & -0.0003 &  0.0078\\
78 & $-$0.0075 &  -0.0079 &  0.0042  & -0.0141 & -0.0094 & -0.0099 & -0.0111  & -0.0095 & -0.0233 & -0.0127 & -0.0076 & -0.0128\\
79 & $-$0.0022 &   0.0017 &  0.0037  &  0.0071 &  0.0014 &  0.0022 &  0.0030  & -0.0030 & -0.0037 & -0.0147 & -0.0012 & -0.0032\\
80 & $-$0.0016 &   0.0022 &  0.0021  &  0.0078 &  0.0016 &  0.0022 &  0.0041  &  0.0008 & -0.0046 & -0.0049 & -0.0026 &  0.0005\\
82 & $-$0.0021 &   0.0022 &  0.0022  &  0.0087 & -0.0005 &  0.0012 &  0.0058  & -0.0010 & -0.0051 & -0.0047 &  0.0005 &  0.0017\\
83 & $-$0.0024 &   0.0012 &  0.0029  &  0.0075 &  0.0012 &  0.0016 &  0.0039  & -0.0013 & -0.0049 & -0.0151 &  0.0003 & -0.0004\\
90 & $+$0.0035 &   0.0071 &  0.0020  &  0.0117 &  0.0074 &  0.0068 &  0.0115  &  0.0037 & -0.0002 & -0.0062 &  0.0062 &  0.0038\\
\end{tabular}
\label{tabdeltabis}
\end{table*}

\begin{table*}
\caption[delta]{The mean residual selection bias $<\Delta\delta>$ for Cases
29 - 38.}
\begin{tabular}{rrrrrrrrrrr}
\noalign{\smallskip}
CAN & 29  & 30  & 31 & 32 & 33 & 34 & 35 & 36 & 37 & 38  \\
 & +d$a$ & -d$a$ & +d$b$ & -d$b$ & +d\meanRe & -d\meanRe & +d\meanSBe & -d\meanSBe & +d\meansig & -d\meansig  \\
\noalign{\smallskip}
1  &   0.0059 & -0.0027  &  0.0037 & -0.0003 &  0.0267 & -0.0233  & -0.0176 &  0.0209 & -0.0103 &  0.0125\\
2  &  -0.0320 & -0.0422  & -0.0341 & -0.0401 & -0.0121 & -0.0621  & -0.0564 & -0.0177 & -0.0510 & -0.0225\\
3  &  -0.0108 & -0.0225  & -0.0149 & -0.0183 &  0.0084 & -0.0416  & -0.0360 &  0.0028 & -0.0327 & -0.0001\\
4  &  -0.0052 & -0.0139  & -0.0077 & -0.0113 &  0.0154 & -0.0345  & -0.0288 &  0.0097 & -0.0221 &  0.0026\\
5  &  -0.0091 & -0.0209  & -0.0119 & -0.0184 &  0.0108 & -0.0413  & -0.0354 &  0.0049 & -0.0199 & -0.0104\\
6  &   0.0062 &  0.0008  &  0.0047 &  0.0024 &  0.0285 & -0.0214  & -0.0159 &  0.0230 & -0.0134 &  0.0206\\
7  &  -0.0118 & -0.0232  & -0.0151 & -0.0197 &  0.0076 & -0.0424  & -0.0367 &  0.0020 & -0.0328 & -0.0014\\
8  &   0.0065 &  0.0002  &  0.0059 &  0.0010 &  0.0284 & -0.0216  & -0.0161 &  0.0229 & -0.0108 &  0.0190\\
9  &   0.0095 &  0.0075  &  0.0077 &  0.0094 &  0.0343 & -0.0173  & -0.0114 &  0.0284 & -0.0078 &  0.0228\\
10 &   0.0043 & -0.0008  &  0.0031 &  0.0006 &  0.0268 & -0.0232  & -0.0176 &  0.0212 & -0.0152 &  0.0197\\
11 &  -0.0044 & -0.0102  & -0.0054 & -0.0092 &  0.0178 & -0.0322  & -0.0268 &  0.0123 & -0.0223 &  0.0076\\
12 &   0.0063 &  0.0010  &  0.0045 &  0.0029 &  0.0287 & -0.0213  & -0.0157 &  0.0232 & -0.0134 &  0.0212\\
13 &   0.0091 &  0.0030  &  0.0069 &  0.0053 &  0.0311 & -0.0189  & -0.0132 &  0.0255 & -0.0079 &  0.0208\\
14 &  -0.0002 & -0.0029  & -0.0008 & -0.0022 &  0.0235 & -0.0265  & -0.0208 &  0.0177 & -0.0172 &  0.0146\\
15 &   0.0137 &  0.0086  &  0.0127 &  0.0097 &  0.0362 & -0.0138  & -0.0081 &  0.0305 & -0.0016 &  0.0233\\
16 &   0.0048 &  0.0009  &  0.0032 &  0.0027 &  0.0279 & -0.0221  & -0.0164 &  0.0223 & -0.0127 &  0.0188\\
17 &   0.0011 & -0.0056  & -0.0010 & -0.0033 &  0.0228 & -0.0272  & -0.0214 &  0.0171 & -0.0159 &  0.0110\\
18 &  -0.0012 & -0.0053  & -0.0028 & -0.0036 &  0.0218 & -0.0282  & -0.0226 &  0.0162 & -0.0192 &  0.0133\\
19 &   0.0019 & -0.0009  &  0.0018 & -0.0007 &  0.0274 & -0.0264  & -0.0203 &  0.0213 & -0.0159 &  0.0191\\
20 &   0.0089 &  0.0053  &  0.0077 &  0.0066 &  0.0321 & -0.0179  & -0.0122 &  0.0264 & -0.0098 &  0.0245\\
21 &   0.0057 &  0.0029  &  0.0049 &  0.0037 &  0.0293 & -0.0207  & -0.0151 &  0.0237 & -0.0119 &  0.0204\\
22 &   0.0049 &  0.0022  &  0.0047 &  0.0025 &  0.0286 & -0.0214  & -0.0157 &  0.0229 & -0.0128 &  0.0191\\
23 &   0.0047 & -0.0008  &  0.0030 &  0.0009 &  0.0269 & -0.0231  & -0.0175 &  0.0214 & -0.0151 &  0.0187\\
24 &   0.0042 & -0.0012  &  0.0025 &  0.0006 &  0.0265 & -0.0235  & -0.0179 &  0.0210 & -0.0155 &  0.0181\\
25 &  -0.0054 & -0.0124  & -0.0072 & -0.0106 &  0.0161 & -0.0339  & -0.0283 &  0.0105 & -0.0241 &  0.0064\\
26 &  -0.0035 & -0.0109  & -0.0061 & -0.0083 &  0.0178 & -0.0322  & -0.0266 &  0.0122 & -0.0221 &  0.0078\\
27 &   0.0149 &  0.0119  &  0.0130 &  0.0137 &  0.0383 & -0.0117  & -0.0060 &  0.0326 &  0.0005 &  0.0259\\
29 &  -0.0045 & -0.0121  & -0.0050 & -0.0114 &  0.0168 & -0.0333  & -0.0275 &  0.0110 & -0.0215 &  0.0063\\
30 &  -0.0133 & -0.0215  & -0.0155 & -0.0193 &  0.0084 & -0.0432  & -0.0373 &  0.0025 & -0.0338 &  0.0001\\
31 &   0.0109 &  0.0082  &  0.0097 &  0.0094 &  0.0348 & -0.0157  & -0.0103 &  0.0294 & -0.0075 &  0.0275\\
32 &   0.0048 & -0.0013  &  0.0030 &  0.0006 &  0.0268 & -0.0232  & -0.0176 &  0.0212 & -0.0153 &  0.0190\\
33 &  -0.0021 & -0.0076  & -0.0032 & -0.0066 &  0.0201 & -0.0299  & -0.0243 &  0.0146 & -0.0200 &  0.0088\\
34 &  -0.0045 & -0.0136  & -0.0066 & -0.0114 &  0.0160 & -0.0340  & -0.0283 &  0.0102 & -0.0231 &  0.0036\\
35 &   0.0053 & -0.0006  &  0.0034 &  0.0015 &  0.0274 & -0.0226  & -0.0170 &  0.0219 & -0.0143 &  0.0189\\
36 &  -0.0045 & -0.0123  & -0.0067 & -0.0101 &  0.0166 & -0.0334  & -0.0278 &  0.0110 & -0.0247 &  0.0073\\
37 &   0.0070 &  0.0027  &  0.0060 &  0.0038 &  0.0299 & -0.0201  & -0.0145 &  0.0243 & -0.0116 &  0.0213\\
38 &   0.0064 & -0.0003  &  0.0046 &  0.0015 &  0.0281 & -0.0219  & -0.0162 &  0.0224 & -0.0121 &  0.0184\\
39 &   0.0125 &  0.0078  &  0.0117 &  0.0088 &  0.0352 & -0.0148  & -0.0091 &  0.0296 & -0.0043 &  0.0252\\
40 &   0.0022 & -0.0026  & -0.0003 & -0.0001 &  0.0248 & -0.0252  & -0.0194 &  0.0191 & -0.0163 &  0.0149\\
42 &  -0.0018 & -0.0082  & -0.0036 & -0.0063 &  0.0200 & -0.0300  & -0.0245 &  0.0146 & -0.0202 &  0.0101\\
43 &  -0.0151 & -0.0264  & -0.0178 & -0.0237 &  0.0043 & -0.0457  & -0.0399 & -0.0015 & -0.0332 & -0.0085\\
44 &  -0.0274 & -0.0409  & -0.0307 & -0.0376 & -0.0091 & -0.0591  & -0.0533 & -0.0148 & -0.0482 & -0.0190\\
45 &   0.0131 &  0.0080  &  0.0118 &  0.0093 &  0.0356 & -0.0144  & -0.0089 &  0.0301 & -0.0057 &  0.0262\\
46 &   0.0018 & -0.0046  & -0.0004 & -0.0023 &  0.0236 & -0.0263  & -0.0206 &  0.0179 & -0.0150 &  0.0130\\
48 &   0.0035 & -0.0030  &  0.0017 & -0.0011 &  0.0253 & -0.0247  & -0.0190 &  0.0195 & -0.0134 &  0.0150\\
49 &  -0.0009 & -0.0094  & -0.0027 & -0.0074 &  0.0199 & -0.0301  & -0.0244 &  0.0142 & -0.0205 &  0.0119\\
50 &   0.0096 &  0.0045  &  0.0079 &  0.0062 &  0.0321 & -0.0179  & -0.0123 &  0.0264 & -0.0085 &  0.0226\\
51 &   0.0109 &  0.0044  &  0.0085 &  0.0067 &  0.0329 & -0.0176  & -0.0120 &  0.0272 & -0.0076 &  0.0222\\
52 &   0.0192 &  0.0135  &  0.0171 &  0.0155 &  0.0413 & -0.0087  & -0.0032 &  0.0359 &  0.0028 &  0.0293\\
53 &  -0.0173 & -0.0290  & -0.0209 & -0.0254 &  0.0019 & -0.0481  & -0.0425 & -0.0037 & -0.0368 & -0.0092\\
55 &   0.0026 &  0.0001  &  0.0032 & -0.0005 &  0.0264 & -0.0236  & -0.0183 &  0.0210 & -0.0155 &  0.0195\\
56 &  -0.0280 & -0.0378  & -0.0310 & -0.0348 & -0.0079 & -0.0579  & -0.0521 & -0.0136 & -0.0452 & -0.0202\\
57 &   0.0048 & -0.0015  &  0.0030 &  0.0003 &  0.0267 & -0.0234  & -0.0176 &  0.0210 & -0.0121 &  0.0148\\
58 &  -0.0037 & -0.0115  & -0.0053 & -0.0098 &  0.0175 & -0.0325  & -0.0269 &  0.0118 & -0.0226 &  0.0079\\
59 &   0.0010 & -0.0054  & -0.0008 & -0.0034 &  0.0229 & -0.0271  & -0.0214 &  0.0172 & -0.0179 &  0.0136\\
60 &  -0.0107 & -0.0204  & -0.0124 & -0.0185 &  0.0095 & -0.0405  & -0.0348 &  0.0038 & -0.0306 &  0.0006\\
61 &   0.0075 &  0.0049  &  0.0064 &  0.0060 &  0.0312 & -0.0188  & -0.0133 &  0.0257 & -0.0106 &  0.0215\\
62 &   0.0075 &  0.0004  &  0.0054 &  0.0026 &  0.0290 & -0.0210  & -0.0153 &  0.0233 & -0.0107 &  0.0180\\
63 &  -0.0060 & -0.0093  & -0.0061 & -0.0092 &  0.0186 & -0.0340  & -0.0281 &  0.0126 & -0.0237 &  0.0094\\
64 &  -0.0173 & -0.0264  & -0.0194 & -0.0242 &  0.0032 & -0.0468  & -0.0411 & -0.0026 & -0.0358 & -0.0080\\
65 &   0.0109 &  0.0050  &  0.0088 &  0.0072 &  0.0330 & -0.0170  & -0.0113 &  0.0273 & -0.0073 &  0.0241\\
\end{tabular}
\label{tabdelta2938}
\end{table*}

\addtocounter{table}{-1}
\begin{table*}
\caption[delta]{Continued.}
\begin{tabular}{rrrrrrrrrrr}
\noalign{\smallskip}
CAN & 29  & 30  & 31 & 32 & 33 & 34 & 35 & 36 & 37 & 38  \\
 & +d$a$ & -d$a$ & +d$b$ & -d$b$ & +d\meanRe & -d\meanRe & +d\meanSBe & -d\meanSBe & +d\meansig & -d\meansig  \\
\noalign{\smallskip}
66 &   0.0095 &  0.0054  &  0.0082 &  0.0068 &  0.0325 & -0.0175  & -0.0120 &  0.0270 & -0.0086 &  0.0238 \\
67 &  -0.0027 & -0.0112  & -0.0055 & -0.0083 &  0.0181 & -0.0319  & -0.0263 &  0.0125 & -0.0223 &  0.0081 \\
68 &   0.0014 & -0.0031  & -0.0000 & -0.0016 &  0.0241 & -0.0258  & -0.0202 &  0.0185 & -0.0140 &  0.0119 \\
69 &  -0.0152 & -0.0310  & -0.0204 & -0.0257 &  0.0054 & -0.0515  & -0.0454 & -0.0007 & -0.0400 & -0.0060 \\
70 &   0.0042 & -0.0002  &  0.0033 &  0.0009 &  0.0271 & -0.0229  & -0.0173 &  0.0214 & -0.0130 &  0.0168 \\
71 &   0.0170 &  0.0121  &  0.0157 &  0.0135 &  0.0399 & -0.0106  & -0.0051 &  0.0344 &  0.0001 &  0.0291 \\
72 &  -0.0048 & -0.0123  & -0.0070 & -0.0101 &  0.0164 & -0.0336  & -0.0280 &  0.0109 & -0.0206 &  0.0054 \\
73 &   0.0131 &  0.0078  &  0.0106 &  0.0104 &  0.0357 & -0.0148  & -0.0090 &  0.0299 & -0.0043 &  0.0257 \\
74 &  -0.0310 & -0.0398  & -0.0334 & -0.0374 & -0.0102 & -0.0607  & -0.0549 & -0.0160 & -0.0494 & -0.0226 \\
75 &  -0.0017 & -0.0068  & -0.0034 & -0.0051 &  0.0208 & -0.0292  & -0.0235 &  0.0150 & -0.0174 &  0.0093 \\
76 &  -0.0216 & -0.0308  & -0.0237 & -0.0287 &  0.0004 & -0.0528  & -0.0467 & -0.0057 & -0.0420 & -0.0102 \\
77 &   0.0160 &  0.0116  &  0.0138 &  0.0138 &  0.0388 & -0.0112  & -0.0056 &  0.0333 & -0.0034 &  0.0323 \\
78 &  -0.0053 & -0.0105  & -0.0073 & -0.0085 &  0.0171 & -0.0329  & -0.0272 &  0.0115 & -0.0241 &  0.0084 \\
79 &   0.0044 & -0.0010  &  0.0020 &  0.0014 &  0.0267 & -0.0233  & -0.0177 &  0.0211 & -0.0153 &  0.0191 \\
80 &   0.0042 &  0.0001  &  0.0029 &  0.0016 &  0.0272 & -0.0228  & -0.0172 &  0.0216 & -0.0149 &  0.0201 \\
82 &   0.0043 & -0.0000  &  0.0027 &  0.0017 &  0.0272 & -0.0228  & -0.0172 &  0.0216 & -0.0144 &  0.0190 \\
83 &   0.0032 & -0.0009  &  0.0019 &  0.0006 &  0.0262 & -0.0238  & -0.0181 &  0.0205 & -0.0148 &  0.0175 \\
90 &   0.0087 &  0.0054  &  0.0074 &  0.0069 &  0.0321 & -0.0179  & -0.0122 &  0.0264 & -0.0097 &  0.0237 \\
\end{tabular}
\label{tabdelta2938bis}
\end{table*}
\clearpage

\begin{table*}
\caption[delta]{The mean residual selection bias $<\delta>$ for Cases 39-47.}
\begin{tabular}{rrrrrrrrrr}
\noalign{\smallskip}
CAN & 39 & 40 & 41 & 42 & 43 & 44 & 45 & 46 & 47\\
 & +d\sigone & -d\sigone & +d\sigtwo & -d\sigtwo & +d\sigthree & -d\sigthree & UG & GU & UU\\
\noalign{\smallskip}
1  &  0.0021 &  0.0012  &  0.0018 &  0.0015 &  0.0057 & -0.0031 &  -0.0066 & -0.0131 & -0.0072 \\
2  & -0.0377 & -0.0365  & -0.0368 & -0.0373 & -0.0334 & -0.0416 &  -0.0250 & -0.0355 & -0.0236 \\
3  & -0.0172 & -0.0160  & -0.0165 & -0.0167 & -0.0137 & -0.0203 &  -0.0077 & -0.0115 & -0.0185 \\
4  & -0.0093 & -0.0098  & -0.0095 & -0.0096 & -0.0059 & -0.0139 &  -0.0071 & -0.0050 & -0.0022 \\
5  & -0.0146 & -0.0159  & -0.0152 & -0.0154 & -0.0075 & -0.0227 &   0.0077 & -0.0026 & -0.0002 \\
6  &  0.0042 &  0.0030  &  0.0036 &  0.0035 &  0.0047 &  0.0020 &   0.0041 & -0.0027 &  0.0063 \\
7  & -0.0179 & -0.0169  & -0.0173 & -0.0175 & -0.0140 & -0.0217 &  -0.0156 & -0.0138 & -0.0203 \\
8  &  0.0040 &  0.0029  &  0.0036 &  0.0032 &  0.0058 &  0.0004 &  -0.0003 & -0.0037 & -0.0026 \\
9  &  0.0103 &  0.0069  &  0.0084 &  0.0086 &  0.0086 &  0.0083 &   0.0115 & -0.0003 &  0.0129 \\
10 &  0.0024 &  0.0013  &  0.0019 &  0.0017 &  0.0029 &  0.0003 &   0.0042 &  0.0040 &  0.0068 \\
11 & -0.0066 & -0.0079  & -0.0071 & -0.0074 & -0.0054 & -0.0097 &  -0.0111 & -0.0007 &  0.0035 \\
12 &  0.0044 &  0.0031  &  0.0037 &  0.0037 &  0.0048 &  0.0022 &   0.0069 &  0.0063 &  0.0080 \\
13 &  0.0070 &  0.0053  &  0.0061 &  0.0061 &  0.0087 &  0.0029 &   0.0045 &  0.0099 &  0.0108 \\
14 & -0.0003 & -0.0027  & -0.0015 & -0.0016 & -0.0008 & -0.0026 &   0.0086 &  0.0081 &  0.0052 \\
15 &  0.0123 &  0.0101  &  0.0113 &  0.0110 &  0.0135 &  0.0083 &   0.0062 &  0.0074 &  0.0042 \\
16 &  0.0039 &  0.0021  &  0.0029 &  0.0029 &  0.0042 &  0.0013 &   0.0029 &  0.0031 &  0.0069 \\
17 & -0.0016 & -0.0027  & -0.0022 & -0.0022 &  0.0005 & -0.0055 &  -0.0017 & -0.0065 &  0.0023 \\
18 & -0.0023 & -0.0040  & -0.0032 & -0.0032 & -0.0021 & -0.0047 &   0.0038 &  0.0082 &  0.0110 \\
19 &  0.0019 & -0.0007  &  0.0006 &  0.0004 &  0.0011 & -0.0003 &   0.0232 &  0.0050 &  0.0141 \\
20 &  0.0081 &  0.0062  &  0.0071 &  0.0071 &  0.0080 &  0.0058 &   0.0037 &  0.0052 &  0.0046 \\
21 &  0.0056 &  0.0032  &  0.0043 &  0.0043 &  0.0050 &  0.0033 &   0.0126 &  0.0096 &  0.0070 \\
22 &  0.0047 &  0.0025  &  0.0037 &  0.0035 &  0.0042 &  0.0027 &   0.0022 &  0.0066 &  0.0136 \\
23 &  0.0025 &  0.0015  &  0.0021 &  0.0018 &  0.0031 &  0.0003 &   0.0017 &  0.0025 & -0.0033 \\
24 &  0.0021 &  0.0011  &  0.0016 &  0.0015 &  0.0027 & -0.0001 &  -0.0026 &  0.0035 &  0.0009 \\
25 & -0.0086 & -0.0091  & -0.0087 & -0.0090 & -0.0067 & -0.0117 &  -0.0101 & -0.0099 & -0.0046 \\
26 & -0.0068 & -0.0076  & -0.0072 & -0.0072 & -0.0051 & -0.0099 &   0.0012 &  0.0155 & -0.0006 \\
27 &  0.0149 &  0.0119  &  0.0132 &  0.0134 &  0.0139 &  0.0125 &   0.0192 &  0.0217 &  0.0290 \\
29 & -0.0080 & -0.0085  & -0.0080 & -0.0086 & -0.0050 & -0.0121 &   0.0031 & -0.0047 & -0.0078 \\
30 & -0.0173 & -0.0175  & -0.0173 & -0.0175 & -0.0151 & -0.0204 &  -0.0145 & -0.0050 &  0.0020 \\
31 &  0.0109 &  0.0084  &  0.0095 &  0.0096 &  0.0101 &  0.0088 &   0.0107 &  0.0030 &  0.0147 \\
32 &  0.0023 &  0.0014  &  0.0019 &  0.0017 &  0.0030 &  0.0002 &  -0.0029 &  0.0001 &  0.0058 \\
33 & -0.0045 & -0.0051  & -0.0048 & -0.0050 & -0.0032 & -0.0070 &  -0.0051 &  0.0007 &  0.0016 \\
34 & -0.0091 & -0.0089  & -0.0088 & -0.0092 & -0.0054 & -0.0134 &  -0.0191 & -0.0027 & -0.0088 \\
35 &  0.0029 &  0.0020  &  0.0025 &  0.0024 &  0.0038 &  0.0006 &  -0.0011 & -0.0023 &  0.0036 \\
36 & -0.0083 & -0.0085  & -0.0083 & -0.0085 & -0.0066 & -0.0107 &  -0.0077 & -0.0039 & -0.0033 \\
37 &  0.0058 &  0.0041  &  0.0049 &  0.0048 &  0.0059 &  0.0036 &   0.0043 & -0.0018 &  0.0006 \\
38 &  0.0037 &  0.0025  &  0.0032 &  0.0029 &  0.0051 &  0.0004 &   0.0052 &  0.0066 & -0.0045 \\
39 &  0.0112 &  0.0093  &  0.0103 &  0.0101 &  0.0119 &  0.0081 &   0.0139 &  0.0021 &  0.0042 \\
40 &  0.0006 & -0.0009  & -0.0003 & -0.0001 &  0.0013 & -0.0022 &   0.0021 & -0.0023 &  0.0024 \\
42 & -0.0044 & -0.0055  & -0.0049 & -0.0050 & -0.0031 & -0.0073 &   0.0070 &  0.0053 & -0.0007 \\	       
43 & -0.0212 & -0.0202  & -0.0205 & -0.0210 & -0.0157 & -0.0266 &  -0.0248 & -0.0258 & -0.0338 \\
44 & -0.0353 & -0.0329  & -0.0338 & -0.0344 & -0.0290 & -0.0402 &  -0.0308 & -0.0333 & -0.0279 \\
45 &  0.0114 &  0.0098  &  0.0106 &  0.0105 &  0.0117 &  0.0090 &   0.0022 &  0.0018 &  0.0116 \\
46 & -0.0007 & -0.0020  & -0.0013 & -0.0014 &  0.0012 & -0.0045 &  -0.0037 &  0.0031 & -0.0038 \\
48 &  0.0012 & -0.0006  &  0.0003 &  0.0002 &  0.0029 & -0.0030 &  -0.0013 & -0.0017 & -0.0052 \\
49 & -0.0051 & -0.0051  & -0.0049 & -0.0053 & -0.0025 & -0.0084 &  -0.0077 & -0.0076 & -0.0032 \\
50 &  0.0079 &  0.0063  &  0.0071 &  0.0070 &  0.0087 &  0.0049 &   0.0078 &  0.0026 &  0.0032 \\
51 &  0.0083 &  0.0070  &  0.0077 &  0.0076 &  0.0094 &  0.0053 &   0.0087 &  0.0089 &  0.0002 \\
52 &  0.0175 &  0.0153  &  0.0163 &  0.0163 &  0.0181 &  0.0140 &  -0.0004 &  0.0160 &  0.0217 \\
53 & -0.0240 & -0.0223  & -0.0230 & -0.0232 & -0.0187 & -0.0284 &  -0.0216 & -0.0233 & -0.0286 \\
55 &  0.0025 &  0.0004  &  0.0014 &  0.0013 &  0.0020 &  0.0005 &   0.0045 &  0.0070 &  0.0057 \\
56 & -0.0332 & -0.0325  & -0.0328 & -0.0329 & -0.0279 & -0.0387 &  -0.0435 & -0.0214 & -0.0294 \\
57 &  0.0025 &  0.0009  &  0.0017 &  0.0016 &  0.0042 & -0.0015 &   0.0029 &  0.0029 &  0.0087 \\
58 & -0.0075 & -0.0076  & -0.0074 & -0.0078 & -0.0050 & -0.0108 &  -0.0091 & -0.0129 & -0.0035 \\
59 & -0.0017 & -0.0025  & -0.0021 & -0.0022 & -0.0002 & -0.0046 &   0.0012 & -0.0034 & -0.0018 \\
60 & -0.0158 & -0.0153  & -0.0153 & -0.0158 & -0.0124 & -0.0194 &  -0.0189 & -0.0160 & -0.0153 \\
61 &  0.0074 &  0.0051  &  0.0062 &  0.0062 &  0.0068 &  0.0054 &   0.0186 &  0.0001 &  0.0102 \\
62 &  0.0046 &  0.0035  &  0.0041 &  0.0039 &  0.0066 &  0.0008 &  -0.0008 &  0.0053 &  0.0061 \\
63 & -0.0064 & -0.0089  & -0.0076 & -0.0078 & -0.0071 & -0.0085 &  -0.0132 &  0.0112 & -0.0030 \\
64 & -0.0218 & -0.0219  & -0.0216 & -0.0221 & -0.0185 & -0.0260 &  -0.0264 & -0.0251 & -0.0169 \\
65 &  0.0087 &  0.0074  &  0.0080 &  0.0079 &  0.0097 &  0.0058 &   0.0082 &  0.0020 &  0.0035 \\
\end{tabular}          
\label{tabdelta3947}
\end{table*}

\addtocounter{table}{-1}
\begin{table*}
\caption[delta]{Continued}
\begin{tabular}{rrrrrrrrrr}
\noalign{\smallskip}
CAN & 39 & 40 & 41 & 42 & 43 & 44 & 45 & 46 & 47\\
 & +d\sigone & -d\sigone & +d\sigtwo & -d\sigtwo & +d\sigthree & -d\sigthree & UG & GU & UU\\
\noalign{\smallskip}
66 &  0.0084 &  0.0067  &  0.0076 &  0.0074 &  0.0086 &  0.0060 &   0.0045 &  0.0064 &  0.0052 \\
67 & -0.0067 & -0.0070  & -0.0068 & -0.0070 & -0.0042 & -0.0103 &  -0.0216 & -0.0057 & -0.0036 \\
68 &  0.0000 & -0.0016  & -0.0008 & -0.0009 &  0.0008 & -0.0030 &   0.0025 &  0.0037 &  0.0033 \\
69 & -0.0239 & -0.0223  & -0.0229 & -0.0231 & -0.0182 & -0.0291 &   0.0003 & -0.0312 & -0.0166 \\
70 &  0.0029 &  0.0013  &  0.0021 &  0.0020 &  0.0037 & -0.0000 &  -0.0012 &  0.0113 &  0.0023 \\
71 &  0.0156 &  0.0137  &  0.0147 &  0.0145 &  0.0161 &  0.0126 &   0.0194 &  0.0039 &  0.0136 \\
72 & -0.0080 & -0.0091  & -0.0085 & -0.0086 & -0.0052 & -0.0126 &  -0.0001 & -0.0072 & -0.0075 \\
73 &  0.0116 &  0.0095  &  0.0104 &  0.0105 &  0.0120 &  0.0085 &   0.0099 &  0.0021 &  0.0008 \\
74 & -0.0356 & -0.0352  & -0.0353 & -0.0356 & -0.0321 & -0.0395 &  -0.0398 & -0.0257 & -0.0391 \\
75 & -0.0035 & -0.0049  & -0.0042 & -0.0042 & -0.0022 & -0.0067 &   0.0035 &  0.0057 &  0.0031 \\
76 & -0.0259 & -0.0265  & -0.0260 & -0.0264 & -0.0238 & -0.0293 &  -0.0233 & -0.0155 & -0.0208 \\
77 &  0.0148 &  0.0129  &  0.0138 &  0.0138 &  0.0147 &  0.0126 &   0.0136 &  0.0057 &  0.0128 \\
78 & -0.0073 & -0.0084  & -0.0078 & -0.0079 & -0.0065 & -0.0097 &  -0.0041 &  0.0037 &  0.0077 \\
79 &  0.0023 &  0.0012  &  0.0017 &  0.0017 &  0.0029 &  0.0001 &   0.0079 &  0.0045 &  0.0029 \\
80 &  0.0030 &  0.0015  &  0.0022 &  0.0022 &  0.0031 &  0.0010 &   0.0054 &  0.0052 &  0.0055 \\
82 &  0.0030 &  0.0015  &  0.0022 &  0.0022 &  0.0033 &  0.0006 &   0.0080 &  0.0039 &  0.0030 \\
83 &  0.0021 &  0.0003  &  0.0012 &  0.0012 &  0.0024 & -0.0005 &   0.0031 &  0.0082 &  0.0045 \\
90 &  0.0083 &  0.0060  &  0.0071 &  0.0071 &  0.0079 &  0.0059 &   0.0069 &  0.0031 &  0.0088 \\

\end{tabular}          
\label{tabdelta3947bis}
\end{table*}
\clearpage}

\subsubsection{The estimation of the mean motions}
\label{meanmotions}

Given the large mean distance of the EFAR clusters, and the precision
of the FP distance estimator (20\% for the single galaxy, see Paper
VII), the present dataset does not allow the determination of the
peculiar motions of single clusters. Nevertheless, the mean motions of
the EFAR clusters can be constrained. In the following we investigate
this point using the subsample of clusters identified in Paper VII for
the study of the bulk motions. To this belong the 50 EFAR clusters
with 3 or more galaxies (hereafter Peculiar Velocity, PV, clusters), 
redshifts less than 15000 km/s, and peculiar
velocity errors less than 1800 km/s. We compute the average mean velocities 
without applying the bias corrections investigated in
\S\ref{resbias} (although this is done in Paper VII), to
stress that our results are not affected much by these residual
systematic uncertainties, and that random errors dominate the noise.

\begin{figure}
\plotone{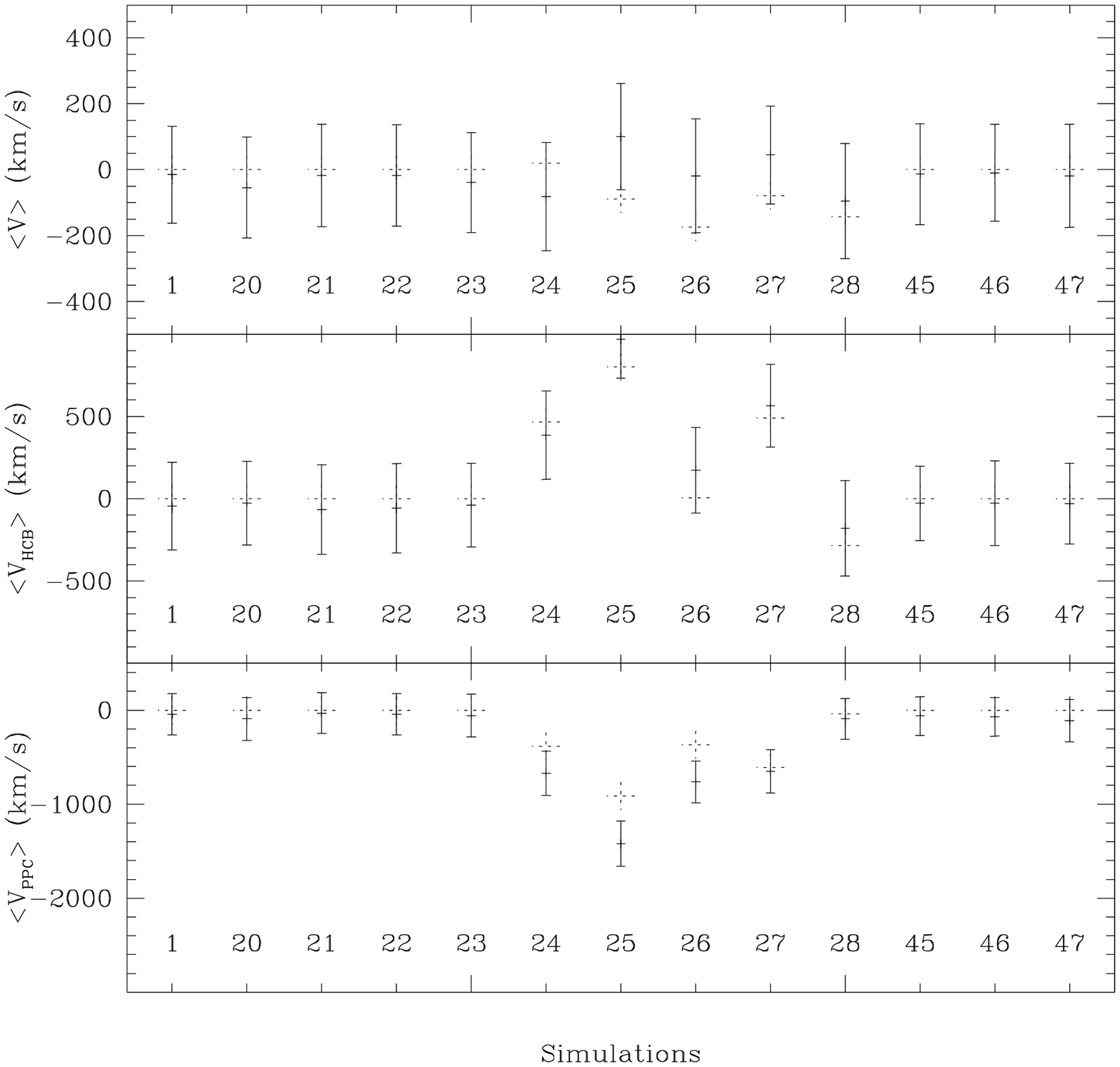}
\caption{The simulations of the mean peculiar velocities $<V>$
(mean velocity of the sample of 50 Clusters defined in Paper VII, top panel),
$<V_{HCB}>$ (mean velocity of the clusters with positive Galactic
longitude, the HCB sample, middle panel) and $<V_{PPC}>$ (mean
velocity of the clusters with negative Galactic longitude, the PPC
sample, bottom panel). The points show the mean values (over 99
simulations) and their rms for Cases 1, 20 to 28, and 45 to 47 of Table
\ref{tabsimumean}. The dotted bars show the input values.}
\label{figvmean}
\end{figure}

We begin by examining the mean peculiar velocities $<V>$, $<V_{HCB}>$
and $<V_{PPC}>$ of the sample. These are the mean over the PV
clusters, over the PV clusters with positive (the
Hercules Corona Borealis, HCB,
sample), and over the ones with negative Galactic longitude (the Perseus Pisces
Cetus, PPC, sample). 
Fig. \ref{figvmean} shows that these average quantities are
recovered with statistical errors of about 200 km/s. In general these
errors are larger than the systematic differences from the input
values. In particular, errors on the selection function parameters do
not affect the result much (Cases 20 to 23) and uniform distributions 
of the FP parameters, of
the errors or both do not change this conclusion (Cases 45 to 47). 
The largest systematic
difference is observed for Cases 24 to 26 and $V_{PPC}$, where the
input values are ill-defined, since the rms of the cluster peculiar
velocities is large (see Figs. \ref{figpecinput} and
\ref{figlpsmacinput}). 

\begin{figure}
\plotone{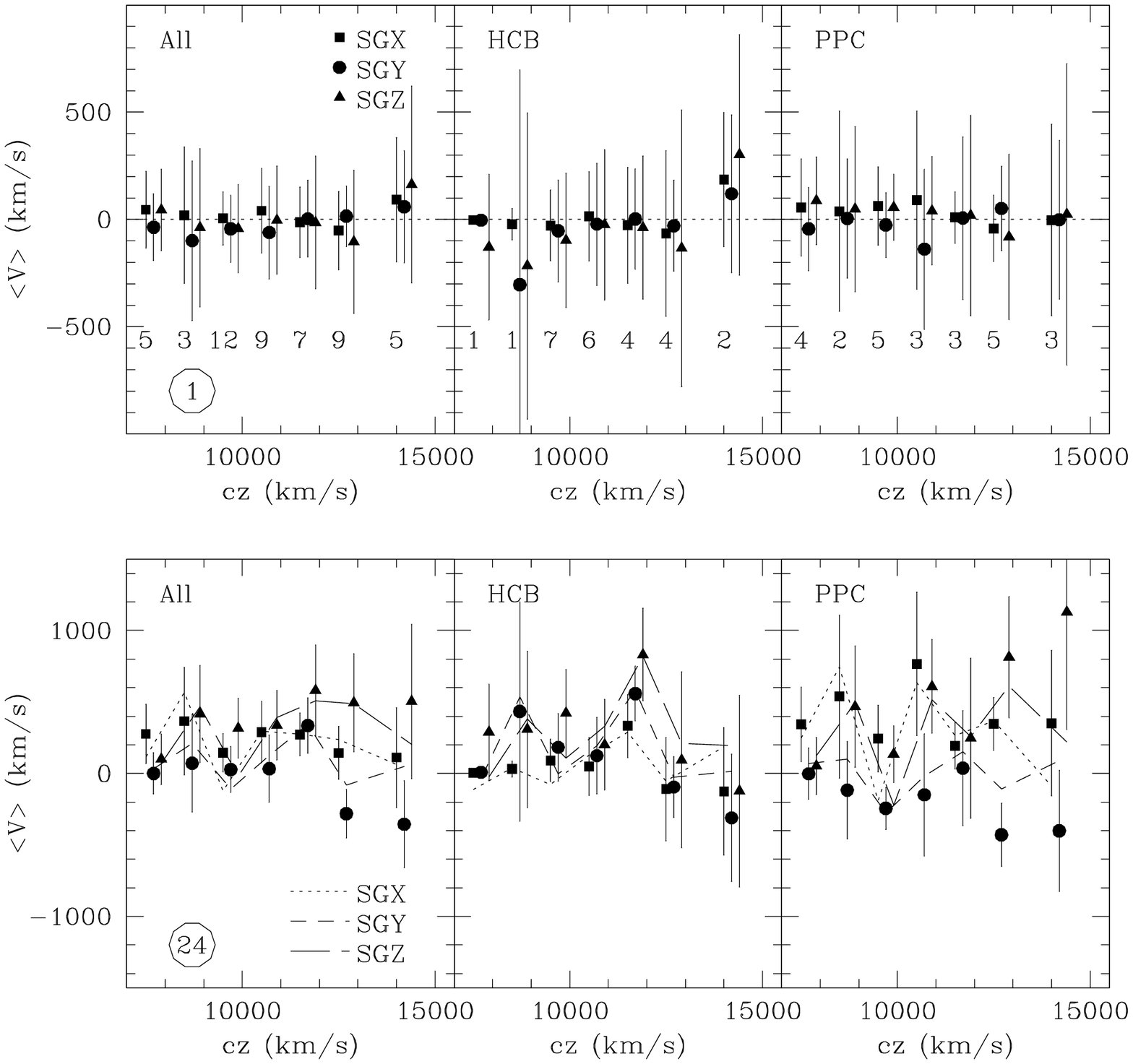}
\caption{The simulations of the mean peculiar velocities in radial
shells. The clusters are grouped in 7 redshift ranges: the first is
4000-8000 km/s, the next five cover 8000 km/s to 13000 km/s in 1000
km/s steps, and the last is 13000-15000 km/s. The left panel shows the
whole sample of 50 PV clusters, the middle panel shows the HCB clusters,
and the right panel shows the PPC clusters. 
The points show the mean values (over 99
simulations) of the peculiar velocities and their rms for Cases 1
(top)  and 24 (bottom)  of Table
\ref{tabsimumean}. The Supergalactic X, Y and
Z components are shown as filled squares, circles and triangles
respectively (with small offsets in redshift for clarity).  
In the bottom panels the dotted, short dashed and long dashed lines  show
the input values. The number of clusters in each redshift range is
indicated at the bottom of each top panel.}
\label{figxyz}
\end{figure}

As a second test we consider the Supergalactic X, Y and Z components
of the mean peculiar velocities in radial shells. Fig. \ref{figxyz} 
shows the averages of 99 simulations obtained for Case 1 and 24 of
Table \ref{tabsimumean}. Again, the input
values are recovered with systematic differences smaller than the
statistical errors.

\begin{figure}
\plotone{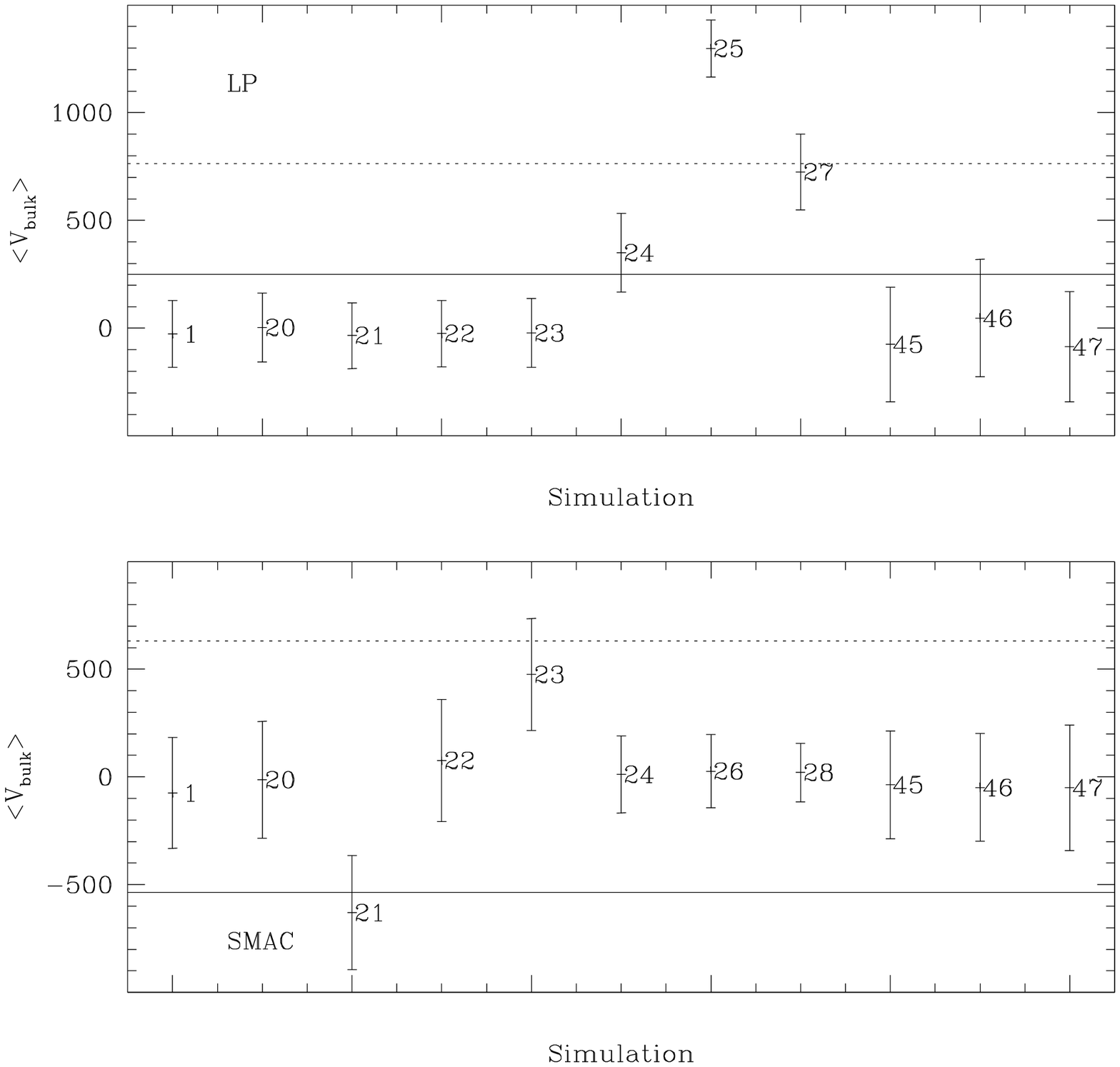}
\caption{The simulations of the bulk flow motions of 
Lauer and Postman (1994, top) and
Hudson et al. (1998, bottom). The points show the mean values (over 99
simulations) of $V_{bulk}$ derived using the 50 clusters used in Paper
VII for the determination of the EFAR mean peculiar velocities, 
for 11 Cases identified by the
numbers in Table \ref{tabsimumean}. The errorbars show the rms. 
The dotted lines show the values of $V_{bulk}$ as derived by
LP and SMAC. The full lines show the values of $V_{bulk}$ as derived
in Paper VII.}
\label{figlpsmac}
\end{figure}

We determine now how well the EFAR cluster subsample can constrain the
bulk flow motions determined by Lauer and Postman (1994) and Hudson et
al. (1999).  We fitted the peculiar velocities of the 50 clusters used
in Paper VII with a bulk flow model $V_{bulk} \cos \theta$, where
$\theta$ is the angle between the dipole and cluster direction.
determining $V_{bulk}$ in a least squares sense. Fig. 15 of Paper VII
shows these fits to the EFAR data.
Fig. \ref{figlpsmac} shows that when simulations with no peculiar
velocities are considered (Cases 1 and 20 to 23 of Table
\ref{tabsimumean}), the recovered mean $V_{bulk}$ is compatible with
zero, with statistical errors of about 160 km/s. Errors on the
selection function parameters do not affect the result much (Cases 20
to 23), as uniform distributions of the FP parameters, of the errors or
both (Cases 45 to 47). Similarly, when the simulations with pure LP or
SMAC dipole motions are considered (Cases 27 and 28
respectively), the recovered mean $V_{bulk}$ is consistent with the
input values within the statistical uncertainties. Fig. 16 of Paper VII
shows the histograms of the recovered bulk flow amplitudes for 500 
simulations. Only one out of these simulations of the LP flow, and
none of SMAC yields a $V_{bulk}$ less than the value measured from the
EFAR data in Paper VII. In contrast, the simulations with a random
peculiar velocity component (Case 24) give mean $V_{bulk}$ values
similar to the observed values, both for the LP and the SMAC
directions. Simulations with a random component on top of the LP and
SMAC bulk flows (Cases 25 and 26) produce large mean $V_{bulk}$ along
the LP direction, and little net motion along the SMAC direction.

\begin{figure}
\plotone{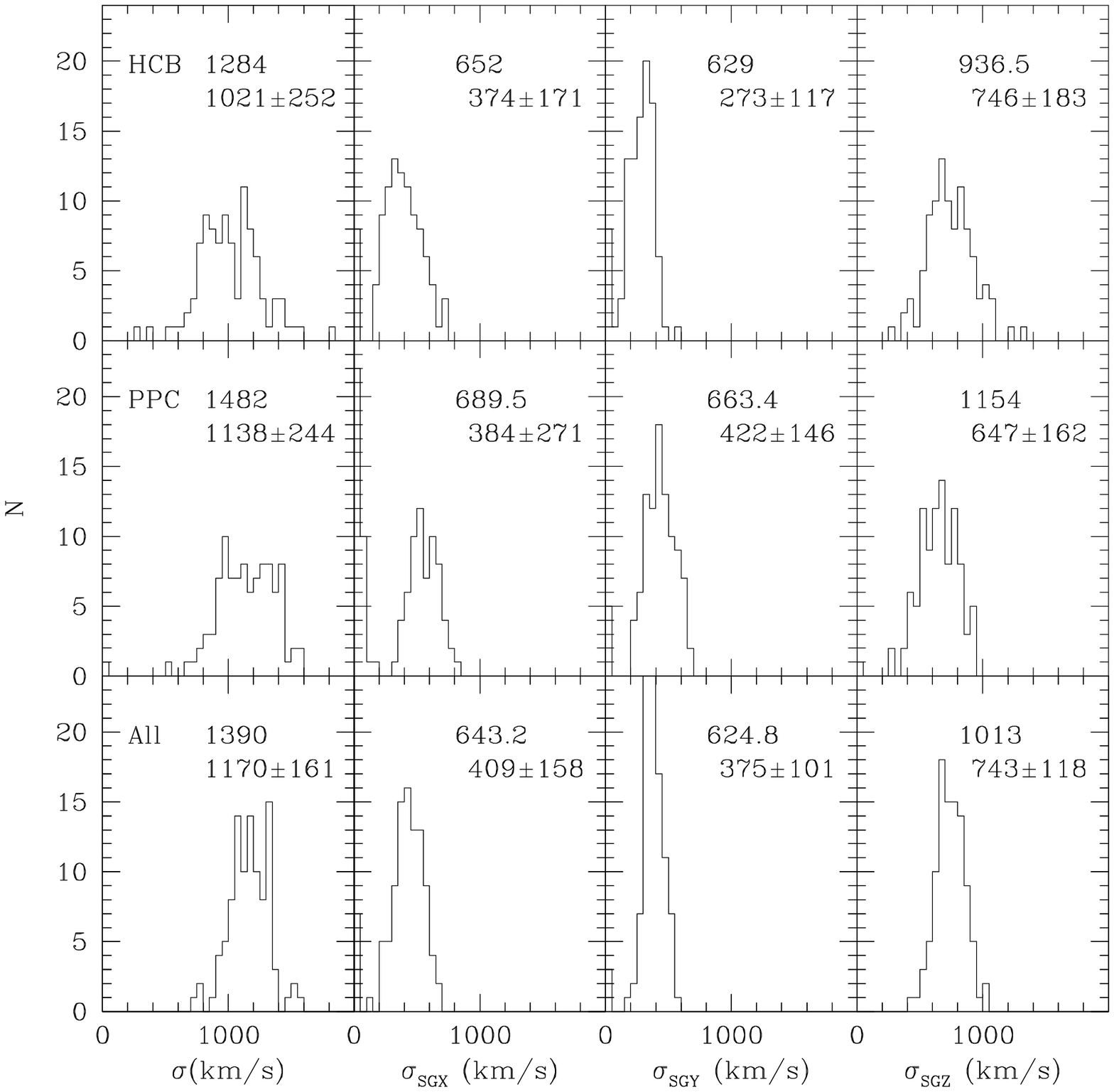}
\caption{The histograms of the peculiar velocity dispersions derived
from the one-dimensional ML analysis of the 99 simulations of Case 24
of Table \ref{tabsimumean}. The dispersions are shown separately for
the HCB, PPC and PV cluster samples, for the radial peculiar
velocities and their projections along the supergalactic X, Y and Z
projections. The labels give the input dispersions, the results
averaged over the 99 simulations and their rms.}
\label{figrms}
\end{figure}

Finally, we investigate whether not only the mean motions, but also
their global intrinsic fluctuations could be determined with the EFAR
cluster sample. To this purpose we use the one-dimensional ML
algorithm of \S\ref{onedim}: we consider the 
peculiar velocities derived for the 99 simulations of Case 24 of Table
\ref{tabsimumean} and their projections along the supergalactic X, Y
and Z coordinates with their estimated errors, separately for the whole
sample of PV clusters, and the northern HCB and southern PPC subsets. 
For each of these datasets the ML algorithm estimates the intrinsic
(gaussian) dispersion. Fig. \ref{figrms} shows the results. The
derived rms are systematically underestimated, up to more than a
factor two, when the input values are of the order of $\approx 600$
km/s, and by $\approx 25$\% if larger input values are considered. We
conclude that cluster sample is too small to allow an accurate
determination of these quantities.

To summarize, the EFAR cluster sample allows to measure the mean
motions of the HCB and PPC regions with 200 km/s precision, and to
constrain the bulk flow along the LP dipole direction strongly. It is
less suited to constrain the SMAC result and does not allow the
accurate determination of the random peculiar motion component. 
\clearpage

\section{Conclusions}
\label{conclusions}

We have described the maximum likelihood gaussian algorithm we
developed to investigate the correlations between the parameters of
the EFAR database (the \mgmg\ relation of Paper II, the \mgbpsig\ and
\mgtwosig\ relations of Paper V, the Fundamental Plane of Paper VII)
and determine the cluster peculiar velocities of Paper VII. 

We performed extensive testing based on mock catalogues of
the EFAR sample. We find that ``canonical'' methods based on a
least-squares approach cannot cope with the challenge of a sample
with a spread of a factor of two in redshift, with sizable selection effects,
non-negligible and non-uniform errors, and explicit cuts. 
We quantify the size of the systematic biases these methods introduce.
In contrast, the maximum likelihood gaussian algorithm 
takes into account errors, selection effects and the
presence of explicit cuts, determining nearly-unbiased estimates of the slopes
of the correlations and their intrinsic and parallel 
spread. Ten to thirty percent
of the analyzed simulations have mean likelihoods larger than that
of the EFAR sample, justifying the use of gaussian modeling.
We derive the analytical solution of the maximum
likelihood gaussian problem in N dimensions in the presence of small errors. 
We show that the residual systematic biases are always smaller
than the statistical errors. We investigate in detail the effects of
cluster sample selection, errors in the selection function parameters,
and selection cuts. We explore the cases of uniform distributions of
parameters and errors. We conclude that the mean peculiar
motions of the EFAR clusters can be determined reliably. In
particular, the large amplitude of the dipole motion measured by Lauer
and Postman (1994) can be strongly constrained. 

\section*{Acknowledgments} 
RPS acknowledges the support by DFG grant SFB 375. 
GW is grateful to the SERC and Wadham College for a
year's stay in Oxford, and to the Alexander von Humboldt-Stiftung for
making possible a visit to the Ruhr-Universit\"at in Bochum. MMC
acknowledges the support of a Lindemann Fellowship, a DIST Collaborative
Research Grant and an Australian Academy of Science/Royal Society
Exchange Program Fellowship. This work was partially supported by NSF
Grant AST90-16930 to DB, AST90-17048 and AST93-47714 to GW,
AST90-20864 to RKM. The entire
collaboration benefitted from NATO Collaborative Research Grant 900159
and from the hospitality and monetary support of Dartmouth College,
Oxford University, the University of Durham and Arizona State
University. Support was also received from PPARC visitors grants to
Oxford and Durham Universities and a PPARC rolling grant:
``Extragalactic Astronomy and Cosmology in Durham 1994-98''.

\appendix
\section{1-dimensional Model, limiting case}
\label{onedimlim}

Neglecting the presence of cuts, the maximization of Eq. (\ref{eqonedim})
 leads to:
\begin{equation}
\label{eqonemean}
\mu=\frac{\sum_i \frac{x_i}{S_i(\sigma_i^2+\sigma^2)}}
{\sum_i\frac{1}{S_i(\sigma_i^2+\sigma^2)}},
\end{equation}
\begin{equation}
\label{eqonesigma}
\sum_i \frac{\sigma_i^2+\sigma^2-(x_i-\mu)^2}{S_i(\sigma_i^2+\sigma^2)^2}=0.
\end{equation}
Here we solve Eqs. \ref{eqonemean} and \ref{eqonesigma} in 
the limiting case of small measurement errors ($\sigma_i\sigma$). 
We get:
\begin{equation}
\label{eqonemularge}
\mu=\overline{x}+(\overline{x}-\overline{x}_{err})
\frac{\sigma_{err}^2}{s^2},
\end{equation}
\begin{equation}
\label{eqonesigmalarge}
\sigma^2=s^2-\sigma_{err}^2
+2\sigma_{err}^2\left(1-\frac{s_{err}^2}{s^2}\right),
\end{equation}
where $\overline{x}=(\sum_i x_i/S_i)/(\sum 1/S_i)$ is the selection
weighted mean, $\overline{x}_{err}=(\sum_i x_i\sigma_i^2/S_i)/(\sum_i
\sigma^2_i/S_i)$ is the error and selection weighted mean, $\sigma^2_{err}=
(\Sigma \sigma^2_i/S_i)/(\Sigma 1/S_i)$ is the selection weighted mean
square error, $s^2=(\sum (x-\overline{x})^2/S_i)/(\sum_i 1/S_i)$ the
selection weighted rms and $s^2_{err}=(\sum
\sigma_i^2(x-\overline{x})^2/S_i)/(\sum_i \sigma_i^2/S_i)$ 
the selection and error weighted rms. Therefore the mean and rms are
obtained by correcting the usual formulae with a term taking into
account the spread in errors.

\section{The 2-dimensional Model, limiting case}
\label{twodimlim}

Here we present the  solution which minimizes Eq. \ref{eqlikelihood}
in the limiting case of small errors (with a diagonal error matrix), 
no peculiar velocities and no cuts, considering expansions to first
order in the errors, for the two dimensional case. 
In Appendix \ref{derivation} we derive the  general result in the N
dimensional case.
To simplify the notation in the
following we call the components $(x_1,x_2)$ as $(x,y)$. For the mean
values we find:
\begin{equation}
\label{eqxmeantwo}
\overline{x}=x^{(0)}+\sigma_{ex}^2(V_{22}^{(0)}(x^{(0)}-x^{(1)}_x)-V_{12}^{(0)}(y^{(0)}-y^{(1)}_x))/\Delta^{(0)},
\end{equation}
\begin{equation}
\label{eqymeantwo}
\overline{y}=y^{(0)}+\sigma_{ey}^2(-V_{12}^{(0)}(x^{(0)}-x^{(1)}_y)+V_{11}^{(0)}(y^{(0)}-y^{(1)}_y))/\Delta^{(0)},
\end{equation}
where $x^{(0)}=(\sum_i x_i/S_i)/S$, 
$y^{(0)}=(\sum_i y_i/S_i)/S$, $S=\sum_i 1/S_i$,
$x^{(1)}_x=(\sum_i x_i\sigma_{xi}^2/S_i)/(S\sigma_{ex}^2)$, 
$x^{(1)}_y=(\sum_i x_i\sigma_{yi}^2/S_i)/(S\sigma_{ey}^2)$, 
$\sigma_{ex}^2=(\sum_i \sigma_{xi}^2/S_i)/S$,
$\sigma_{ey}^2=(\sum_i \sigma_{yi}^2/S_i)/S$,
$y^{(1)}_x=(\sum_i y_i\sigma_{xi}^2/S_i)/(S\sigma_{ex}^2)$, 
$y^{(1)}_y=(\sum_i y_i\sigma_{yi}^2/S_i)/(S\sigma_{ey}^2)$, 
$V_{11}^{(0)}=(\sum_i(x_i-x^{(0)})^2/S_i)/S$, 
$V_{22}^{(0)}=(\sum_i(y_i-y^{(0)})^2/S_i)/S$, 
$V_{12}^{(0)}=(\sum_i(x_i-x^{(0)})(y_i-y^{(0)})/S_i)/S$, and
$\Delta^{(0)}=V_{11}^{(0)}V_{22}^{(0)}-V_{12}^{(0)2}$. 
The covariance matrix $V$ is:
\begin{equation}
\label{eqvarxx}
\begin{array}{ccl}
V_{11}&=&V_{11}^{(0)}+V_{11}^{(1)}\\
&=&V_{11}^{(0)}-\sigma_{ex}^2\\
&&+2\sigma_{ex}^2
(1-(V_{22}^{(0)}\sigma^2_{xx,ex}-V_{12}^{(0)}\sigma_{xy,ex})/\Delta^{(0)}),
\end{array}
\end{equation}
\begin{equation}
\label{eqvaryy}
\begin{array}{ccl}
V_{22}&=&V_{22}^{(0)}+V_{22}^{(1)}\\
&=&V_{22}^{(0)}-\sigma_{ey}^2\\
 &&+2\sigma_{ey}^2
(1-(V_{11}^{(0)}\sigma^2_{yy,ey}-V_{12}^{(0)}\sigma_{xy,ey})/\Delta^{(0)}),
\end{array}
\end{equation}
\begin{equation}
\label{eqvarxy}
\begin{array}{ccccl}
V_{12}&=&V_{12}^{(0)}&+&V_{12}^{(1)}\\
&=&V_{12}^{(0)}&-&\sigma_{ex}^2
(V_{22}^{(0)}\sigma_{xy,ex}-V_{12}^{(0)}\sigma_{yy,ex}^2)/\Delta^{(0)}\\
&&&-&\sigma_{ey}^2
(V_{11}^{(0)}\sigma_{xy,ey}-V_{12}^{(0)}\sigma_{xx,ey}^2)/\Delta^{(0)},
\end{array}
\end{equation}
where $\sigma_{xx,ex}^2=(\sum_i (x^i-x^{(0)})^2\sigma_{xi}^2/S_i)/(S\sigma_{ex}^2)$,
$\sigma_{yy,ey}^2=(\sum_i (y_i-y^{(0)})^2\sigma_{yi}^2/S_i)/(S\sigma_{ey}^2)$,
$\sigma_{xy,ex}^2=(\sum_i (x_i-x^{(0)})(y_i-y^{(0)})\sigma_{xi}^2/S_i)/(S\sigma_{ex}^2)$,
$\sigma_{xy,ey}^2=(\sum_i (x_i-x^{(0)})(y_i-y^{(0)})\sigma_{yi}^2/S_i)/(S\sigma_{ey}^2)$.
As in the 1-dim case, we recognise the usual (error corrected)
zeroth-order plus a term taking into account the spread in the
errors. With no spread in the errors we have $V_{12}^{(1)}=0$ and we
recover the {\it ansatz} of Akritas and Bershady (1996). 
To first order Eq. \ref{eqao} then reads:
\begin{equation}
\label{eqafirst}
a_o\approx a_o^{(0)}+a_o^{(1)}= a_o^{(0)}
\left(1+\frac{V_{22}^{(1)}-V_{11}^{(1)}}
{\sqrt{(V_{22}^{(0)}-V_{11}^{(0)})^2+4V_{12}^{(0)2}}}
-\frac{V_{12}^{(1)}}{V_{12}^{(0)}}\frac{V_{22}^{(0)}-V_{11}^{(0)}}
{\sqrt{(V_{22}^{(0)}-V_{11}^{(0)})^2+4V_{12}^{(0)2}}}\right),
\end{equation}
where $a_o^{(0)}$ is Eq. \ref{eqao} evaluated using $V^{(0)}$.
With no spread in the errors:
\begin{equation}
\label{eqasecond} 
a_{+}=a^{(0)}_o\left[1+(\sigma^2_{ex}-\sigma^2_{ey})/
\sqrt{(V_{22}^{(0)}-V_{11}^{(0)})^2+4V_{12}^{(0)2}}\right].
\end{equation}
 Therefore
$a_o^{(0)}$ {\it underestimates} the true slope if the error in the $x$
direction is larger than the one in $y$.  

\section{Derivation of the N-dimensional Case}
\label{derivation}

Let $V$ be the intrinsic covariance matrix and $\Lambda$ its inverse, 
$E_i$ the error matrix of each point, $V_i=V+E_i$ the observed
covariance matrix of each point, with $E_i<<V$. Then we have:
\begin{equation}
\label{eqct}
\Lambda_i=(V+E_i)^{-1}=[V(I+V^{-1}E_i)]^{-1}=(I+\Lambda E_i)^{-1}\Lambda
\sim (I-\Lambda E_i)\Lambda=\Lambda-\Lambda E_i\Lambda,
\end{equation}
where the first order approximation $(I+\epsilon )^{-1}\sim I-\epsilon$
was used. We get:
\begin{equation}
\label{eqdetV}
det V_i=det V det(1+\Lambda E_i)=det V [1+tr(\Lambda E_i)]=
det V [1+\Sigma_{jk} \Lambda_{jk} E_{i,jk}],
\end{equation}
where the first order approximation $det (I+\epsilon)\sim
1+tr(\epsilon)$ was used. Therefore to first order we get:
\begin{equation}
\label{eqdetC}
det \Lambda_i = det \Lambda [1-\Sigma_{jk} \Lambda_{jk} E_{i,jk}].
\end{equation}
The equation for the likelihood reads:
\begin{equation}
\label{eqlikeapprox}
{\cal L}=\Pi_{i=1}^m \left\{ \frac{\sqrt{ det \Lambda}}{(2\pi)^{n/2}}
(1-0.5 \Sigma_{jk} \Lambda_{jk} E_{i,jk})
\exp \left [-\hat{\vec{x}}_i^T(\Lambda-\Lambda E_i \Lambda)\hat{\vec{x}}_i/2
\right ]\right \}^{1/S_i},
\end{equation}
where $\hat{\vec{x}}_i=\vec{x}_i-\overline{\vec{x}}$, $\vec{x}_i$ 
are the n-dimensional m
vectors of datapoints, and $\overline{\vec{x}}$ the vector of the mean values.
Taking the logarithm we get:
\begin{equation}
\label{eqlnlikeapprox}
\ln {\cal L} = \Sigma_{i=1}^m \frac{1}{S_i} \left
[ -\frac{n}{2}\ln(2\pi)+\frac{1}{2}\ln det \Lambda 
-\frac{1}{2}\Sigma_{jk} \Lambda_{jk} E_{i,jk}
-\frac{1}{2}\hat{\vec{x}}_i^T(\Lambda-\Lambda E_i \Lambda)\hat{\vec{x}}_i.
\right ]
\end{equation}
The equation for the vector of the mean values reads:
\begin{equation}
\label{eqmeanval}
\frac{\partial \ln {\cal L}}{\partial\overline{\vec{x}}_k}=\Sigma_i
(\Lambda-\Lambda E_i\Lambda)(\vec{x}_i-\overline{\vec{x}})_k/S_i=0.
\end{equation}
Setting $\overline{\vec{x}}=\vec{x}^{(0)}+\vec{x}^{(1)}$, 
the zero-th order term is:
\begin{equation}
\label{eqzeroth}
\vec{x}^{(0)}=(\Sigma_i \vec{x}_i/S_i)/(\Sigma_i 1/S_i).
\end{equation}
The first order term is then:
\begin{equation}
\label{eqfirst}
-\Sigma_i \Lambda \vec{x}^{(1)}/S_i=\Sigma_i \Lambda E_i 
\Lambda (\vec{x}_i-\vec{x}^{(0)})/S_i,
\end{equation} 
and therefore:
\begin{equation}
\label{eqfirstfin}
\vec{x}^{(1)}=(\Sigma_iE_i\Lambda (\vec{x}^{(0)}-\vec{x}_i)/S_i)/(\Sigma_i 1/S_i),
\end{equation}
which is equivalent to Eqs. \ref{eqxmeantwo}-\ref{eqymeantwo} 
for $E_i$ diagonal and taking the zeroth order for $\Lambda$ (see below).

Let now  consider the derivatives with respect to the $\Lambda$ components:
\begin{equation}
\label{eqdlambda}
\frac{\partial \ln {\cal L}}{\partial \Lambda_{lm}}=
\Sigma_i\frac{1}{S_i}\left \{
\frac{1}{2 det \Lambda}\frac{\partial det \Lambda}{\partial\Lambda_{lm}}
-\frac{1}{2}\Sigma_{jk}E_{i,jk}\frac{\Lambda_{jk}}{\partial \Lambda_{lm}}
-\frac{1}{2}
\hat{\vec{x}}_i^T(\frac{\partial \Lambda}{\partial\Lambda_{lm}})\hat{\vec{x}}_i
+\frac{1}{2}\hat{\vec{x}}_i^T(\frac{\partial \Lambda E_i\Lambda}{\partial
\Lambda_{lm}})\hat{\vec{x}}_i
\right \}=0.
\end{equation}

Let first note that:
\begin{equation}
\label{eqpartialdet}
\frac{1}{det A}\frac{\partial det A}{\partial A_{lm}}=
\frac{(-1)^{l+m}det A_{(l,m)}}{det A}=A^{-1}_{lm},
\end{equation}
where $A_{(l,m)}$ is the matrix obtained eliminating the row $l$ and the
column $m$. Since $\Lambda$ is symmetric we get:
\begin{equation}
\label{eqpardetlm}
\frac{1}{det \Lambda}\frac{\partial det \Lambda}{\partial\Lambda_{lm}}=2V_{lm}
\end{equation}
if $l\ne m$, and:
\begin{equation}
\label{eqpardetll}
\frac{1}{det \Lambda}\frac{\partial det \Lambda}{\partial\Lambda_{ll}}=V_{ll}
\end{equation}
otherwise. Similarly, we get $\partial \Lambda_{jk}/\partial
\Lambda_{lm}=\delta_{jl}\delta_{km}+\delta_{jm}\delta_{kl}$ if $l\ne m$ and 
$\partial \Lambda_{jk}/\partial \Lambda_{ll}=\delta_{jl}\delta_{kl}$. 
Finally, we derive:
\begin{equation}
\label{eqlellm}
\left( \frac{\partial \Lambda E_i \Lambda}{\partial \Lambda_{lm}}\right)_{jk}=
\Sigma_\mu E_{i,l\mu}(\Lambda_{\mu k}\delta_{jm}+\Lambda_{\mu j}\delta_{km})+
\Sigma_\nu E_{i,m\nu}(\Lambda_{\nu k}\delta_{jl}+\Lambda_{\nu j}\delta_{kl}),
\end{equation}
if $l\ne m$, and:
\begin{equation}
\label{eqlelll}
\left( \frac{\partial \Lambda E_i \Lambda}{\partial\Lambda_{ll}}\right)_{jk}=
\Sigma_\mu \delta_{kl}\Lambda_{j\mu}E_{i,\mu l}+\Sigma_\nu
\delta_{jl}E_{i,l\nu}\Lambda_{\nu k},
\end{equation}
otherwise. 
Writing the covariance matrix as $V=V^{(0)}+V^{(1)}$, the zeroth order reads
$V^{(0)}_{jk}=(\sum_i
(\vec{x}_i-\vec{x}^{(0)})_j(\vec{x}_i-\vec{x}^{(0)})_k/ S_i)
/(\Sigma_i 1/S_i)$. 
The first order comes from the terms of Eq. \ref{eqdlambda} linear in $E_i$:
\begin{equation}
\label{eqv1ll}
V^{(1)}_{ll}=E_{ll}-2\Sigma_{j\mu} \sigma_{jl,\mu l}
\Lambda^{(0)}_{j\mu}E_{\mu l},
\end{equation}
\begin{equation}
\label{eqv1lm}
V^{(1)}_{lm}=E_{lm}
-\Sigma_{k\mu} \sigma_{mk,l\mu} \Lambda^{(0)}_{\mu k} E_{\mu l}
-\Sigma_{k\nu} \sigma_{lk,m\nu} \Lambda^{(0)}_{\nu k} E_{\nu m}
\end{equation}
where $E=(\Sigma_i E_i/S_i)/(\Sigma_i 1/S_i)$ and
$\sigma_{jl,\mu\nu}=(\Sigma_i (\vec{x}_i-\vec{x}^{(0)})_j(\vec{x}_i-\vec{x}^{(0)})_l
E_{i,\mu \nu}/S_i)/(\Sigma_i E_{i,\mu \nu}/S_i)$.
Note that $V^{(1)}=-E$ if there is no spread in the errors (i.e.,
$E_i=constant$). 

\section{The photometric error matrix}
\label{photerrmatrix}

The photometric part of the error matrix $E_i$ of Eq. \ref{eqerror} can be
derived as follows. Paper IV shows that the errors in the
half-luminosity radii $\delta r_i$ and average surface brightness
$\delta u_i$ derived from the fits to the photometric profiles are
correlated, with $\delta FP_i=\delta r_i-\alpha \delta u_i$ and
$\alpha\approx 0.3$, and
calibrates the values of $\delta r_i$ and $\delta FP_i$ in terms of
the quality parameter Q. Therefore, we consider the rotated
coordinates $(y,z)=T(\log R_{e,i},\langle S\!B_{e,i}\rangle)$, where
$y=(\log R_{e,i}-\alpha \langle S\!B_{e,i}
\rangle)/\sqrt{1+\alpha^2}$ and 
$z=(\alpha\log R_{e,i}+\langle S\!B_{e,i}
\rangle)/\sqrt{1+\alpha^2}$, and $T$ the corresponding rotation
matrix. In these coordinates the error matrix
$E_{yz}$ is
diagonal, with $E_{yz,11}=\sigma_y^2$ and $E_{yz,22}=\sigma_z^2$. 
From the definition of $y$ it follows $\sigma^2_y=\delta
FP_i^2/(1+\alpha^2)$. From $\log R_{e,i}=(y+\alpha
z)/\sqrt{1+\alpha^2}$
and the fact that the errors on $y$ and $z$ are uncorrelated, we get
$\delta r_i^2=(\sigma_y^2+\alpha^2 \sigma_z^2)/(1+\alpha^2)$ and finally
$\sigma_z^2=((1+\alpha^2)\delta r_i^2-\delta FP_i^2/(1+\alpha^2))/\alpha^2$.
The matrix $T^TE_{yz}T$ gives then:
\begin{equation}
\label{eqe11}
E_{i,11}=\frac{\sigma_y^2+\alpha^2\sigma_z^2}{1+\alpha^2}=\delta r_i^2,
\end{equation}
\begin{equation}
\label{eqe13}
E_{i,13}=E_{i,31}=\frac{\alpha}{1+\alpha^2}(\sigma_z^2-\sigma_y^2)=
\frac{(1+\alpha^2)\delta r_i^2-\delta FP_i^2}{\alpha(1+\alpha^2)},
\end{equation}
\begin{equation}
\label{eq33}
E_{i,33}=\frac{\alpha^2\sigma_y^2+\sigma_z^2}{1+\alpha^2}=
\frac{(\alpha^2-1)\delta FP^2_i+(1+\alpha^2)\delta
r^2}{\alpha^2(1+\alpha^2)}.
\end{equation}

\section{The normalization of the 3-dim distribution function}
\label{norm3dim}

The integral of Eq. \ref{eqnorm} can be performed by defining the set
of rotated coordinates $(w_1,w_2,w_3)=Q(\hat{\vec{x}})$:
\begin{eqnarray}
\label{eqrotw}
w_1 & = & \frac{\hat x_1-b \hat x_3}{\sqrt{1+b^2}},\nonumber \\
w_2 & = & \hat x_2, \\
w_3 & = & \frac{b \hat x_1+\hat x_3}{\sqrt{1+b^2}}.\nonumber \\
\end{eqnarray}
Writing $M=Q^T\Lambda_iQ$, integration over $w_3$ leads to the following integral:
\begin{equation}
\label{eqtwodimint}
f_{3i}=\frac{|W|}{2\pi}\int_{w_{1c}}^\infty\int_{w_{2c}}^\infty
\exp(-\vec{w}^TW\vec{w}/2)dw_1 dw_2,
\end{equation}
where the $2\times 2$ matrix $W$ is
\begin{equation}
\label{eqwmatrix}
W_{i,j}=M_{i,j}-\frac{M_{i,3}M_{j,3}}{M_{3,3}},
\end{equation}
and $w_{1c}=(\FPcut-\overline{FP})/\sqrt{1+\alpha^2}$, where 
$\FPcut=0.78\log D_{Wcut}-0.61$ (Paper III, $\alpha=0.3\approx b$),
 $\overline{FP}=\overline{\log R_e}-\delta_j-\alpha\overline{\langle
SB_e\rangle})$ and $w_{2c}=\log\sigma_{cut} -\overline{\log \sigma}$. 
Finally one
gets:
\begin{equation}
\label{eqlhkrho}
f_{3i}=L(h,k,\rho),
\end{equation}
where $L(h,k,\rho)$ is the bivariate probability integral (Abramovitz
\& Stegun 1971), $h=w_{1c}\sqrt{W_{2,2}(1-\rho^2)}$, 
$k=w_{2c}\sqrt{W_{1,1}(1-\rho^2)}$ and $\rho=-W_{1,2}/\sqrt{W_{1,1}W_{2,2}}$.

\end{document}